\documentclass[12pt]{article}
\usepackage[utf8]{inputenc}
\usepackage{amsmath}
\usepackage{amssymb}
\usepackage{amsthm}
\usepackage[authoryear]{natbib}
\usepackage{booktabs}
\usepackage{graphicx}
\usepackage{epstopdf}
\usepackage{caption}
\usepackage{bm}
\usepackage{bbm}
\usepackage{xcolor}
\usepackage{pdflscape}
\usepackage{longtable}
\usepackage{threeparttablex}
\usepackage{mathrsfs}
\usepackage[english]{babel}
\usepackage{subcaption}
\usepackage[title]{appendix}
\usepackage{float}
\usepackage[]{threeparttable}
\usepackage{multirow}
\usepackage[onehalfspacing]{setspace}
\usepackage[a4paper,left=1.25in,right=1.25in,top=1.25in,lines=32,vcentering]{geometry}

\usepackage[hypertexnames=false]{hyperref}
\definecolor{lightblue}{rgb}{0, 0.4, 0.6}
\hypersetup{
    breaklinks=true,
	colorlinks = true,
	citecolor = {lightblue},
	linkcolor = {lightblue},
	urlcolor = {lightblue}
}

\usepackage[capitalise,nameinlink,noabbrev]{cleveref}
\crefname{equation}{}{}
\crefname{appsec}{Appendix}{Appendices}
\crefname{appsubsec}{Appendix}{Appendices}
\usepackage{autonum} 

\usepackage[inline,shortlabels]{enumitem}
\newlist{assenumerate*}{enumerate*}{4} 
\setlist[assenumerate*]{label=(\roman*), ref=\theassumption.\roman*}
\crefname{assenumerate*i}{Assumption}{Assumptions}

\newlist{assenumerate}{enumerate}{4} 
\setlist[assenumerate]{label=(\roman*), ref=\theassumption.\roman*}
\crefname{assenumeratei}{Assumption}{Assumptions}

\usepackage{tikz}
\usetikzlibrary{shapes}
\usetikzlibrary{decorations.pathreplacing}

\newtheorem{counter}{counter} 
\newtheorem{theorem}[counter]{Theorem}
\crefname{theorem}{Theorem}{Theorems}

\crefname{conjecture}{Conjecture}{Conjectures}
\newtheorem{lemma}[counter]{Lemma}
\crefname{lemma}{Lemma}{Lemmas}

\crefname{definition}{Definition}{Definitions}
\newtheorem{assumption}{Assumption}
\renewcommand{\theassumption}{A\arabic{assumption}} 
\crefname{assumption}{Assumption}{Assumptions}

\crefname{corollary}{Corollary}{Corollaries}
\theoremstyle{definition}

\newcommand{\bs}[1]{\boldsymbol{#1}}

\DeclareMathOperator{\tr}{tr}
\DeclareMathOperator{\E}{E}
\DeclareMathOperator{\var}{var}
\let\Pr\relax
\DeclareMathOperator{\Pr}{P}

\DeclareMathOperator{\corr}{corr}
\DeclareMathOperator{\lambdamax}{\lambda_{\max}}
\DeclareMathOperator{\lambdamin}{\lambda_{\min}}
\DeclareMathOperator{\rank}{rank}

\let\vec\relax
\DeclareMathOperator{\vec}{vec}

\newcommand{\toas}{\overset{a.s.}{\rightarrow}}
\newcommand{\toprob}{\overset{p}{\rightarrow}}
\newcommand{\todist}{\rightsquigarrow}
\newcommand{\R}{\mathbb{R}}
\newcommand{\nmax}{n_{\max}}
\newcommand{\1}{\mathbbm{1}}
\renewcommand{\th}[1][th]{^{\text{#1}}} 
\newcommand{\Normal}{N}

\newcommand{\citepos}[1]{\citeauthor{#1}'s (\citeyear{#1})} 

\title{Inference in clustered IV models with many and weak instruments}

\author{Johannes W. Ligtenberg\thanks{Erasmus University Rotterdam, Burgemeester Oudlaan 50, 3062 PA Rotterdam, The Netherlands.   \texttt{ligtenberg@ese.eur.nl}.\\
I thank Stanislav Anatolyev, Tom Boot, Mikkel S\o lvsten, Tom Wansbeek, Tiemen Woutersen and participants at conferences and seminars for insightful comments and discussions. This paper previously circulated under the title \textit{Inference in IV models with clustered dependence, many instruments and weak identification}. In the preparation of this paper I used the AI tool copilot. I take full responsibility for the content of the paper.}
}

\date{\today}

\begin{document}

\maketitle


\begin{abstract}
    \noindent 
    This paper studies inference in instrumental variable models with many, weak instruments and clustered data. It first examines how clustering affects identification strength, showing how to adapt the concentration parameter to account for within-cluster dependence. The adapted concentration parameter may be smaller and may grow more slowly asymptotically than the concentration parameter that ignores the dependence, thereby weakening identification. Next, the paper extends the identification-robust jackknife Anderson--Rubin and jackknife score statistics for independent data to clustered data by removing clusters rather than individual observations from the statistics. These statistics converge jointly to a normal distribution under many-instrument asymptotics, such that scaling with cluster cross-fit variance estimators yields identification-robust tests. Simulations show that the cluster jackknife tests achieve close to nominal size and have good power. A reanalysis of a study on the effect of queenly reign on war shows the empirical relevance of the new tests.
    \\[1ex]
\end{abstract}

\section{Introduction}
The widespread use of clustered standard errors in instrumental variable (IV) models shows that studies that use IVs to estimate a coefficient on an endogenous regressor often use data that may be dependent within clusters. This paper shows how such dependence affects the identification strength of the model. In particular, the concentration parameter in \citepos{mikusheva2021inference} measure of identification needs to be scaled with the dependence. Consequently, if the within-cluster dependence is unrestricted, the concentration parameter is lower in finite samples and it can grow more slowly asymptotically, making the IV model less strongly identified.

The paper next considers tests for the parameters on the endogenous regressors in possibly weakly identified IV models with clustered data. Identification in IV models depends on the number of instruments and their strength, with particularly low identification if there are many, weak instruments. The tests I therefore consider allow for many instruments in the sense that the number of instruments grows proportionally to sample size \citep{bekker1994alternative} and weak instruments in the sense that their first-stage coefficients may be zero.

Recently, several tests that are robust against such types of many, weak instruments have been proposed \citep{crudu2021inference,mikusheva2021inference,matsushita2020jackknife,dovi2022ridge,lim2024conditional,lim2026dimension}. These tests, however, assume independent observations and therefore cannot be applied to clustered observations. This paper extends the jackknife Anderson--Rubin \citep[AR,][]{anderson1949estimation} test by \citet{crudu2021inference} and \citet{mikusheva2021inference} and the jackknife score test by \citet{matsushita2020jackknife} to clustered data. The jackknife tests remove from the AR and score statistics the terms that have nonzero expectation. For independent data these are the terms that involve squares of observations. For clustered data, due to the dependence between observations, there are additional terms that have nonzero expectation, which can be removed with a cluster jackknife. After rescaling, this yields the cluster jackknife AR and cluster jackknife score statistics. I derive their limiting distribution by generalising the central limit theorem (CLT) in Lemma A2 in \citet{chao2012asymptotic} to clustered data.

A Monte Carlo experiment assesses the performance of the cluster-robust tests. The cluster jackknife tests attain close to nominal size in a range of settings. When there are many instruments, the two tests also have superior power relative to the cluster adaptations of the AR and score \citep{kleibergen2005testing} tests.

To illustrate the practical relevance of the new tests, I revisit \citepos{dube2020queens} study on the effect of having a queen on the probability that a state is at war. The effect is identified using two instruments based on the family composition of European monarchs and is estimated using a data set of 18 polities observed during part of the period 1480 to 1913, resulting in $3\,586$ polity-year observations. The data are clustered in reigns, which yields 176 clusters of varying size. In the appendix \citet{dube2020queens} propose three additional instruments based on interactions of the two instruments with exogenous regressors. The authors note, however, that these interactions are relatively weak instruments. I continue the analysis with the new cluster robust tests and find that having a queen significantly increases the likelihood of an armed conflict.

This paper relates to several recent papers. The first strand of literature studies IV models with many instruments and clustered data. Following an earlier version of this paper, \citet{frandsen2023cluster} used the cluster jackknife to make the jackknife IV estimator \citep[JIVE,][]{phillips1977bias,angrist1999jackknife} handle clustered data. However, \citepos{frandsen2023cluster} cluster JIVE (CJIVE) is not guaranteed to be robust against many and weak instruments, nor is its asymptotic distribution derived.

\citet{chao2023jackknife} do derive the asymptotic distribution of three estimators in an IV model with many instruments and clustered data. However, their model assumes that the clustered dependence is fully captured by a large number of fixed effects. Conditional on the fixed effects, the observations are independent and hence a CLT in the spirit of \citet{chao2012asymptotic} is applicable. This paper assumes a more general form of clustered dependence and therefore requires a further extension of this CLT.

The many and weak instrument robust generalised methods of moments (GMM) AR test by \citet{boot2023identification} allows for clustered data. This test requires invariance of the moment conditions, in the sense that the moment conditions are symmetrically distributed around zero. For IV models this restricts the distribution of the instruments and the second-stage errors. If one is unwilling to make these assumptions, the test cannot be applied. This paper makes no invariance assumption and therefore, within the family of IV models, the resulting tests are more widely applicable.

\citet{kolesar2026cluster} study inference on quadratic forms with clustered data, which is a more general problem than the one this paper studies. In an example they show how their results yield an identification-robust score statistic when applied to an IV model with many instruments and heterogeneous treatment effects. This paper, on the other hand, considers constant treatment effects and therefore an AR statistic can be derived as well \citep{yap2024inference}. Similar to this paper's CLT, the CLT for quadratic forms in \citet{kolesar2026cluster} follows from a martingale difference array CLT \citep[see e.g.][]{hallheyde1980}. Because of the more general setup, the conditions for the theorem are also stated more generally, whereas the conditions in this paper are tailored to the IV model.

The CLTs in the aforementioned papers are all based on a martingale difference array CLT. The CLT for quadratic forms with clustered data in \citet{mikusheva2025estimation}, on the other hand, is based on Stein's method as also applied in \citet{solvsten2020robust}. The resulting requirements for asymptotic normality are, however, largely similar, albeit non-nested. A notable difference is that this paper states the assumptions in terms of maximal cluster size, and in that way as a worst-case bound, whereas \citet{mikusheva2025estimation} state the assumptions in terms of eigenvalues of the covariance matrix of the errors, letting the assumptions depend on the degree of within-cluster dependence.

Finally, a recent literature has examined the need to account for clustering in the data. See, for example, \citet{barrios2012clustering}, \citet{chetverikov2023standard} and \citet{abadie2023should}. This paper abstracts from this question and instead assumes that clustering at the specified level is necessary and adequate.

The remainder of the paper is structured as follows. In \cref{sec:model} I present the linear IV model for clustered data and show the effect of clustering on the identification strength. \Cref{sec:cljar,sec:cljs} discuss the cluster jackknife AR and the cluster jackknife score tests. Variance estimation is discussed in \cref{sec:var}. \Cref{sec:MC,sec:application} proceed with the simulation results and the empirical application. Finally, \cref{sec:conclusion} concludes.

Throughout I use the following notation. $\bs\iota$ is a vector of ones and $\bs e_i$ is $i^{\text{th}}$ unit vector. $\bs A_{(i)}=\bs A\bs e_i$ denotes the $i^{\text{th}}$ column of $\bs A$. I use the projection matrices $\bs P_{A}=\bs A(\bs A'\bs A)^{-1}\bs A'$ and $\bs M_{A}=\bs I-\bs P_{A}$. For $\odot$ the Hadamard product, I write $\bs D_{A}=\bs I\odot\bs A$ for a square matrix $\bs A$. $\dot{\bs A}=\bs A-\bs D_{A}$ is the matrix with the diagonal elements set to zero. For a square matrix $\bs A$, $\lambdamin(\bs A)$ and $\lambdamax(\bs A)$ are the minimum and maximum eigenvalues of $\bs A$. For $\bs A$ not necessarily square $\|\bs A\|_2=\sqrt{\lambdamax(\bs A'\bs A)}$ is the spectral norm. Let $\otimes$ be the Kronecker product and $\vec(\bs A)$ be the column vectorisation of a matrix. For a partition of indices $\{1,\dots,n\}$ into $G$ clusters, let $[g]$ be the indices corresponding to cluster $g$, $g=1,\dots,G$. Denote a union of indices as $[gh]=[g]\cup[h]$. For any $n\times m$ matrix $\bs A$, let $\bs A_{[g]}$ be the $n_g\times m$ matrix with only the rows of $\bs A$ indexed by $[g]$ selected. Also if $m=n$, let $\bs A_{[g,h]}$ be the $n_g\times n_h$ matrix with only the columns of $\bs A_{[g]}$ indexed by $[h]$ selected. Furthermore, when the indices are ordered by cluster, that is, when each cluster only contains consecutive indices, then for an $n\times n$ matrix $\bs A$, $\bs B_{A}$ is the $n\times n$ block diagonal matrix with the blocks $\bs A_{[g,g]}$ on its diagonal. Also, denote $\ddot{\bs A}=\bs A-\bs B_{A}$ for the matrix with the blocks on its diagonal set to zero. I denote convergence in distribution by $\rightsquigarrow$, convergence in probability by $\toprob$ and almost sure (a.s.) convergence by $\toas$. $\sum_{g\neq h}$ can denote the double sum $\sum_{g=1}^G\sum_{h=1,h\neq g}^H$, and similarly for other types of sums. Whether this is the case and what the maxima of the sums are, will be clear from the context. Finally, $C$ denotes a finite positive scalar that is not necessarily the same in each appearance. 

\section{Model}\label{sec:model}
Consider the linear IV model
\begin{equation}\label{eq:model individual}
    \begin{split}
        y_i&=\bs X_i'\bs\beta+\varepsilon_i\\
        \bs X_i&=\bs\Pi'\bs Z_i+\bs\eta_i.
    \end{split}
\end{equation}
Here $y_i$ is a scalar outcome, $\bs X_i\in\R^{p}$ is a vector of endogenous variables, $\bs Z_i\in\R^k$ is a vector of IVs and $\varepsilon_i\in\R$ and $\bs\eta_i\in\R^p$ are the second and first stage errors, $i=1,\dots,n$. Assume that the exogenous covariates are small in number and have been partialled out. Stack the observations in $\bs y=(y_1,\dots,y_n)'$, and similarly for the other variables. The model in \cref{eq:model individual} then becomes
\begin{equation}\label{eq:model stacked}
    \begin{split}
        \bs y&=\bs X\bs\beta+\bs\varepsilon\\
        \bs X&=\bs Z\bs\Pi+\bs\eta.
    \end{split}
\end{equation}

The data are divided into $G$ clusters with sizes $n_g$, $g=1,\dots, G$, that are not necessarily equal. Assume that the observations have been ordered per cluster. The following assumption formalises the cluster structure. It requires the observations between clusters to be independent, but allows for dependence of observations within clusters.

\begin{assumption}\label{ass:model}
Conditional on $\bs Z$, $\{\bs\varepsilon_{[g]},\bs\eta_{[g]}\}_{g=1}^G$ is independent with mean zero, $\E(\bs\varepsilon_{[g]}\bs\varepsilon_{[h]}'|\bs Z)\allowbreak=\bs\Sigma_{g}$, $\E(\bs\eta_{(i),[g]}\bs\eta_{(j),[h]}'|\bs Z)=\bs\Omega^{(i,j)}_g$, and $\E(\bs\eta_{(i),[h]}\bs\varepsilon_{[g]}'|\bs Z)=\bs\Xi^{(i)}_g$ if $g=h$ and zero otherwise, $g,h=1,\dots, G$, $i,j=1,\dots p$.
\end{assumption}

It is the dependence between observations within a cluster that affects the identification in IV models. To build some intuition for this result: a clustered sample generally contains less information than an independent sample of the same size. In the extreme case that there is perfect dependence within clusters, one effectively has only $G$ observations instead of $n$. Thus one could say clustering decreases the effective sample size down from the number of observations towards the number of clusters. 

For independent data, many instruments are modelled as the number of instruments being non-negligible relative to sample size. The reduced effective sample size can imply that a number of instruments that is negligible relative to the sample size is non-negligible relative to the effective sample size and hence that there is a many-instrument problem in a clustered sample, but not in an independent sample. Therefore, if there is strong within-cluster dependence, the number of instruments relative to the number of clusters may give a give a better indication of whether there are many instruments, than the number of instruments relative to the number of observations.

Similarly, for independent data weak instruments are modelled as the first stage coefficient $\bs\Pi$ in \cref{eq:model individual} shrinking to zero at the rate $1/\sqrt{n}$. Accordingly, the reduced effective sample size means that in clustered samples in which $\bs\Pi$ shrinks more slowly than $1/\sqrt{n}$, the instruments can nevertheless be weak.

This intuition that within-cluster dependence changes how identification in IV models should be measured can be formalised by generalising Theorem 1 in \citet{mikusheva2021inference}. Their theorem states that for model \cref{eq:model stacked} with a single endogenous regressor and where the errors are independent and identically normally distributed with known covariance matrix, there exists no consistent test for $\beta$ as long as the scaled concentration parameter $\mu_{\text{ind}}^2/\sqrt{k}=\bs\Pi'\bs Z'\bs Z\bs\Pi/(\sigma_{\eta}^2\sqrt{k})$ stays bounded, where $\sigma_{\eta}^2=\var(\eta_i|\bs Z)$. Hence, even in the relatively favourable setting of the theorem, the concentration parameter needs to grow at least with rate $\sqrt{k}$ to make consistent testing possible. Consistent testing in more general models requires weakly higher growth rates.

The theorem below considers a setup similar to \citet{mikusheva2021inference}, but with a specific type of clustering.

\begin{theorem}[Theorem 1 of \citet{mikusheva2021inference} with a clustered sample]\label{thm:MS2021 T1}
    Assume we have a sample from \cref{eq:model stacked} with a single endogenous regressor $p=1$. Consider the reduced-form errors $(\bs u_{[g]},\bs\eta_{[g]})$ where $\bs u_{[g]}=\bs y_{[g]}-\beta\bs\pi'\bs Z_{[g]}$. Assume the reduced-form errors are independently drawn from a Gaussian distribution $(\bs u_{[g]}',\bs\eta_{[g]})'\sim\Normal(\bs0,\bs\Omega\otimes\bs V_{g})$ with a known nonsingular covariance matrix $\bs\Omega\otimes\bs V_{g}$ where $\bs\Omega$ is $2\times 2$ and constant over clusters and $\bs V_{g}$ is $n_g\times n_g$ and can vary over clusters. Assume that the $k\times k$ matrix $\bs Z'\bs Z$ has rank $k$ and $k\to\infty$ as $n\to\infty$. For any sample of size $n$ let $\Psi_n$ be the class of all tests of size $\alpha$ for testing the hypothesis $H_0\colon\beta=\beta_0$, that is any $\psi\in\Psi_{n}$ is a measurable function from $\{(y_i,X_i,\bs Z_i),i=1,\dots,n\}$ to the interval $[0,1]$ such that $\E_{\beta_0,\Pi}(\psi)\leq\alpha$ for any value of $\bs\Pi\in\R^k$. Then for any $\beta^*\neq\beta_0$ we have
    \begin{equation}
        \begin{split}
            \lim\sup_{n\to\infty}\max_{\psi\in\Psi_{n}}\bigg(\min_{\Pi:\frac{\Pi'Z'V^{-1}Z\Pi}{\sigma_{\eta}^2\sqrt{k}}=C}\E_{\beta^*,\Pi}(\psi|\bs Z)\bigg)<1,
        \end{split}
    \end{equation}
    where $\sigma_{\eta}^2$ is the bottom right entry of $\bs\Omega$ and $\bs V$ is the block diagonal matrix with the $\bs V_{g}$, $g=1,\dots, G$, on its diagonal.
\end{theorem}

\begin{proof}
The proof is given in \cref{sec:proof MS2021 T1}
\end{proof}

The theorem shows that in a sample with this specific type of clustering, the concentration parameter that needs to grow faster than $\sqrt{k}$ for consistent testing to be possible is $\mu^2_{\text{cl}}=\bs\Pi'\bs Z'\bs V^{-1}\bs Z\bs\Pi/\sigma_{\eta}^2$. Again, for models with more general clustering, the required growth rate of the concentration parameter is weakly higher. 

The result of the theorem is an asymptotic impossibility result. There are different ways of translating this result to practical implications. A first way is to take a finite sample view and to say that a model is more strongly identified when its concentration parameter is larger. Taking this view, the simple example below shows that when moving from an independent sample to a clustered sample with unrestricted dependence, one can, except in an exceptional case, always find a within-cluster dependence structure such that the concentration parameter decreases. In that sense an IV model with clustered data is more weakly identified than one with independent data.

Consider the setting of the theorem with first-stage errors that have equal variance, $\sigma_{\eta}^2$,  and with constant correlation within clusters, $\rho_{g}$. That is, $\bs V_{g}=(1-\rho_g)\bs I_{n_g}+\rho_g\bs\iota_{n_g}\bs\iota_{n_g}'$. Note that then $\bs V^{-1}_{g}=\bs P_{\iota_{n_g}}/\big(1+(n_{g}-1)\rho_g\big)+\bs M_{\iota_{n_g}}/(1-\rho_g)$. Decompose the first stage signal in two orthogonal components as $\bs Z_{[g]}\bs\Pi=\bs P_{\iota_{n_g}}\bs Z_{[g]}\bs\Pi+\bs M_{\iota_{n_g}} \bs Z_{[g]}\bs\Pi$. The first component captures the signal in the cluster mean direction, whereas the second component captures the signal in the other directions. Denote $\bs\pi_{g}=\bs Z_{[g]}\bs\Pi/s_g$, where $s_g=\|\bs Z_{[g]}\bs\Pi\|$. Then the concentration parameters for independent data and clustered data can be written as
\begin{equation}\label{eq:concentration parameters}
    \begin{split}
        \sigma_{\eta}^2\mu_{\text{ind}}^2
        &=\sum_{g=1}^Gs_g^2(\bs\pi'\bs P_{\iota_{n_g}}\bs\pi+\bs\pi'\bs M_{\iota_{n_g}}\bs\pi)\\
        \sigma_{\eta}^2\mu_{\text{cl}}^2&=\sum_{g=1}^Gs_g^2(\bs\pi'\bs P_{\iota_{n_g}}\bs V_{g}^{-1}\bs P_{\iota_{n_g}}\bs\pi+2\bs\pi'\bs P_{\iota_{n_g}}\bs V_{g}^{-1}\bs M_{\iota_{n_g}}\bs\pi+\bs\pi'\bs M_{\iota_{n_g}}\bs V_{g}^{-1}\bs M_{\iota_{n_g}}\bs\pi)\\
        &=\sum_{g=1}^Gs_g^2\bigg(\frac{\bs\pi'\bs P_{\iota_{n_g}}\bs\pi}{1+(n_g-1)\rho_{g}}+\frac{\bs\pi'\bs M_{\iota_{n_g}}\bs\pi}{1-\rho_{g}}\bigg).
    \end{split}
\end{equation}
A sufficient condition for $\mu_{\text{cl}}^2<\mu_{\text{ind}}^2$ is $(\bs\pi'\bs P_{\iota_{n_g}}\bs\pi)/\big(1+(n_g-1)\rho_{g}\big)+(\bs\pi'\bs M_{\iota_{n_g}}\bs\pi)/(1-\rho_{g})<\bs\pi'\bs P_{\iota_{n_g}}\bs\pi+\bs\pi'\bs M_{\iota_{n_g}}\bs\pi$ for all $g$, which reduces to $\rho_{g}(\rho_{g}-r_g)<0$ where $r_g=\big((n_g-1)\bs\pi'\bs P_{\iota_{n_g}}\bs\pi-\bs\pi'\bs M_{\iota_{n_g}}\bs\pi\big)/\big((n_g-1)\bs\pi'\bs P_{\iota_{n_g}}\bs\pi+\bs\pi'\bs M_{\iota_{n_g}}\bs\pi\big)$ and thus that $\rho_g$ and $r_g$ must have the same sign and $|\rho_g|<|r_g|$.

Consider first the case that $(n_g-1)\bs\pi'\bs P_{\iota_{n_g}}\bs\pi>\bs\pi'\bs M_{\iota_{n_g}}\bs\pi$. Then $r_g>0$, such that for $0<\rho_g<h_g$ it holds that $\mu_{\text{cl}}^2<\mu_{\text{ind}}^2$. The interpretation is that in this case the signal in the cluster mean direction, $(n_g-1)\bs\pi'\bs P_{\iota_{n_g}}\bs\pi$, is larger than the signal in the other directions, $\bs\pi'\bs M_{\iota_{n_g}}\bs\pi$. Positive within-cluster correlation makes it more difficult to distinguish the signal in the cluster mean direction from the errors. Since this is, in this case, the dominant part of the signal, the concentration parameter decreases.

The opposite holds in the case that $(n_g-1)\bs\pi'\bs P_{\iota_{n_g}}\bs\pi<\bs\pi'\bs M_{\iota_{n_g}}\bs\pi$ for which $h_g<0$. Then the signal in the cluster mean direction is smaller than in the other directions, such that a negative correlation makes it harder to distinguish the dominant part of the signal from the errors. Therefore, in this case $\mu_{\text{cl}}^2<\mu_{\text{ind}}^2$ for $h_g<\rho_g<0$.

When $(n_g-1)\bs\pi'\bs P_{\iota_{n_g}}\bs\pi=\bs\pi'\bs M_{\iota_{n_g}}\bs\pi$, the effects of increasing or decreasing $\rho_g$ offset each other, such that $\mu_{\text{cl}}^2$ cannot be smaller than $\mu_{\text{ind}}^2$. However, by setting $\rho_g=0$ for all $g$, we recover the independent case and thus $\mu_{\text{cl}}^2=\mu_{\text{ind}}^2$. Therefore, the concentration parameter weakly decreases and in most cases strictly decreases when going from independent observations to clustered data.

A second way of translating the asymptotic impossibility result from \cref{thm:MS2021 T1} to practical implications, is to take a more asymptotic view and analyse what the concentration parameters do when the sample grows. From \cref{eq:concentration parameters} it is clear that for $\rho_g>0$ the concentration parameter in clustered data generating processes (DGP) is smaller for larger values of $n_g$. Consequently, if $n_g$ grows with the sample size, the clustered concentration parameter grows more slowly than the independent concentration parameter. Depending on the growth rates, $\mu_{\text{ind}}^2/\sqrt{k}$ could therefore diverge, whereas $\mu_{\text{cl}}^2/\sqrt{k}$ stays bounded. In that case identification is strong in independent samples and weak in clustered samples.

The next sections propose two tests that are robust against many and weak instruments and that can handle clustered data. To be robust against weak or even irrelevant instruments, I consider identification robust tests that test the null hypothesis $H_0\colon\bs\beta=\bs\beta_0$ against the alternative $H_1\colon\bs\beta\neq\bs\beta_0$. To be robust against many instruments, I study their behaviour under many instrument asymptotics, in which the number of instruments diverges and can be proportional to the sample size.

\section{Cluster jackknife AR}\label{sec:cljar}
The first test I propose that is robust against many and weak instruments and clustered data is the cluster jackknife AR test. To explain its construction, it is useful to start with the original AR statistic.

For a given value of $\bs\beta^*$ that is not necessarily equal to $\bs\beta$, let $\bs\varepsilon(\bs\beta^*)=\bs y-\bs X\bs\beta^*$. Then, under certain conditions, the original AR statistic for the linear IV model converges as
\begin{equation}
    \begin{split}
        AR(\bs\beta_0)=\frac{1}{n}\bs\varepsilon(\bs\beta_0)'\bs Z\bs V_{AR}^{-1}(\bs\beta_0)\bs Z'\bs\varepsilon(\bs\beta_0)\rightsquigarrow\chi^2_k,
    \end{split}
\end{equation}
where $\bs V_{AR}(\bs\beta_0)=\var\big(\bs Z'\bs\varepsilon(\bs\beta_0)/\sqrt{n}|\bs Z\big)$ captures the variance of the moment conditions.

The original AR statistic is robust against weak instruments, but it has some problems when the number of instruments is large. For homoskedastic data and if the number of instruments is a non-negligible fraction of the sample size, the test based on the feasible version of this statistic has low power \citep{kleibergen2002pivotal} and can be oversized \citep{anatolyev2011specification}. Furthermore, if in addition to many instruments, the data are heteroskedastic, then the estimation of the covariance matrix, $\bs V_{AR}(\bs\beta_0)$, is complex due to the large number of variances and covariances. These estimation problems can add to the size and power problems. 

\citet{crudu2021inference} and \citet{mikusheva2021inference} therefore adapt the AR statistic in two ways to make it suited for many instruments and heteroskedasticity when the data are independent. First, they substitute $\bs V_{AR}(\bs\beta_0)$ by the homoskedasticity-inspired weighting matrix $\bs Z'\bs Z$. This matrix does not depend on the variances and covariances of the errors and is therefore easier to handle in the asymptotic approximations.

Second, \citet{crudu2021inference} and \citet{mikusheva2021inference} note that the $\chi^2_k$ distribution has many degrees of freedom in case there are many instruments. This makes the distribution spread out and causes low power. The $\chi^2_k$ distribution, however, can be changed to a normal distribution if the statistic is properly centred at zero. To do so the terms in the AR statistic that involve squares of the errors, and therefore have nonzero expectation, need to be removed. This is done efficiently by jackknifing, which boils down to setting the diagonal elements of the matrix that weighs the errors, $\bs P_{Z}=\bs Z(\bs Z'\bs Z)^{-1}\bs Z'$, to zero. 

After rescaling and under certain assumptions, \citet{crudu2021inference} and \citet{mikusheva2021inference} then show that \citepos{chao2012asymptotic} central limit theorem can be applied to their jackknife AR statistic, such that it converges as
\begin{equation}
    \begin{split}
        \frac{AR_{J}(\bs\beta_0)}{\sqrt{V_{J}(\bs\beta_0)}}=\frac{1}{\sqrt{V_{J}(\bs\beta_0)k}}\bs\varepsilon(\bs\beta_0)'\dot{\bs P}_{Z}\bs \varepsilon(\bs\beta_0)\rightsquigarrow N(0,1),
    \end{split}
\end{equation}
where $V_{J}(\bs\beta_0)=\var\big(\bs\varepsilon(\bs\beta_0)'\dot{\bs P}_{Z}\bs \varepsilon(\bs\beta_0)/\sqrt{k}|\bs Z\big)$.

To extend the jackknife AR test to clustered data, note that when the data are clustered, not only the terms involving squared errors but also all terms involving products of errors belonging to the same cluster have nonzero expectation. Hence, to centre the AR statistic it is no longer sufficient to remove only the squared errors. Rather, all products of errors within a cluster need to be removed, which is done by substituting $\dot{\bs P}_{Z}$ by the cluster jackknife projection matrix $\ddot{\bs P}_{Z}$. Then $\bs\varepsilon(\bs\beta_0)'\ddot{\bs P}_{Z}\bs\varepsilon(\bs\beta_0)/k$ is centred at zero. I need the following assumption to derive its distribution.

\begin{assumption}\label{ass:cl jackknife AR}
    Conditional on $\bs Z$ and with probability one for all $n$ large enough, it holds that
    \begin{assenumerate*}
        \item\label{assit:cljar rank} $\rank(\bs P_{Z})=k$;
        \item\label{assit:cljar bound P} $\|\bs P_{Z,[g,g]}\|_2^2\leq C<1$ for all $g$;
        \item\label{assit:cljar many iv} $k\to\infty$ as $G\to\infty$;
        \item\label{assit:cljar nmax} $\nmax^4/k\toas 0$ for $\nmax=\max_{g=1,\dots,G}n_g$;
        
        \item\label{assit:cljar eigval} $0<\allowbreak 1/C\leq\allowbreak\lambdamin(\bs\Sigma_{g})\leq\allowbreak\lambdamax(\bs\Sigma_{g})\leq\allowbreak n_gC$
    a.s. for all $g$;
    \item\label{assit:cljar moment} $\E(\varepsilon_i^4|\bs Z)\leq C<\infty$ a.s. for all $i$. 
    \end{assenumerate*}
\end{assumption}

\Cref{assit:cljar rank} requires $\bs Z$ to have full rank. \Cref{assit:cljar bound P} generalises \citepos{crudu2021inference} and \citepos{mikusheva2021inference} requirement that the diagonal elements of $\bs P_Z$ are less than one, which is a common assumption in the many instrument literature. It ensures that the quadratic form of the errors with respect to the projection matrix does not become degenerate after deleting the block diagonal. Furthermore, together with the other assumptions, it prevents one cluster from dominating the others and is in that sense related to the measure of cluster leverage in the first stage regression by \citet{mackinnon2022leverage}. \Cref{assit:cljar many iv} specifies the asymptotic sampling scheme. The condition on $k$ requires the number of instruments and therefore the rank of the projection matrix to increase. \citet{mikusheva2025estimation} and \citet{kolesar2026cluster} similarly require the rank of the matrix in the quadratic form to increase, but since they work with more general matrices than projection matrices the condition cannot be reduced to a condition on the number of instruments. Also note that \cref{assit:cljar many iv} allows for, but does not require the number of instruments to increase proportionally to the sample size. The fourth item of \cref{ass:cl jackknife AR} bounds the relative cluster sizes and ensures that there is not one cluster that dominates the sample. Such a requirement is common in the literature on clustered data. See for example \citet{djogbenou2019asymptotic} and \citet{hansen2019asymptotic}. \citet{mikusheva2025estimation} and \citet{kolesar2026cluster} also impose bounds on the contribution of each cluster to the statistic and its variance. Here, the bounds are worst-case bounds and stated in terms of maximum cluster sizes, whereas \citet{mikusheva2025estimation} and \citet{kolesar2026cluster} allow for a trade-off between within-cluster dependence and maximum cluster size by stating the assumption in terms of eigenvalues of $\bs\Sigma$. When there is strong dependence within clusters, the rates they impose on the maximum cluster sizes match the one here. Finally, \cref{assit:cljar eigval,assit:cljar moment} are regularity conditions that require the errors to have uniformly non-singular covariance matrices with maximum eigenvalues that do not grow faster than the cluster size and require the errors to have uniform finite fourth moments. Under these assumptions, the following result holds.

\begin{theorem}\label{thm:cl jackknife AR}
    Under \cref{ass:model,ass:cl jackknife AR} the cluster jackknife AR converges as
    \begin{equation}\label{eq:cljar}
        \begin{split}
            \frac{AR_{CLJ}(\bs\beta_0)}{\sqrt{V^{AR}_{CLJ}}}=\frac{1}{\sqrt{V_{CLJ}^{AR}k}}\bs\varepsilon(\bs\beta_0)'\ddot{\bs P}_{Z}\bs \varepsilon(\bs\beta_0)\rightsquigarrow N(0,1),
        \end{split}
    \end{equation}
    with conditional variance 
    \begin{equation}\label{eq:V clj AR}
        \begin{split}
            V_{CLJ}^{AR}=\var\bigg(\frac{1}{\sqrt{k}}\bs\varepsilon'\ddot{\bs P}_{Z}\bs\varepsilon\bigg|\bs Z\bigg)=\frac{2}{k}\sum_{g\neq h}\tr(\bs\Sigma_{g}\bs P_{Z,[g,h]}\bs\Sigma_{h}\bs P_{Z,[h,g]}).
        \end{split}
    \end{equation}
\end{theorem}

\begin{proof}
    The result can be shown by adapting the proof of \cref{thm:clj AR S} below.
\end{proof}

\section{Cluster jackknife score}\label{sec:cljs}
The original AR test can have low power in overidentified IV models even when the number of IVs remains fixed. \citet{kleibergen2002pivotal} therefore proposes a score test, which, similarly to the AR test, is robust against weak instruments, but is more powerful. \citet{matsushita2020jackknife} show that jackknifing can also be used to make the score test robust against many instruments and heteroskedasticity. Similarly, the cluster jackknife from the previous section can be used to derive a cluster jackknife score test. For this I require the following assumption.
\begin{assumption}\label{ass:cljs}
Conditional on $\bs Z$ and with probability one for all $n$ large enough,
\begin{assenumerate}
    \item\label{assit:cljs eigval} For any $\bs v\in\R^p$ such that $\bs v'\bs v=1$ and 
    \begin{equation}
        \begin{split}
            \bs M_g(\bs v)=\E\bigg(\begin{bmatrix}
                \bs\varepsilon_{[g]}\bs\varepsilon_{[g]}' & \bs\varepsilon_{[g]}\bs v'\bs\eta_{[g]}' \\
                \bs\eta_{[g]}\bs v\bs\varepsilon_{[g]}' & \bs\eta_{[g]}\bs v\bs v'\bs\eta_{[g]}' 
            \end{bmatrix}\bigg|\bs Z\bigg),
        \end{split}
    \end{equation}
    it holds that $0<1/C\leq \lambdamin\big(\bs M_g(\bs v)\big)\leq\lambdamax\big(\bs M_g(\bs v)\big)\leq n_gC$;

    \item\label{assit:cljs moment} $\E(\|\bs\eta_i\|^4|\bs Z)\leq C<\infty$;

    \item\label{assit:cljs Z} $\lambdamax(\bs\Pi'\bs Z_{[g]}'\bs Z_{[g]}\bs\Pi/n_g)\leq C<\infty$.
\end{assenumerate}
\end{assumption}

\cref{assit:cljs eigval} is a stronger version of \cref{assit:cljar eigval}. It requires the variances to be positive and to not grow faster than the cluster sizes. Furthermore, it ensures that the linear combination of the first stage errors picked by the score statistic defined below and this linear combination together with the second stage errors are nondegenerate. \Cref{assit:cljs moment} requires finite fourth moments of the first stage errors. \Cref{assit:cljs Z} ensures convergence of the scaled instruments.


Then, define the cluster jackknife score statistic as
\begin{equation}
    \begin{split}
        \bs S_{CLJ}(\bs\beta_0)=\frac{1}{\sqrt{n}}\bs X'\ddot{\bs P}_{Z}\bs\varepsilon(\bs\beta_0).
    \end{split}
\end{equation}
Write its conditional variance under the null hypothesis as
\begin{equation}\label{eq:V clj S}
    \begin{split}
        \bs V^{S}_{CLJ}&=\frac{1}{n}\bigg[\bs\Pi'\bs Z'\ddot{\bs P}_{Z}\bs\Sigma\ddot{\bs P}_{Z}\bs Z\bs\Pi+\frac{1}{2}\E\bigg(\sum_{g\neq h}\bs\eta_{[g]}'\bs P_{Z,[g,h]}\bs\varepsilon_{[h]}\bs\varepsilon_{[h]}'\bs P_{Z,[h,g]}\bs\eta_{[g]}\\
        &\quad+2\bs\eta_{[h]}'\bs P_{Z,[h,g]}\bs\varepsilon_{[g]}\bs\varepsilon_{[g]}'\bs P_{Z,[g,h]}\bs\eta_{[h]}
        +\bs\eta_{[h]}'\bs P_{Z,[h,g]}\bs\varepsilon_{[g]}\bs\varepsilon_{[h]}'\bs P_{Z,[h,g]}\bs\eta_{[g]}\bigg|\bs Z\bigg)\bigg],
    \end{split}
\end{equation}
where $\bs\Sigma=\E(\bs\varepsilon\bs\varepsilon'|\bs Z)$. Also, write the covariance between the cluster jackknife AR statistic and the cluster jackknife score statistic as
\begin{equation}\label{eq:C clj}
    \begin{split}
        \bs C_{CLJ}&=\frac{2}{\sqrt{nk}}\E\bigg(\sum_{g\neq h}\bs\eta_{[g]}'\bs P_{Z,[g,h]}\bs\varepsilon_{[h]}\bs\varepsilon_{[g]}'\bs P_{Z,[g,h]}\bs\varepsilon_{[h]}\bigg|\bs Z\bigg),
    \end{split}
\end{equation}
with $i^{\text{th}}$ element $C_{CLJ,i}=2\sum_{g\neq h}\tr(\bs\Xi_{g}^{(i)\prime}\bs P_{Z,[g,h]}\bs\Sigma_{h}\bs P_{Z,[h,g]})/\sqrt{nk}$ and gather the variances and covariances in
\begin{equation}
    \begin{split}
        \bs V_{CLJ}=\begin{bmatrix}
            V_{CLJ}^{AR}& \bs C_{CLJ}' \\
            \bs C_{CLJ} & \bs V_{CLJ}^S
        \end{bmatrix}.
    \end{split}
\end{equation}

Then the joint limiting distribution of the cluster jackknife AR and cluster jackknife score is given by the following theorem.
\begin{theorem}\label{thm:clj AR S}
    Under \cref{ass:model,ass:cl jackknife AR,ass:cljs} and if the cluster jackknife AR and cluster jackknife score are not perfectly correlated, then
    \begin{equation}
        \begin{split}
            \bs V_{CLJ}^{-1/2}\begin{bmatrix}
                AR_{CLJ}(\bs\beta_0)\\
                \bs S_{CLJ}(\bs\beta_0)
            \end{bmatrix}\todist N(\bs 0,\bs I).
        \end{split}
    \end{equation}
\end{theorem}
\begin{proof}
    The proof is given in \cref{sec:proof clj AR S}
\end{proof}

\section{Variance estimation}\label{sec:var}
The result from the previous section requires a variance estimator to be used for testing. The most straightforward approach is to estimate $\bs\Sigma_{g}$ by $\bs\varepsilon(\bs\beta_0)_{[g]}\bs\varepsilon(\bs\beta_0)_{[g]}$, $\bs\Xi_{g}^{(i)}$ by $\bs X_{(i),[g]}\bs\varepsilon_{[g]}$ and $\bs Z_{[g]}\bs\Pi_{(j)}\bs\Pi_{(i)}'\bs Z_{[g]}'+\bs\Omega^{(i,j)}$ by $\bs X_{(i),[g]}\bs X_{(j),[g]}$. This yields
\begin{equation}\label{eq:Vhat clj}
    \begin{split}
        \hat{V}_{CLJ}^{AR}(\bs\beta_0)&=\frac{2}{k}\sum_{g\neq h}\bs\varepsilon(\bs\beta_0)_{[g]}'\bs P_{Z,[g,h]}\bs\varepsilon(\bs\beta_0)_{[h]}\bs\varepsilon(\bs\beta_0)_{[h]}'\bs P_{Z,[h,g]}\bs\varepsilon(\bs\beta_0)_{[g]};\\
        \hat{\bs V}^{S}_{CLJ}(\bs\beta_0)&=\frac{1}{n}\big(\bs X'\ddot{\bs P}_{Z}\bs B_{\varepsilon(\beta_0)\varepsilon(\beta_0)'}\ddot{\bs P}_{Z}\bs X+\sum_{g\neq h}\bs X_{[g]}'\bs P_{Z,[g,h]}\bs\varepsilon(\bs\beta_0)_{[h]}\bs\varepsilon(\bs\beta_0)_{[g]}'\bs P_{Z,[g,h]}\bs X_{[h]}\big);\\
        \hat{\bs C}_{CLJ}(\bs\beta_0)&=\frac{2}{\sqrt{nk}}\sum_{g\neq h}\bs X_{[g]}'\bs P_{Z,[g,h]}\bs\varepsilon(\bs\beta_0)_{[h]}\bs\varepsilon(\bs\beta_0)_{[g]}'\bs P_{Z,[g,h]}\bs\varepsilon(\bs\beta_0)_{[h]}.
    \end{split}
\end{equation}
\Cref{thm:V cljar} in \cref{app:proof V cljar CF} shows that these are unbiased and consistent estimators of $\bs V^{CLJ}$ under \cref{ass:model,ass:cl jackknife AR,ass:cljs}. 

However, \citet{mikusheva2021inference} note that the jackknife AR test for independent data by \citet{crudu2021inference} that uses the equivalent of $\hat{V}_{CLJ}^{AR}(\bs\beta_0)$ has low power against alternatives far from $\bs\beta$, due to a bias in the residuals that makes the variance estimate unnecessarily large. Therefore \citet{mikusheva2021inference} propose an alternative estimator that removes this bias by using cross-fitted residuals to improve the power \citep{newey2018cross,kline2020leave}.

Cross-fit variance estimators for clustered data can be obtained by defining $\tilde{\bs P}(g,h)=\bs Z(\bs Z'_{-[gh]}\bs Z_{-[gh])})^{-1}\bs Z'_{-[gh]}$, where a matrix indexed by $-[gh]$ has all rows corresponding to $[gh]=[g]\cup[h]$ set to zero. Pre-multiplying $\bs \varepsilon(\bs\beta_0)_{-[gh]}$ with $\tilde{\bs P}(g,h)$ therefore gives the leave-cluster-$(g,h)$-out fitted values of $\bs\varepsilon(\bs\beta_0)$ on $\bs Z$. Denote this by $\tilde{\bs\varepsilon}(\bs\beta_0;g,h)=\tilde{\bs P}(g,h)\bs\varepsilon(\bs\beta_0)_{-[gh]}$. The cross-fit variance estimators can then be written as
\begin{equation}\label{eq:Vhat cljcf}
    \begin{split}
        &\hat{V}_{CF}^{AR}(\bs\beta_0)\\
        &=\frac{2}{k}\sum_{g\neq h}\big(\bs\varepsilon(\bs\beta_0)-\tilde{\bs\varepsilon}(\bs\beta_0;g,h)\big)_{[g]}'\bs P_{Z,[g,h]}\bs\varepsilon(\bs\beta_0)_{[h]}\big(\bs\varepsilon(\bs\beta_0)-\tilde{\bs\varepsilon}(\bs\beta_0;g,h)\big)_{[h]}'\bs P_{Z,[h,g]}\bs\varepsilon(\bs\beta_0)_{[g]};\\
        &\hat{\bs V}^{S}_{CF}(\bs\beta_0)=\frac{1}{n}\sum_{\substack{g,h,j=1\\g\neq h\neq j}}^G\bs X_{[g]}'\bs P_{Z,[g,h]}\big(\bs\varepsilon(\bs\beta_0)-\tilde{\bs\varepsilon}(\bs\beta_0;h)\big)_{[h]}\bs\varepsilon(\bs\beta_0)_{[h]}(\bs\beta_0)'\bs P_{Z,[h,j]}\bs X_{[j]}\\
        &\quad+\sum_{g\neq h}\bs X_{[g]}'\bs P_{Z,[g,h]}\bs\varepsilon(\bs\beta_0)_{[h]}\big(\bs\varepsilon(\bs\beta_0)-\tilde{\bs\varepsilon}(\bs\beta_0;g,h)\big)_{[g]}'\bs P_{Z,[g,h]}\bs X_{[h]};\\
        &\hat{\bs C}_{CF}(\bs\beta_0)\\
        &=\frac{2}{\sqrt{nk}}\sum_{g\neq h}\bs X_{[g]}'\bs P_{Z,[g,h]}\big(\bs\varepsilon(\bs\beta_0)-\tilde{\bs\varepsilon}(\bs\beta_0;g,h)\big)_{[h]}\big(\bs\varepsilon(\bs\beta_0)-\tilde{\bs\varepsilon}(\bs\beta_0;g,h)\big)_{[g]}'\bs P_{Z,[g,h]}\bs\varepsilon(\bs\beta_0)_{[h]}.
    \end{split}
\end{equation}

Since $g\neq h$ and $\tilde{\bs\varepsilon}(\bs\beta;g,h)$ depends neither on observations in cluster $g$ nor on observations in cluster $h$, the terms with $\tilde{\bs\varepsilon}(\bs\beta;g,h)$ in \cref{eq:Vhat cljcf} have expectation zero. Consequently, the cross-fit variance estimators are unbiased. The theorem below shows that these variance estimators are also consistent under the slightly stronger version of \cref{ass:cl jackknife AR}.

\begin{assumption}\label{ass:cl jackknife AR alt}
    Conditional on $\bs Z$ and with probability one for all $n$ large enough, it holds that
    \begin{assenumerate*}
        \item\label{assit:cljar alt rank} $\rank(\bs P_{Z})=k$;
        \item\label{assit:cljar alt bound P} $\|\bs P_{Z,[gh,gh]}\|_2^2\leq C<1$ for all $g,h$, where $\bs P_{[gh,gh]}=\bs Z_{[gh]}(\bs Z'\bs Z)\bs Z_{[gh]}'$;
        \item\label{assit:cljar alt many iv} $k\to\infty$ as $G\to\infty$;
        \item\label{assit:cljar alt nmax} $\nmax^4/k\toas 0$ for $\nmax=\max_{g=1,\dots,G}n_g$;
        
        \item\label{assit:cljar alt eigval} $0<\allowbreak 1/C\leq\allowbreak\lambdamin(\bs\Sigma_{g})\leq\allowbreak\lambdamax(\bs\Sigma_{g})\leq\allowbreak n_gC$
    a.s. for all $g$;
    \item\label{assit:cljar alt moment} $\E(\varepsilon_i^8|\bs Z)\leq C<\infty$ a.s. for all $i$. 
    \end{assenumerate*}
\end{assumption}
In particular, \cref{assit:cljar alt bound P} strengthens \cref{assit:cljar bound P} to blocks of two clusters and \cref{assit:cljar alt moment} strengthens \cref{assit:cljar moment} to finite eighth moments.

\begin{theorem}\label{thm:V cljar CF}
    Under \cref{ass:model,ass:cl jackknife AR alt,ass:cljs} it holds that 
    $\E\big(\hat{\bs V}_{CF}(\bs\beta_0)\big|\bs Z\big)=\bs V_{CLJ}$ and, conditional on $\bs Z$, $\hat{\bs V}_{CF}(\bs\beta_0)\toprob \bs V_{CLJ}$, where
    \begin{equation}
        \begin{split}
            \hat{\bs V}_{CF}(\bs\beta_0)=\begin{bmatrix}
                \hat{V}^{AR}_{CF}(\bs\beta_0) & \hat{\bs C}_{CF}(\bs\beta_0)'\\
                \hat{\bs C}_{CF}(\bs\beta_0) & \hat{\bs V}^{S}_{CF}(\bs\beta_0)
            \end{bmatrix}.
        \end{split}
    \end{equation}
\end{theorem}
\begin{proof}
    The proof is given in \cref{app:proof V cljar CF}.
\end{proof}

\section{Simulation results}\label{sec:MC}
This section explores the finite sample performance of the different cluster robust tests relative to each other and to those for independent data. I use a linear IV model with a single endogenous regressor and clustered data as in \citet{boot2023identification}, which is itself an adaptation of the DGP from \citet{hausman2012instrumental} to clustered data.

I generate $n=800$ observations as $y_i=\alpha+\beta x_i+\varepsilon_i$ for $i=1,\dots,n$ and $\alpha=0$. Throughout I test $H_0\colon\beta=0$, but the $\beta$ in the DGP can have a different value when investigating the power. The endogenous regressor $x_i$ relates to a single instrumental variable $\tilde{z}_i$, $x_i=\pi\tilde{z}_{i}+\eta_i$. There are $k$ additional instrumental variables that can be used for inference $\bs Z_i=(1,\tilde{z}_i, \tilde{z}_i^2, \tilde{z}_i^3, \tilde{z}_i^4, \tilde{z}_i D_{i1},\dots, \tilde{z}_iD_{ik-4})'$, where the $D_{ij}$, $j=1,\dots,k-4$, are independently distributed Bernoulli(1/2) random variables. $\varepsilon_i$ is a function of random variables, $\varepsilon_i=\rho\eta_i+\sqrt{(1-\rho^2)/(\phi^2+0.86^4)}(\phi v_{1i}+0.86v_{2i})$, where $\rho=0.3$ is the degree of endogeneity and $\phi=1.38072$ as in \citet{bekker2015jackknife}. The distribution of the other random variables is detailed below.

The $n$ observations are divided over $G=100$ clusters. The clusters can be unbalanced and are determined by first setting $n_g=\max\big\{1, n\exp(\gamma g/G)/\allowbreak\big[\sum_{g=1}^{G-1}\exp(\gamma g/G)+1\big]\big\}$ for $g=1,\dots G-1$ and $n_G=\max\big\{1,n-\sum_{g=1}^{G-1}n_g\big\}$. Next, all cluster sizes are made integer valued by rounding them down. Finally, to get the correct sample size, the first $n-\sum_{g=1}^Gn_g$ clusters are increased by one. Note that $\gamma$ determines the degree of unbalancedness and $\gamma=0$ implies balanced clusters.

The random variables $\tilde{z}_i$, $\eta_i$, $v_{1i}$ and $v_{2i}$ consist of an idiosyncratic and a cluster common component, weighted by $\lambda=1/2$. To be precise, $\tilde{\bs z}_{[g]}=\sqrt{\lambda}\tilde{\bs z}_{[g]}^\text{ind}+\sqrt{1-\lambda}\tilde{z}_{g}^\text{cl}$, $\bs\eta_{[g]}=\sqrt{\lambda}\bs\eta_{[g]}^\text{ind}+\sqrt{1-\lambda}\eta_{g}^\text{cl}$, $\bs v_{1,[g]}=\sqrt{\lambda}\bs v_{1,[g]}^\text{ind}+\sqrt{1-\lambda}v_{1,g}^\text{cl}$ and $\bs v_{2,[g]}=\sqrt{\lambda}\bs v_{2,[g]}^\text{ind}+\sqrt{1-\lambda}v_{2,g}^\text{cl}$. The cluster common and idiosyncratic components of $z_{i}$ and $\eta_i$ follow standard normal distributions. I draw $v_{1i}^\text{ind}\sim N\big(0,(\tilde{z}_{i}^\text{ind})^2\big)$ and $v_{1h}^\text{cl}\sim N\big(0,(\tilde{z}_{h}^\text{cl})^2\big)$. The cluster common and idiosyncratic components of $v_{2i}$ come from a normal distribution with mean zero and variance $0.86^2$. 

\subsection{Size}
To assess the importance of accounting for clustering, consider the size of the AR, score, jackknife AR and jackknife score tests for independent data and a $t$-test based on JIVE with heteroskedasticity-robust standard errors and normal critical values in the DGP above with $\pi=0.1$, $\gamma=1$ such that the smallest cluster contains $4$ observations and the largest $12$, and $k$ ranging from $2$ to $90$. 

\begin{figure}[t]
    \centering
    \caption{Size of tests for independent and clustered data when the data are clustered.}
    \label{fig:size}
    \includegraphics{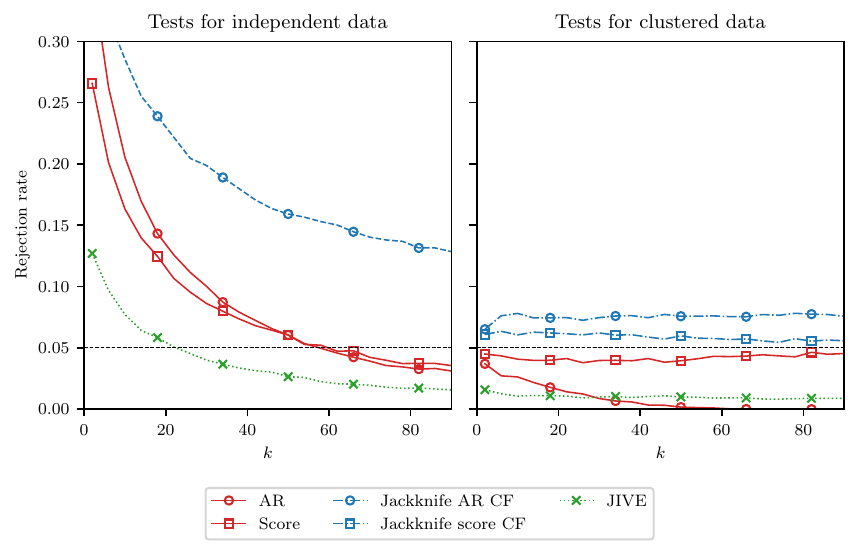}
    \begin{minipage}{0.95\textwidth}
    \footnotesize
    \textit{Note:} The left panel shows the rejection rates of the AR test, score test, jackknife AR test with cross-fit (CF) variance, jackknife score test with cross-fit variance and a $t$-test based on JIVE and heteroskedasticity robust standard errors when testing $\beta=0$ at $5\%$ significance level. The right panel shows the rejection rates for the cluster adaptations of the same tests when testing the same hypothesis. The data are clustered with first-stage coefficient $\pi=0.1$, the number of clusters $G=100$ and the cluster balancedness governed by $\gamma=1$. $k$ is the number of instruments. The DGP is described in \cref{sec:MC}.
    \end{minipage}
\end{figure}

The left panel of \cref{fig:size} shows the rejection rates based on $10\,000$ draws when testing $H_0\colon\beta=0$ at a $5\%$ significance level. Clearly, as these tests do not take the clustered dependence into account the tests are size distorted. For smaller values of $k$ all tests are oversized. When $k$ increases, the identification robust tests become closer to size correct. For the AR and score tests this can partly be explained by them being conservative for large $k$. A closer inspection of how $\bs P_{Z}$ changes with $k$ showed that the fall in the rejection rates of the jackknife AR and the jackknife score is due to a reduction in the true variance not captured by the variance estimator. Such a reduction is observed in the current DGP, but there are no guarantees that this holds more generally. Moreover, although the jackknife tests become closer to size correct, they are still oversized for large values of $k$.

Next, in the right panel of the same figure, I apply the cluster robust versions of the AR, score, jackknife AR, and jackknife score tests and the $t$-test based on JIVE with clustered standard errors to the same data. Observe that the $t$-test based on JIVE is conservative for all values of $k$. For small values of $k$, the AR and score tests are size correct. When $k$ increases, however, the rejection rate of the AR test drops. The jackknife score test is size correct regardless of the number of instruments. The jackknife AR test, on the other hand, slightly overrejects. This is likely due to the cross-fit estimator underestimating the variance. \Cref{fig:size CF} in \cref{app:extra MC} compares the jackknife AR and jackknife score tests with cross-fit variance estimators to those without cross-fit variance estimators. This figure shows that the jackknife AR test without cross-fit variance has rejection rates close to 5\%.

\cref{app:extra MC} investigates the size of the cluster robust tests further. By varying the number of clusters it confirms the intuition that with clustering in the data, the number of instruments relative to the number of clusters, rather than the number of observations, matters for many instrument problems. Furthermore, the appendix gives results on the robustness of the cluster jackknife AR and cluster jackknife score tests to the arguably most stringent assumption, \cref{assit:cljar nmax}, that limits the size of the largest cluster. It shows that with a dominant cluster, the cluster jackknife AR breaks down and becomes oversized, whereas the cluster jackknife score is surprisingly robust.

\subsection{Power}
\begin{figure}[tp]
    \centering
    \caption{Power for clustered data.}
    \label{fig:power clustered}
    \includegraphics{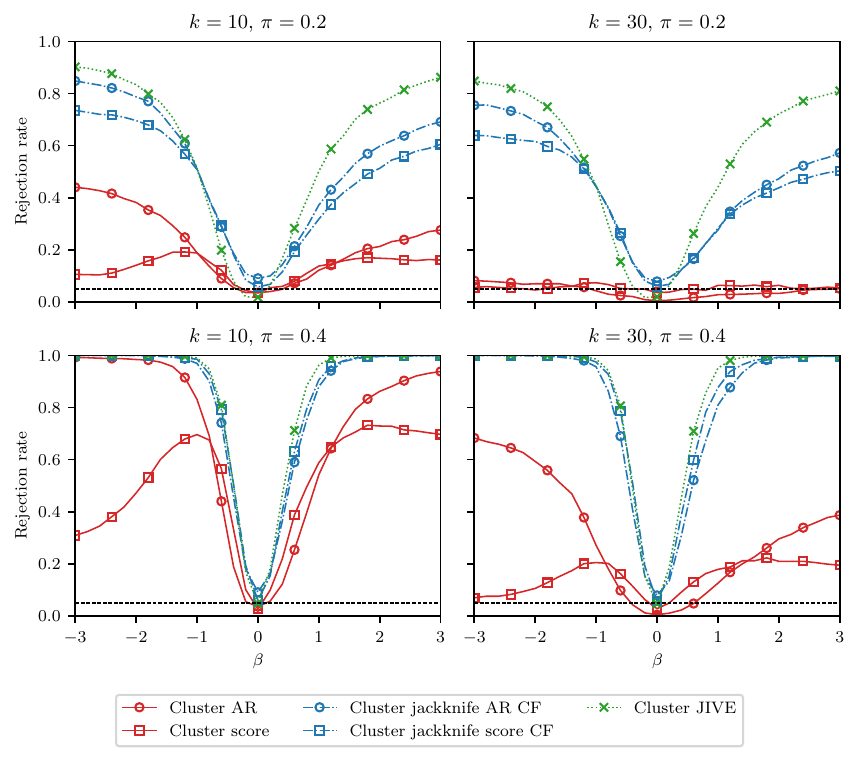}
    \begin{minipage}{0.95\textwidth}
    \footnotesize
    \textit{Note:} Rejection rates of the cluster AR test, cluster score test, cluster jackknife AR test with cross-fit (CF) variance, cluster jackknife score test with cross-fit variance and a $t$-test based on CJIVE and cluster robust standard errors when testing $\beta=0$ at $5\%$ significance level when the true value of $\beta$ in the DGP varies. $k$ is the number of instruments and $\pi$ is the first stage coefficient. The DGP is described in \cref{sec:MC}.
    \end{minipage}
\end{figure}

Next, consider the power of the tests that were also considered in the previous section. \Cref{fig:power clustered} shows the rejection rates based on $1000$ draws for tests of $H_0\colon\beta=0$ when the value of $\beta$ in the DGP varies from $-3$ to $3$. $k$, furthermore, varies between 10 and 30 to show the effect of the number of instruments on the power. Similarly, different instrument strengths are considered by setting $\pi\in\{0.2,0.4\}$. The other parameters in the DGP are as for \cref{fig:size}.

Observe that although the CJIVE-based test has low power for true values of $\beta$ close to the tested value $\beta=0$, it has the highest power when the true value is far from the tested value. Furthermore, in each of the four panels the cluster jackknife AR and cluster jackknife score tests have higher power than their counterparts not adapted for many instruments. Moreover, whereas the cluster AR and cluster score tests show a clear drop in power when $k$ increases from 10 to 30, the power of the cluster jackknife AR test and the power of the cluster jackknife score test are almost unaffected.

Also note that in most panels the cluster score test has higher power than the cluster AR test when the true value of $\beta$ is close to the tested value of $0$. As is often observed for score tests, the power of the cluster score test drops below that of the cluster AR for true values of $\beta$ farther from $0$. The advantage of the cluster score test over the cluster AR test is mainly observed for a moderate number of strong instruments.

For the cluster jackknife tests the ordering in power is clearer. In the four DGPs of \cref{fig:power clustered} the cluster jackknife AR test outperforms the cluster jackknife score test when the instruments are weak as shown in the top panels. In the bottom panels, where the instruments are stronger, the difference in power is marginal.

\begin{figure}[tp]
    \centering
    \caption{Power for clustered data and high heteroskedasticity.}
    \label{fig:power clustered kappa}
    \includegraphics{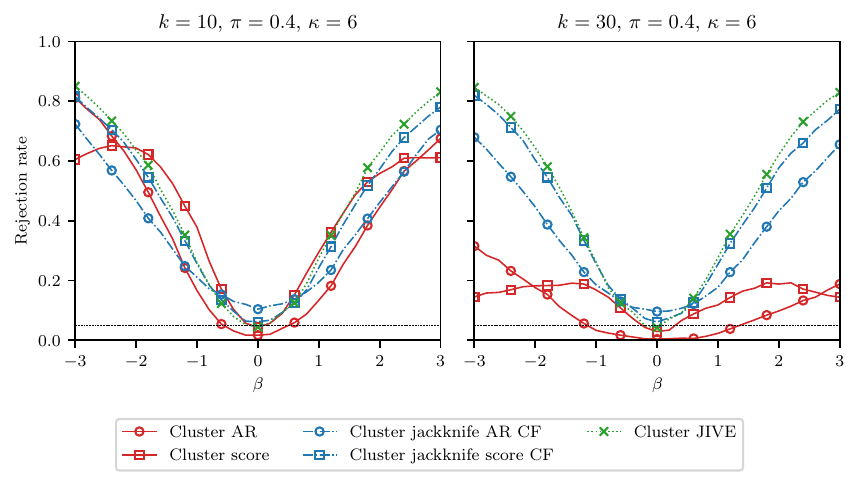}
    \begin{minipage}{0.95\textwidth}
    \footnotesize
    \textit{Note:} See the note to \cref{fig:power clustered}. $\kappa$ is an additional heteroskedasticity parameter.
    \end{minipage}
\end{figure}

However, it need not always be the case that the power of the cluster jackknife AR test is greater than or equal to the power of the cluster jackknife score test. Consider for example an adaptation of the DGP also featured in \citet{boot2023identification}, in which there is high heteroskedasticity. In this DGP there is an additional heteroskedasticity parameter, $\kappa$, used to scale $v_{1i}^\text{ind}\sim N\big(0,(\tilde{z}_{i}^\text{ind})^\kappa\big)$ and $v_{1h}^\text{cl}\sim N\big(0,(\tilde{z}_{h}^\text{cl})^\kappa\big)$. Previously $\kappa=2$. \cref{fig:power clustered kappa} shows the rejection rates of the different tests over $1000$ draws when $\kappa=6$. In this case the cluster jackknife score test clearly outperforms the cluster jackknife AR test.

\Cref{fig:power clustered CF} in \cref{app:extra MC} compares the power of the cluster jackknife tests with and without cross-fit variance. It shows that the cross-fit variance is especially beneficial for the score test when the instruments are weak and, as expected, the tested value of $\beta$ is far from the true value.

\section{Empirical application}\label{sec:application}
To further illustrate the cluster robust tests I revisit the study by \citet{dube2020queens} on the effect of female leadership on the probability that a country is at war. As the probability that a woman ascends the throne and comes into power might be higher or lower depending on whether the polity is at war, the gender of the leader might be endogenous. \citet{dube2020queens} therefore propose instruments for it based on historic succession laws for monarchs in Europe. The authors note that a vacant throne was less likely to be taken by a woman if the previous ruler had a male firstborn child and more likely to be taken by a woman if the previous ruler had a sister. Neither the firstborn's gender nor whether the previous ruler had a sister seem to directly affect the involvement in armed conflict, which justifies the identification strategy.

To estimate the effect of queenly reign on war, the authors use a data set on 18 European polities, each observed during part of the period 1480 to 1913, yielding an unbalanced panel with 3\,586 polity-year observations in total. Throughout, the data are grouped into broadly defined reign clusters, resulting in 176 clusters that vary in size. The largest cluster counts 66 observations and the smallest 1.

The estimates from the main model, which uses the two instruments described above, imply that polities led by a woman are 39 percentage points more likely to be at war in a given year compared to polities led by a man. As a robustness check, models with additional instruments based on interactions between the original instruments and exogenous regressors are proposed. Table A5 in \citepos{dube2020queens} appendix shows the 2SLS estimates from these models. The smallest and largest estimates imply that polities led by women are 29 and 50 percentage points more likely to be at war. However, the new instruments are relatively weak, which makes the estimates and their standard errors unreliable and therefore the larger models are not investigated further. 

There could furthermore be a many-instrument problem. Although the total number of possible instruments, $5$, is low compared to the total number of observations, $3\,586$, it is not compared to the number of clusters, $176$.
Reliably extending the analysis to the larger models therefore requires weak and many instrument robust tests suited for clustered data. 

\begin{figure}[tp]
	\centering
	\caption{95\% confidence interval for the effect of queenly reign on war.}\label{fig:DH2020}
	\includegraphics{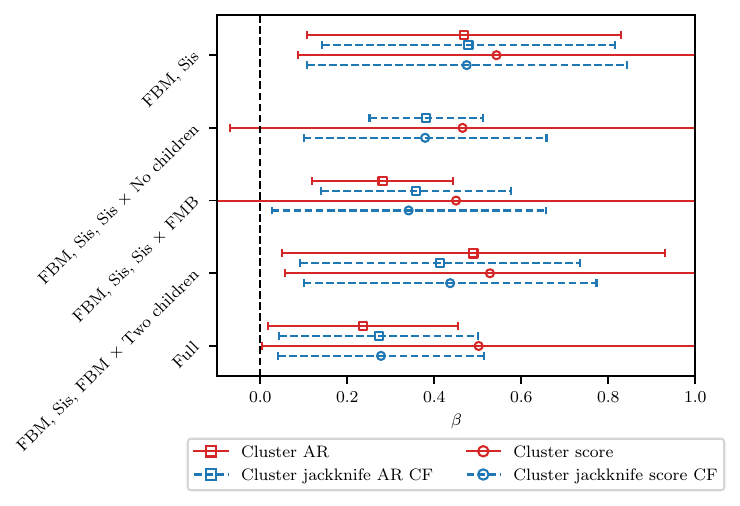}
    \begin{minipage}{0.95\textwidth}
    \footnotesize
    \textit{Note:} 95\% confidence set for the effect of queenly reign on the probability that a polity is at war relative to polities led by a king. The confidence intervals are based on inverting the cluster AR test, the cluster jackknife AR with cross-fit variance (CF), the cluster score and the cluster jackknife score with cross-fit variance. The data come from \citet{dube2020queens} and are clustered into 176 clusters. The instruments are whether the previous sovereign had a firstborn male child (FBM), whether the previous sovereign had a sister (Sis) and interactions of these with each other and indicators of whether the previous sovereign had no or two children. The control variables are the same as in columns 1, 2, 4 and 5 of Table A5 in \citet{dube2020queens}. The full model pools the instruments and control variables of the other models.
    \end{minipage}
\end{figure}

\Cref{fig:DH2020} shows the 95\% confidence sets for $\beta$, which measures the effect of queenly reign on the likelihood of war relative to polities led by kings, obtained by inverting the cluster AR test, the cluster score test, the cluster jackknife AR test and the cluster jackknife score test. I consider candidate values of $\beta$ between -0.1 and 1, as values below -0.1 did not add any information and values above 1 have no sensible interpretation. The groups correspond to the models in columns 1, 2, 4 and 5 of Table A5 in \citet{dube2020queens} and the full model that pooled all the instruments and regressors of the other models.

The figure shows that the cluster score test yields confidence sets that are unbounded from above, and for the third model also from below. \cref{sec:MC} suggests that this can happen when the instruments are relatively weak, in which case it is important to use robust tests. The concern about weak instruments is in line with the results from \citet{dube2020queens} and further confirmed by the relatively wide confidence sets for the cluster jackknife score test.

Furthermore, the tests other than the cluster score test yield for most specifications significantly positive effects of female rule on the likelihood of being at war. The confidence set obtained by inverting the cluster AR test second model is empty, which could be a sign of invalid moment conditions. The exact bounds of the confidence sets are reported in \cref{tab:DH20} in \cref{app:application}.

That there can be a benefit from using tests that are more powerful with many instruments more generally can be seen from the relative lengths of the cluster jackknife AR and cluster AR confidence sets. These are $0.9321$, $1.3457$, $0.7302$ and $1.0481$ for the four models without empty confidence intervals. Although the cluster jackknife AR test does not always yield a narrower confidence set, the reduction can be sizeable, reaching almost 30\% in the fourth model.

\section{Conclusion}\label{sec:conclusion}
This paper argued that weak identification in IV models is more likely when the data are clustered. This makes it particularly important to use tests robust against many and weak instruments when using clustered data. I showed that the many and weak instrument robust jackknife AR and jackknife score tests can be extended by removing clusters of observations, rather than individual observations, when jackknifing. Monte Carlo simulations showed that the cluster jackknife tests are close to size correct and have good power. The last section applied the newly developed tests to the data from \citet{dube2020queens} and showed that queenly reign has a significant positive effect on the probability that a state is at war.


\newpage\begin{appendices}
\crefalias{section}{appsec}
\crefalias{subsection}{appsubsec}
\numberwithin{equation}{section}
\numberwithin{counter}{section}

\section{Proofs}
\subsection{Proof of
\texorpdfstring{\cref{thm:MS2021 T1}}{Theorem 1}}\label{sec:proof MS2021 T1}
The proof follows from the same steps as the proof of \citepos{mikusheva2021inference} Theorem 1.

To derive sufficient statistics for $\beta$ and $\bs\Pi$, let
\begin{equation}
    \begin{split}
        \bs Y_{g}=\begin{bmatrix}\bs y_{[g]}\\\bs X_{[g]}\end{bmatrix}=\begin{bmatrix}\bs Z_{[g]}\bs\Pi\beta\\\bs Z_{[g]}\bs\Pi\end{bmatrix}+\begin{bmatrix}\bs u_{[g]}\\\bs\eta_{[g]}\end{bmatrix}=\begin{bmatrix}\bs Z_{[g]} & \bs 0\\\bs 0& \bs Z_{[g]}\end{bmatrix}\begin{bmatrix}\bs\Pi\beta\\\bs\Pi\end{bmatrix}+\begin{bmatrix}\bs u_{[g]}\\\bs\eta_{[g]}\end{bmatrix}=\bs M_{g}\bs\theta+\begin{bmatrix}\bs u_{[g]}\\\bs\eta_{[g]}\end{bmatrix},
    \end{split}
\end{equation}
Then $\bs Y_{g}\sim\Normal(\bs M_{g}\bs\theta,\bs\Omega\otimes\bs V_{g})$ with density 
\begin{equation}
    \begin{split}
        p_{\theta}(\bs Y)&=\prod_{g=1}^G(2\pi)^{2n_{g}/2}\det(\bs\Omega)^{-1/2}\exp\big(-\frac{1}{2}(\bs Y_{g}-\bs M_{g}\bs\theta)'(\bs\Omega\otimes\bs V_{g})^{-1}(\bs Y_{g}-\bs M_{g}\bs\theta)\big)\\
        &=C\exp\bigg(-\frac{1}{2}\sum_{g=1}^G\bs Y_{g}'(\bs\Omega\otimes\bs V_{g})^{-1}\bs Y_{g}-\bs\theta'\bs M_{g}'(\bs\Omega\otimes\bs V_{g})^{-1}\bs Y_{g}\\
        &\quad-\bs Y_{g}'(\bs\Omega\otimes\bs V_{g})^{-1}\bs M_{g}\bs\theta+\bs\theta'\bs M_{g}'(\bs\Omega\otimes\bs V_{g})^{-1}\bs M_{g}\bs\theta\bigg)\\
        &=C\exp\bigg(-\frac{1}{2}\sum_{g=1}^G\bs Y_{g}'(\bs\Omega\otimes\bs V_{g})^{-1}\bs Y_{g}\bigg)\exp\bigg(\frac{1}{2}\sum_{g=1}^G\bs\theta'\bs M_{g}'(\bs\Omega\otimes\bs V_{g})^{-1}\bs Y_{g}\bigg)\\
        &\quad\exp\bigg(\frac{1}{2}\sum_{g=1}^G\bs Y_{g}'(\bs\Omega\otimes\bs V_{g})^{-1}\bs M_{g}\bs\theta\bigg)\exp\bigg(-\frac{1}{2}\sum_{g=1}^G\bs\theta'\bs M_{g}'(\bs\Omega\otimes\bs V_{g})^{-1}\bs M_{g}\bs\theta\bigg).
    \end{split}
\end{equation}
The factors involving $\bs\theta$ depend on the data only through $\sum_{g=1}^G\bs M_{g}'(\bs\Omega\otimes\bs V_{g})^{-1}\bs Y_{g}$. Given the particular forms of the matrices $\sum_{g=1}^G\bs M_{g}'(\bs\Omega\otimes\bs V_{g})^{-1}\bs Y_{g}=\sum_{g=1}^G(\bs\Omega^{-1}\otimes\bs Z_{[g]}'\bs V^{-1}_{g})\bs Y_{g}$, which therefore only depends on the sufficient statistics $\bs Z'\bs V^{-1}\bs y$ and $\bs Z'\bs V^{-1}\bs X$ and it is enough to only consider tests that depend on these sufficient statistics.

Let $\bs A$ be an upper triangular matrix such that $\bs A\bs\Omega\bs A'=\bs I_{2}$. Then, standardisation with known quantities yields
\begin{equation}\label{eq:standardised sufficient stats}
    \begin{split}
        \begin{bmatrix}
            \bs\xi_1\\\bs\xi_2
        \end{bmatrix}=(\bs A\otimes\bs I_{k})\begin{bmatrix}
            (\bs Z'\bs V^{-1}\bs Z)^{-1/2}\bs Z'\bs V^{-1}\bs y \\ (\bs Z'\bs V^{-1}\bs Z)^{-1/2}\bs Z'\bs V^{-1}\bs X
        \end{bmatrix}\sim\Normal\bigg(\begin{bmatrix}
            \tilde{\beta}\tilde{\bs\Pi}\\\tilde{\bs\Pi}
        \end{bmatrix},\bs I_{2k}\bigg),
    \end{split}
\end{equation}
where $\tilde{\beta}=\sigma_{\eta}(1,0)\bs A(\beta,1)'$ and $\tilde{\bs\Pi}=(\bs Z'\bs V^{-1}\bs Z)^{1/2}\bs\Pi/\sigma_{\eta}$.

Next, the same steps as in \citet{mikusheva2021inference} conclude the proof.


\subsection{Proof of \texorpdfstring{\cref{thm:clj AR S}}{Theorem 3}}\label{sec:proof clj AR S}
\begin{proof} The proof of \cref{thm:clj AR S} requires the following generalisation of Lemma A2 from \citet{chao2012asymptotic}, where I have slightly reordered the conditions and, to avoid a notational conflict with the main text, relabelled $\bs\varepsilon$ to $\bar{\bs\varepsilon}$. Also, as in \citet{chao2012asymptotic}, let a.s.n.\ be short for almost surely for $n$ large enough and let for any matrix $\bs A$, $\|\bs A\|_F=\sqrt{\tr(\bs A'\bs A)}$ be the Frobenius norm.

\begin{lemma}[Generalisation of Lemma A2 from \citet{chao2012asymptotic}]\label{lem:cluster LA2}
    Suppose that, conditional on $\mathcal{Z}$, the following conditions hold a.s.:
    \begin{enumerate}[(i)]
        \item\label{it:P} $\bs P=\bs P(\mathcal{Z})$ is a symmetric, idempotent matrix with $\rank(\bs P)=k$ and $\|\bs P_{[g,g]}\|_2^2\leq C<1$;
        \item $(\bs W_{[1],n}, \bs U_{[1]},\bar{\bs\varepsilon}_{[1]},\dots,\bs W_{[G],n}, \bs U_{[G]},\bar{\bs\varepsilon}_{[G]})$ are independent;
        \item\label{it:ue}
        \begin{enumerate}
            \item $\E(\bs U_{[g]}|\mathcal{Z})=\bs 0$, $\E(\bar{\bs\varepsilon}_{[g]}|\mathcal{Z})=\bs 0$;
            \item For any $\bs v\in\R^p$ with $\bs v'\bs v=1$ and 
            \begin{equation}
                \begin{split}
                    \bs M_g(\bs v)=\E\bigg(\begin{bmatrix}
                        \bar{\bs\varepsilon}_{[g]}\bar{\bs\varepsilon}_{[g]}' & \bar{\bs\varepsilon}_{[g]}\bs v'\bs U_{[g]}' \\
                        \bs U_{[g]}\bs v\bar{\bs\varepsilon}_{[g]}' & \bs U_{[g]}\bs v\bs v'\bs U_{[g]}' 
                    \end{bmatrix}\bigg|\mathcal{Z}\bigg),
                \end{split}
            \end{equation}
            it holds that $\lambdamax\big(\bs M_g(\bs v)\big)\leq n_gC$; and
            \item $\E(\|\bs U_{i}\|^4|\mathcal{Z})\leq C$ and $\E(\bar{\varepsilon}_{i}^4|\mathcal{Z})\leq C$.
        \end{enumerate}
        \item\label{it:w} $\E(\bs W_{[g],n}|\mathcal{Z})=\bs 0$, $\bs D_{n}=\sum_{g=1}^G\sum_{i,j\in[g]}\E(\bs W_{i,n}\bs W_{j,n}'|\mathcal{Z})$ satisfies $\|\bs D_{n}\|_2\leq\nmax C$ a.s.n., $\sum_{g=1}^G\E(\|\bs W_{[g]}'\bs\iota\|^4|\mathcal{Z})\toas 0$; and
        \item\label{it:rate} $K\to\infty$ as $n\to\infty$ and $\nmax^4/k\toas 0$.
    \end{enumerate}
    Then for
    \begin{equation}
        \bar{\bs\Sigma}_{n}=\sum_{g\neq h}\E(\bs U_{[g]}'\bs P_{[g,h]}\bar{\bs\varepsilon}_{[h]}\bar{\bs\varepsilon}_{[g]}'\bs P_{[g,h]}\bs U_{[h]}+\bs U_{[g]}'\bs P_{[g,h]}\bar{\bs\varepsilon}_{[h]}\bar{\bs\varepsilon}_{[h]}'\bs P_{[h,g]}\bs U_{[g]}|\mathcal{Z})/k,
    \end{equation}
    and any sequences $\bs c_{1n}$ and $\bs c_{2n}$ depending on $\mathcal{Z}$ of conformable vectors with $\|\bs c_{1n}\|\leq C$ and $\|\bs c_{2n}\|\leq C$ and $\Xi_{n}=\bs c_{1n}'\bs D_{n}\bs c_{1n}+\bs c_{2n}'\bar{\bs\Sigma}\bs c_{2n}>1/C$ a.s.n., it follows that
    \begin{equation}\label{eq:cluster LA2}
        Y_{n}=\Xi_{n}^{-1/2}\big(\bs c_{1n}'\bs W_{n}'\bs\iota+\bs c_{2n}'\bs U'\ddot{\bs P}\bar{\bs\varepsilon}/\sqrt{k}\big)\todist N(0,1)\quad a.s.
    \end{equation}
\end{lemma}

\begin{proof}
    The proof is given in \cref{app:cluster LA2}.
\end{proof}

To prove \cref{thm:clj AR S}, let for any $\bs t\in\R^{p+1}\setminus\{\bs 0\}$ and $(\bar{c}_1,\bar{\bs c}_2')=(\bs t'\bs t)^{-1/2}\bs t'\bs V_{CLJ}^{-1/2}$
\begin{equation}
    \begin{split}
        (\bar{c}_{1},\bar{\bs c}_{2}')\begin{pmatrix}
            AR(\bs\beta_0)\\\bs S(\bs\beta_0)
        \end{pmatrix}&=\frac{\bar{c}_1}{\sqrt{k}}\bs\varepsilon'\ddot{\bs P}_{Z}\bs\varepsilon+\frac{1}{\sqrt{n}}\bar{\bs c}_2'\bs\Pi'\bs Z'[\bs I-\bs B_{P_Z}]\bs\varepsilon+\frac{1}{\sqrt{n}}\bar{\bs c}_2'\bs\eta'\ddot{\bs P}_{Z}\bs\varepsilon\\
        &=\Xi_{n}^{-1/2}\big(\bs c_{1n}'\bs W_{n}'\bs\iota+\bs c_{2n}'\bs U'\ddot{\bs P}\bs\varepsilon/\sqrt{k}\big)
    \end{split}
\end{equation}
for $\bs P=\bs P_{Z}$, $\bs W_{[g],n}=\bs\Pi'\bs Z_{[g]}'[\bs I-\bs P_{Z,[g,g]}]\bs D_{\varepsilon_{[g]}}/\sqrt{n}$, $\bs U_{[g]}=\big(\bs\varepsilon_{[g]},\sqrt{k/n}\bs\eta_{[g]}\big)$, $\bar{\bs\varepsilon}_{[g]}=\bs\varepsilon_{[g]}$, $\bs c_{1n}=\bar{\bs c}_2$, and $\bs c_{2n}'=(\bar{c}_1,\bar{\bs c}_2')$.

I will now show that the statistic is well-defined and satisfies the conditions from \cref{lem:cluster LA2}, such that it converges to a standard normal distribution. The conclusion of \cref{thm:clj AR S} then follows from the Cramér--Wold device \citep[T29.4]{billingsley1995probability}.

\subsubsection{Statistic well-defined}

For the statistic to be well-defined the variance $\bs V_{CLJ}$ needs to be non-singular. First consider the conditional variance of the AR statistic. By \cref{ass:model},
\begin{equation}
    \begin{split}
        V_{CLJ}^{AR}=\frac{2}{k}\tr(\bs\Sigma\ddot{\bs P}_{Z}\bs\Sigma\ddot{\bs P}_{Z})&\geq\frac{2}{k}\lambdamin^2(\bs\Sigma)\tr(\ddot{\bs P}_{Z}\ddot{\bs P}_{Z})\geq\frac{C}{k}\sum_{g\neq h}\tr(\bs P_{Z,[g,h]}\bs P_{Z,[h,g]})>0,
    \end{split}
\end{equation}
because
\begin{equation}\label{eq:tr PghPhg}
    \begin{split}
        \sum_{g\neq h}\tr(\bs P_{Z,[g,h]}\bs P_{Z,[h,g]})
        &=\sum_{h=1}^G\tr(\bs P_{Z,[h,h]})-\tr(\bs P_{Z,[h,h]}^2)\\
        &\geq\sum_{h=1}^G\tr(\bs P_{Z,[h,h]})-\lambdamax(\bs P_{Z,[h,h]})\tr(\bs P_{Z,[h,h]})\\
        &\geq (1-C)k,
    \end{split}
\end{equation}
since, by \cref{assit:cljar bound P}, $\|\bs P_{Z,[g,g]}\|^2_2=\lambdamax(\bs P_{Z,[g,g]}^2)=\lambdamax^2(\bs P_{Z,[g,g]})<1$ implies that $\lambdamax(\bs P_{Z,[g,g]})<1$.

Next, let $\bs v\in\R^{p}$ such that $\bs v'\bs v=1$ be given and consider a quadratic form of the variance of the score from \cref{eq:V clj S}. The part with $\bs Z$ is greater than or equal to zero by \cref{assit:cljs eigval}. The second part is
\begin{equation}
    \begin{split}
        &\E\bigg(\sum_{g\neq h}\bs v'\bs\eta_{[g]}'\bs P_{Z,[g,h]}\bs\varepsilon_{[h]}\bs\varepsilon_{[h]}'\bs P_{Z,[h,g]}\bs\eta_{[g]}\bs v+\bs v'\bs\eta_{[h]}'\bs P_{Z,[h,g]}\bs\varepsilon_{[g]}\bs\varepsilon_{[g]}'\bs P_{Z,[g,h]}\bs\eta_{[h]}\bs v\\
        &\quad+2\bs v'\bs\eta_{[g]}'\bs P_{Z,[g,h]}\bs\varepsilon_{[h]}\bs\varepsilon_{[g]}'\bs P_{Z,[g,h]}\bs\eta_{[h]}\bs v\bigg|\bs Z\bigg)\\
        &=\sum_{g\neq h}\E\bigg(\begin{bmatrix}
            \bs\varepsilon_{[g]}' & \bs v'\bs\eta_{[g]}'
        \end{bmatrix}\begin{bmatrix}
            \bs 0 & \bs P_{Z,[g,h]} \\
            \bs P_{Z,[g,h]} & \bs 0
        \end{bmatrix}\begin{bmatrix}
            \bs\varepsilon_{[h]} \\
            \bs\eta_{[h]}\bs v
        \end{bmatrix}\begin{bmatrix}
            \bs\varepsilon_{[h]}' & \bs v'\bs\eta_{[h]}'
        \end{bmatrix}\\
        &\quad\begin{bmatrix}
            \bs 0 & \bs P_{Z,[h,g]} \\
            \bs P_{Z,[h,g]} & \bs 0
        \end{bmatrix}\begin{bmatrix}
            \bs\varepsilon_{[g]} \\
            \bs\eta_{[g]}\bs v
        \end{bmatrix}\bigg|\bs Z\bigg)\\
        &=\sum_{g\neq h}\E\bigg(\begin{bmatrix}
            \bs\varepsilon_{[g]}' & \bs v'\bs\eta_{[g]}'
        \end{bmatrix}\begin{bmatrix}
            \bs 0 & \bs P_{Z,[g,h]} \\
            \bs P_{Z,[g,h]} & \bs 0
        \end{bmatrix}\begin{bmatrix}
            \bs\varepsilon_{[h]}\bs\varepsilon_{[h]}' & \bs\varepsilon_{[h]}\bs v'\bs\eta_{[h]}' \\
            \bs\eta_{[h]}\bs v\bs\varepsilon_{[h]}' & \bs\eta_{[h]}\bs v\bs v'\bs\eta_{[h]}'
        \end{bmatrix}\\
        &\quad\begin{bmatrix}
            \bs 0 & \bs P_{Z,[h,g]} \\
            \bs P_{Z,[h,g]} & \bs 0
        \end{bmatrix}\begin{bmatrix}
            \bs\varepsilon_{[g]} \\
            \bs\eta_{[g]}\bs v
        \end{bmatrix}\bigg|\bs Z\bigg)\\
        &\geq \sum_{g\neq h}\E\bigg(\lambdamin\bigg(\begin{bmatrix}
            \bs\varepsilon_{[h]}\bs\varepsilon_{[h]}' & \bs\varepsilon_{[h]}\bs v'\bs\eta_{[h]}' \\
            \bs\eta_{[h]}\bs v\bs\varepsilon_{[h]}' & \bs\eta_{[h]}\bs v\bs v'\bs\eta_{[h]}'
        \end{bmatrix}\bigg)\\
        &\quad\begin{bmatrix}
            \bs\varepsilon_{[g]}' & \bs v'\bs\eta_{[g]}'
        \end{bmatrix}\begin{bmatrix}
            \bs 0 & \bs P_{Z,[g,h]} \\
            \bs P_{Z,[g,h]} & \bs 0
        \end{bmatrix}\begin{bmatrix}
            \bs 0 & \bs P_{Z,[h,g]} \\
            \bs P_{Z,[h,g]} & \bs 0
        \end{bmatrix}\begin{bmatrix}
            \bs\varepsilon_{[g]} \\
            \bs\eta_{[g]}\bs v
        \end{bmatrix}\bigg|\bs Z\bigg).
    \end{split}
\end{equation}
Note that this minimum eigenvalue is bounded away from zero by \cref{assit:cljs eigval}. Then proceed with
\begin{equation}
    \begin{split}
        &\sum_{g\neq h}\E\bigg(\begin{bmatrix}
            \bs\varepsilon_{[g]}' & \bs v'\bs\eta_{[g]}'
        \end{bmatrix}\begin{bmatrix}
            \bs 0 & \bs P_{Z,[g,h]} \\
            \bs P_{Z,[g,h]} & \bs 0
        \end{bmatrix}\begin{bmatrix}
            \bs 0 & \bs P_{Z,[h,g]} \\
            \bs P_{Z,[h,g]} & \bs 0
        \end{bmatrix}\begin{bmatrix}
            \bs\varepsilon_{[g]} \\
            \bs\eta_{[g]}\bs v
        \end{bmatrix}\bigg|\bs Z\bigg)\\
        &=\sum_{g\neq h}\E(\bs v'\bs\eta_{[g]}'\bs P_{Z,[g,h]}\bs P_{Z,[h,g]}\bs\eta_{[g]}\bs v+\bs\varepsilon_{[g]}'\bs P_{Z,[g,h]}\bs P_{Z,[h,g]}\bs\varepsilon_{[g]}|\bs Z)\\
        &=\sum_{g\neq h}\tr(\bs P_{Z,[h,g]}\E(\bs\eta_{[h]}\bs v\bs v'\bs\eta_{[h]}|\bs Z)\bs P_{Z,[g,h]})+\tr(\bs P_{Z,[h,g]}\bs\Sigma_{g}\bs P_{Z,[g,h]})>0,
    \end{split}
\end{equation}
by \cref{assit:cljs eigval,eq:tr PghPhg}.

Finally, by the arguments in \citet{boot2023identification}, $\det(\bs V_{CLJ})=\det(\bs V^{S}_{CLJ})\allowbreak\det(\bs V^{S}_{CLJ}-\bs C_{CLJ}\bs C_{CLJ}' [V_{CLJ}^{AR}]^{-1})$ by Schur complements. The $(i,j)\th$ element in $\bs C_{CLJ}\bs C_{CLJ}' [V_{CLJ}^{AR}]^{-1}$ is the covariance of the AR statistic with $i\th$ and $j\th$ element of the cluster jackknife score divided by the variance of the cluster jackknife AR statistic, which therefore equals the correlation of the cluster jackknife AR statistic with the $i\th$ and $j\th$ element of the score statistic times the standard deviations of the $i\th$ and $j\th$ elements of the cluster jackknife score statistic. Define $\bs\rho$ to be the vector of correlations between the cluster jackknife AR statistic and the cluster jackknife score, $\rho_{i}=\corr\big(AR(\bs\beta_0),S(\bs\beta_0)_{i}\big|\bs Z\big)$. Then
\begin{align}
        \det(\bs V_{CLJ})&=\det(\bs V^{S}_{CLJ})\det(\bs V^{S}_{CLJ}-\bs C_{CLJ}\bs C_{CLJ}' [V_{CLJ}^{AR}]^{-1})
        \\
        &=\det(\bs V_{CLJ}^{S})\det(\bs I+\bs D_{\rho})\det(\bs V_{CLJ}^{S})\det(\bs I-\bs D_{\rho})>0,
\end{align}
since $\rho_i\neq\pm 1$ for all $i$. I conclude that $\bs V_{CLJ}$ is invertible.

\subsubsection{Conditions of \texorpdfstring{\cref{lem:cluster LA2}}{Lemma A.1}}
Next, check the conditions of \cref{lem:cluster LA2}.
\begin{enumerate}[(i)]
    \item The first set of assumptions is satisfied by the full rank assumption on $\bs Z$, the fact that $\bs P_{Z}$ is a projection matrix and \cref{assit:cljar bound P};
    \item Follows from the clustering structure specified in \cref{ass:model};
    \item \begin{enumerate}
        \item Follows from the zero mean assumption of the errors in \cref{ass:model};
        \item Let $\bs v=(v_1,\bs v_{-1}')'$ with $\bs v'\bs v=1$ be given. Consider first the case that $\bs v_{-1}\neq \bs 0$. Then note that,
        \begin{equation}
            \begin{split}
                &\var\bigg(\begin{bmatrix}\bar{\bs\varepsilon}_{[g]} \\ \bs U_{[g]}\bs v\end{bmatrix}\bigg|\mathcal{Z}\bigg)\\
                &=\var\bigg(\begin{bmatrix}\bs\varepsilon_{[g]} \\ v_{1}\bs\varepsilon_{[g]}+\sqrt{k/n}\bs\eta_{[g]}\bs v_{-1}\end{bmatrix}\bigg|\mathcal{Z}\bigg)\\
                &=\var\bigg(\begin{bmatrix}\bs I & \bs 0\\ v_{1}\bs I &\sqrt{k/n}\bs I\end{bmatrix}\begin{bmatrix}\bs\varepsilon_{[g]} \\\bs\eta_{[g]}\bs v_{-1}\end{bmatrix}\bigg|\mathcal{Z}\bigg)\\
                &=\begin{bmatrix}\bs I & \bs 0\\ v_{1}\bs I &\sqrt{k/n}\bs I\end{bmatrix}\var\bigg(\begin{bmatrix}\bs\varepsilon_{[g]} \\\bs\eta_{[g]}\bs v_{-1}\end{bmatrix}\bigg|\mathcal{Z}\bigg)\begin{bmatrix}\bs I & \bs v_{1}\bs I\\ \bs 0 &\sqrt{k/n}\bs I\end{bmatrix}.
            \end{split}
        \end{equation}
        Now $\lambdamax\big(\var\big[(\bs\varepsilon_{[g]}',\bs v_{-1}'\bs\eta_{[g]})'\big|\mathcal{Z}\big]\big)\leq Cn_{g}\|\bs v_{-1}\|$ by \cref{assit:cljs eigval}. Additionally,
        \begin{equation}
            \begin{split}
                &\lambdamax\bigg(\begin{bmatrix}\bs I & \bs 0\\ v_{1}\bs I &\sqrt{k/n}\bs I\end{bmatrix}\begin{bmatrix}\bs I & \bs v_{1}\bs I\\ \bs 0 &\sqrt{k/n}\bs I\end{bmatrix}\bigg)\\
                &=\max_{a:a'a=1}\bs a'\begin{bmatrix}\bs I & \bs 0\\ v_{1}\bs I &\sqrt{k/n}\bs I\end{bmatrix}\begin{bmatrix}\bs I & \bs v_{1}\bs I\\ \bs 0 &\sqrt{k/n}\bs I\end{bmatrix}\bs a\\
                &=\max_{a:a'a=1}\begin{bmatrix}\bs a_{1}' &\bs a_{2}'\end{bmatrix}\begin{bmatrix}\bs I & v_{1}\bs I \\ v_{1}\bs I & \bs I(v_{1}^2+k/n)\end{bmatrix}\begin{bmatrix}\bs a_{1} \\\bs a_{2}\end{bmatrix}\\
                &=\max_{a:a'a=1}\begin{bmatrix}\bs a_{1}' &\bs a_{2}'\end{bmatrix}\begin{bmatrix}\bs a_{1}+v_{1}\bs a_{2}\\ v_{1}\bs a_{1} + \bs a_{2}(v_{1}^2+k/n)\end{bmatrix}\\
                &=\max_{a:a'a=1}\bs a_{1}'\bs a_{1}+v_{1}\bs a_{1}'\bs a_{2}+v_{1}\bs a_{2}'\bs a_{1}+\bs a_{2}'\bs a_{2}(v_{1}^2+k/n)\\
                &\leq 1+2v_{1}+v_{1}^2+k/n\leq C,
            \end{split}
        \end{equation}
        because $k/n\leq C$ and $\bs a_{1}'\bs a_{2}\leq\|\bs a_{1}\|\|\bs a_{2}\|=1$. Hence by \cref{assit:cljs eigval} $\lambdamax\big(\var\big[(\bs\varepsilon_{[g]}',\bs v'\bs U_{[g]}')'\big|\mathcal{Z}\big]\big)\leq Cn_{g}$.

        Next, suppose that $\bs v_{-1}=\bs 0$, such that $v_{1}=1$. Then by \cref{assit:cljs eigval} $\lambdamax\big(\var\big[(\bs\varepsilon_{[g]}',\bs v'\bs U_{[g]}')'\big|\mathcal{Z}\big])=\lambdamax\big(\var\big[(\bs\varepsilon_{[g]}',\bs\varepsilon_{[g]}')'\big|\mathcal{Z}\big]\big)\leq Cn_{g}$;
        \item The finite fourth moment of $\bar{\varepsilon}_i$ follows from \cref{assit:cljar moment}. For the finite fourth moment of $\bs U_{i}$ use \cref{assit:cljar moment,assit:cljs moment} to write, $\E(\|\bs U_i\|^4|\mathcal{Z})=\E\big(\big[(\varepsilon_i, \sqrt{k/n}\bs\eta_i')\allowbreak(\varepsilon_i,\sqrt{k/n}\bs\eta_i')'\big]^2\big|\mathcal{Z}\big)=\E\big(\big[\varepsilon_i^2+(k/n)\|\bs\eta_i\|^2\big]^2\big|\mathcal{Z}\big)\leq C\big[\E(\varepsilon_i^4|\mathcal{Z})+\E(\|\bs\eta_i\|^4|\mathcal{Z})\big]\leq C$;
    \end{enumerate}
    \item The zero mean of $\bs W{[g],n}$ is ensured by the zero mean of $\bs\varepsilon$ in \cref{ass:model}. For the condition on $\bs D_{n}$ write
    \begin{equation}
        \begin{split}
            \|\bs D_{n}\|_{2}\leq\|\bs D_{n}\|_{F}&=\|\frac{1}{n}\bs\Pi'\bs Z'\ddot{\bs P}\bs\Sigma\ddot{\bs P}\bs Z\bs\Pi\|_F\\
            &=\frac{1}{n}\big[\tr(\bs\Pi'\bs Z'\ddot{\bs P}\bs\Sigma\ddot{\bs P}\bs Z\bs\Pi\bs\Pi'\bs Z'\ddot{\bs P}\bs\Sigma\ddot{\bs P}\bs Z\bs\Pi)\big]^{1/2}\\
            &\leq\frac{1}{n}\big[\lambdamax(\bs\Pi'\bs Z'\bs Z\bs\Pi)\lambdamax^2(\bs\Sigma)\lambdamax^4(\ddot{\bs P}_{Z})\tr(\bs\Pi'\bs Z'\bs Z\bs\Pi)\big]^{1/2}\\
            &\leq C\nmax,
        \end{split}
    \end{equation}
    because $\lambdamax(\bs\Sigma)\leq\nmax C$ by 
    \cref{assit:cljar eigval}, $\lambdamax(\ddot{\bs P})\leq 1$ by \cref{assit:cljar bound P} and $\tr(\bs\Pi'\bs Z'\bs Z\bs\Pi)\leq\sum_{g=1}^G\tr(\bs\Pi'\bs Z'_{[g]}\bs Z_{[g]}\bs\Pi)\leq\sum_{g=1}^G\sum_{i=1}^p\allowbreak\lambda_{i}(\bs\Pi'\bs Z'_{[g]}\bs Z_{[g]}\bs\Pi)\leq\sum_{g=1}^Gpn_{g}C\leq nC$ by \cref{assit:cljs Z}.

    For the converging fourth moment write
    \begin{equation}\label{eq:Lyapunov W}
        \begin{split}
            &\sum_{g=1}^G\E(\|\bs W_{[g]}'\bs\iota\|_2^4|\mathcal{Z})\\
            &\leq\sum_{g=1}^G\E(\|\bs W_{[g]}'\bs\iota\|_F^4|\mathcal{Z})\\
            &=\frac{1}{n^2}\sum_{g=2}^G\E\big(\tr^2\big(\bs\Pi'\bs Z_{[g]}'[\bs I-\bs P_{Z,[g,g]}]\bs\varepsilon_{[g]}\bs\varepsilon_{[g]}'[\bs I-\bs P_{Z,[g,g]}]\bs Z_{[g]}\bs\Pi\\
            &\quad\bs\Pi'\bs Z_{[g]}'[\bs I-\bs P_{Z,[g,g]}]\bs\varepsilon_{[g]}\bs\varepsilon_{[g]}'[\bs I-\bs P_{Z,[g,g]}]\bs Z_{[g]}\bs\Pi\big)\big|\bs Z\big)\\
            &=\frac{1}{n^2}\sum_{g=2}^G\E\big(\tr\big([\bs I-\bs P_{Z,[g,g]}]\bs\varepsilon_{[g]}\bs\varepsilon_{[g]}'[\bs I-\bs P_{Z,[g,g]}]\bs Z_{[g]}\bs\Pi\bs\Pi'\bs Z_{[g]}'\big)\\
            &\quad\tr\big([\bs I-\bs P_{Z,[g,g]}]\bs\varepsilon_{[g]}\bs\varepsilon_{[g]}'[\bs I-\bs P_{Z,[g,g]}]\bs Z_{[g]}\bs\Pi\bs\Pi'\bs Z_{[g]}\big)\big|\bs Z'\big)\\
            &=\frac{1}{n^2}\sum_{g=2}^G\E\big(\vec(\bs Z_{[g]}\bs\Pi\bs\Pi'\bs Z_{[g]}')'\big([\bs I-\bs P_{Z,[g,g]}]\otimes[\bs I-\bs P_{Z,[g,g]}]\big)\vec(\bs\varepsilon_{[g]}\bs\varepsilon_{[g]}')\\
            &\quad\vec(\bs\varepsilon_{[g]}\bs\varepsilon_{[g]}')'\big([\bs I-\bs P_{Z,[g,g]}]\otimes[\bs I-\bs P_{Z,[g,g]}]\big)\vec(\bs Z_{[g]}\bs\Pi\bs\Pi'\bs Z_{[g]})\big|\bs Z'\big).
        \end{split}
    \end{equation}
    Now note that by convexity of $\lambdamax$ and Jensen's inequality 
    \begin{equation}
        \begin{split}
            &\lambdamax\big[\E\big(\vec(\bs\varepsilon_{[g]}\bs\varepsilon_{[g]}')\vec(\bs\varepsilon_{[g]}\bs\varepsilon_{[g]}')'\big|\mathcal{Z}\big)\big]\\
            &\leq\E\big(\lambdamax\big[\vec(\bs\varepsilon_{[g]}\bs\varepsilon_{[g]}')\vec(\bs\varepsilon_{[g]}\bs\varepsilon_{[g]}')'\big]\big|\mathcal{Z}\big),
        \end{split}
    \end{equation}
    which, by \cref{assit:cljar moment} can be bounded as
    \begin{equation}\label{eq:eigbound kappa}
        \begin{split}
            \E\big(\lambdamax\big[\vec(\bs\varepsilon_{[g]}\bs\varepsilon_{[g]}')\vec(\bs\varepsilon_{[g]}\bs\varepsilon_{[g]}')'\big]\big|\bs Z\big)&=\E\big(\vec(\bs\varepsilon_{[g]}\bs\varepsilon_{[g]}')'\vec(\bs\varepsilon_{[g]}\bs\varepsilon_{[g]}')\big|\mathcal{Z}\big)\\
            &=\E\big(\tr[\bs\varepsilon_{[g]}\bs\varepsilon_{[g]}'\bs\varepsilon_{[g]}\bs\varepsilon_{[g]}']\big|\mathcal{Z}\big)\\
            &=\E(\bs\varepsilon_{[g]}'\bs\varepsilon_{[g]}\bs\varepsilon_{[g]}'\bs\varepsilon_{[g]}|\mathcal{Z)}\\
            &\leq n_g\sum_{i\in[g]}\E(\varepsilon_{i}^4|\mathcal{Z})\\
            &\leq n_g^2\max_{i\in[g]}\E(\varepsilon_i^4|\mathcal{Z})\\
            &\leq n_g^2 C.
        \end{split}
    \end{equation}
    Similarly by \cref{assit:cljs Z},
    \begin{equation}
        \begin{split}
            \lambdamax(\bs Z_{[g]}\bs\Pi\bs\Pi'\bs Z_{[g]}'\otimes\bs Z_{[g]}\bs\Pi\bs\Pi'\bs Z_{[g]}')&=\lambdamax^2(\bs\Pi'\bs Z_{[g]}'\bs Z_{[g]}\bs\Pi)\leq Cn_{g}^2.
        \end{split}
    \end{equation}
    Hence by \cref{assit:cljar bound P,assit:cljar nmax}, \cref{eq:Lyapunov W} becomes
    \begin{equation}
        \begin{split}
            &\frac{1}{n^2}\sum_{g=2}^G\E\big(\vec(\bs Z_{[g]}\bs\Pi\bs c_2\bs c_{2}'\bs\Pi'\bs Z_{[g]}')'\big([\bs I-\bs P_{Z,[g,g]}]\otimes[\bs I-\bs P_{Z,[g,g]}]\big)\vec(\bs\varepsilon_{[g]}\bs\varepsilon_{[g]}')\\
            &\quad\vec(\bs\varepsilon_{[g]}\bs\varepsilon_{[g]}')'\big([\bs I-\bs P_{Z,[g,g]}]\otimes[\bs I-\bs P_{Z,[g,g]}]\big)\vec(\bs Z_{[g]}\bs\Pi\bs c_2\bs c_{2}'\bs\Pi'\bs Z_{[g]})\big|\bs Z\big)\\
            &\leq\frac{C\nmax^2}{n^2}\sum_{g=2}^G\vec(\bs Z_{[g]}\bs\Pi\bs c_2\bs c_{2}'\bs\Pi'\bs Z_{[g]}')'\big([\bs I-\bs P_{Z,[g,g]}]\otimes[\bs I-\bs P_{Z,[g,g]}]\big)\\
            &\quad\big([\bs I-\bs P_{Z,[g,g]}]\otimes[\bs I-\bs P_{Z,[g,g]}]\big)\vec(\bs Z_{[g]}\bs\Pi\bs c_2\bs c_{2}'\bs\Pi'\bs Z_{[g]})\\
            &\leq\frac{C\nmax^2}{n^2}\sum_{g=2}^G\vec(\bs Z_{[g]}\bs\Pi\bs c_2\bs c_{2}'\bs\Pi'\bs Z_{[g]}')'\big([\bs I-\bs P_{Z,[g,g]}][\bs I-\bs P_{Z,[g,g]}]\\&
            \quad\otimes[\bs I-\bs P_{Z,[g,g]}][\bs I-\bs P_{Z,[g,g]}]\big)
            \vec(\bs Z_{[g]}\bs\Pi\bs c_2\bs c_{2}'\bs\Pi'\bs Z_{[g]})\\
            &\leq\frac{C\nmax^2}{n^2}\sum_{g=2}^G\vec\big([\bs I-\bs P_{Z,[g,g]}][\bs I-\bs P_{Z,[g,g]}]\big)'(\bs Z_{[g]}\bs\Pi\bs c_2\bs c_{2}'\bs\Pi'\bs Z_{[g]}'\\
            &\quad\otimes\bs Z_{[g]}\bs\Pi\bs c_2\bs c_{2}'\bs\Pi'\bs Z_{[g]})
            \vec\big([\bs I-\bs P_{Z,[g,g]}][\bs I-\bs P_{Z,[g,g]}]\big)\\
            &\leq\frac{C\nmax^4}{n^2}\sum_{g=2}^G\tr\big([\bs I-\bs P_{Z,[g,g]}][\bs I-\bs P_{Z,[g,g]}][\bs I-\bs P_{Z,[g,g]}][\bs I-\bs P_{Z,[g,g]}]\big)\\
            &\leq\frac{C\nmax^4}{n^2}\sum_{g=2}^G\tr\big([\bs I-\bs P_{Z,[g,g]}][\bs I-\bs P_{Z,[g,g]}]\big)\\
            &\leq\frac{C\nmax^4(n-k)}{n^2}\toas 0.
        \end{split}
    \end{equation}

    \item Holds by \cref{assit:cljar nmax}.
\end{enumerate}

Finally note that $\|\bs c_{1n}\|\leq C$ and $\|\bs c_{2n}\|\leq C$. Also,
\begin{equation}
    \begin{split}
        \bs D_{n}&=\sum_{g=1}^G\sum_{i,j\in[g]}\E(\bs W_{i,n}\bs W_{j,n}'|\mathcal{Z})\\
        &=\frac{1}{n}\sum_{g=1}^G\sum_{i,j\in[g]}\bs\Pi'\bs Z'[\bs I-\bs B_{P_Z}]\bs e_{i}\E(\varepsilon_{i}\varepsilon_{j}|\mathcal{Z})\bs e_j'[\bs I-\bs B_{P_Z}]\bs Z\bs\Pi\\
        &=\frac{1}{n}\bs\Pi'\bs Z'\ddot{\bs P}\bs\Sigma\ddot{\bs P}\bs Z\bs\Pi,
    \end{split}
\end{equation}
and
\begin{equation}
    \begin{split}
        \bar{\bs\Sigma}_{n}&=\sum_{g\neq h}\E(\bs U_{[g]}'\bs P_{[g,h]}\bar{\bs\varepsilon}_{[h]}\bar{\bs\varepsilon}_{[g]}'\bs P_{[g,h]}\bs U_{[h]}+\bs U_{[g]}'\bs P_{[g,h]}\bar{\bs\varepsilon}_{[h]}\bar{\bs\varepsilon}_{[h]}'\bs P_{[h,g]}\bs U_{[g]}|\mathcal{Z})/k\\
        &=\sum_{g\neq h}\E\bigg(\bigg(\frac{1}{\sqrt{k}}\bs\varepsilon_{[g]},\frac{1}{\sqrt{n}}\bs\eta_{[g]}\bigg)'\bs P_{[g,h]}\bs\varepsilon_{[h]}\bs\varepsilon_{[g]}'\bs P_{[g,h]}\bigg(\frac{1}{\sqrt{k}}\bs\varepsilon_{[h]},\frac{1}{\sqrt{n}}\bs\eta_{[h]}\bigg)\\
        &\quad+\bigg(\frac{1}{\sqrt{k}}\bs\varepsilon_{[g]},\frac{1}{\sqrt{n}}\bs\eta_{[g]}\bigg)'\bs P_{[g,h]}\bs\varepsilon_{[h]}\bs\varepsilon_{[h]}'\bs P_{[h,g]}\bigg(\frac{1}{\sqrt{k}}\bs\varepsilon_{[g]},\frac{1}{\sqrt{n}}\bs\eta_{[g]}\bigg)\bigg|\mathcal{Z}\bigg),
    \end{split}
\end{equation}
such that
\begin{equation}\label{eq:Xi}
    \begin{split}
        \Xi&=\bs c_{1n}'\bs D_{n}\bs c_{1n}+\bs c_{2n}'\bar{\bs\Sigma}\bs c_{2n}\\&=\begin{pmatrix}\bar{c}_{1}&\bar{\bs c}_{2}'\end{pmatrix}\begin{pmatrix}V_{CLJ}^{AR} & \bs C_{CLJ}'\\\bs C_{CLJ}&\bs V_{CLJ}^{S}\end{pmatrix}\begin{pmatrix}\bar{c}_{1}\\\bar{\bs c}_{2}\end{pmatrix}\\
        &=(\bs t'\bs t)^{-1/2}\bs t'\bs V^{-1/2}_{CLJ}\bs V_{CLJ}\bs V^{-1/2\prime}_{CLJ}\bs t(\bs t'\bs t)^{-1/2}=1.
    \end{split}
\end{equation}

\end{proof}

\subsection{Proof of \texorpdfstring{\cref{thm:V cljar CF}}{Theorem 4}}\label{app:proof V cljar CF}The proof begins with the following lemma, which proves unbiasedness and consistency of the plug-in variance estimators.

\begin{lemma}\label{thm:V cljar}
    Under \cref{ass:model,ass:cl jackknife AR,ass:cljs} it holds that $\E(\hat{\bs V}_{CLJ}(\bs\beta_0)|\bs Z)=\bs V_{CLJ}$ and, conditional on $\bs Z$, $\hat{\bs V}_{CLJ}(\bs\beta_0)\toprob \bs V_{CLJ}$.
\end{lemma}
\begin{proof}
The proof is given in \cref{app:plug-in variance}.
\end{proof}

Unbiasedness of the cross-fit estimators follows from the argument in the main text.

Consistency is shown for $\hat{V}^{AR}_{CF}(\bs\beta)$, $\hat{\bs C}_{CF}(\bs\beta)'$ and $\hat{\bs V}^{S}_{CF}(\bs\beta)$ separately. Each part can be written as the estimator from \cref{thm:V cljar}, which is thus consistent, plus remainder terms, which are negligible.

For ease of notation, suppress the dependence of $\bs P_{Z}$ and $\bs M_{Z}$ on $\bs Z$. Further, define with slight abuse of notation $\bs P_{[g,h],-kl}=\bs Z_{[g]}(\bs Z'\bs Z-\bs Z_{k}'\bs Z_{k}-\bs Z_{l}\bs Z_{l})^{-1}\bs Z_{[h]}'$ for $k\neq l$ and $\bs P_{[g,h],-kk}=\bs P_{[g,h],-k}=\bs Z_{[g]}(\bs Z'\bs Z-\bs Z_{k}'\bs Z_{k})^{-1}\bs Z_{[h]}'$.

\subsubsection{Variance of AR}
Decompose the cross-fit variance estimator for the cluster jackknife AR as
\begin{equation}\label{eq:CF var AR decomp}
    \begin{split}
        \hat{V}^{AR}_{CF}(\bs\beta_0)&=\frac{2}{k}\sum_{g=1}^G\sum_{h\neq g}\bs\varepsilon_{[g]}'\bs P_{[g,h]}\bs\varepsilon_{[h]}\bs\varepsilon_{[h]}'\bs P_{[h,g]}\bs\varepsilon_{[g]}\\
        &\quad-\frac{2}{k}\sum_{g=1}^G\sum_{h\neq g}\sum_{l\notin\{g,h\}}\bs\varepsilon_{[l]}'\bs P_{[l,g],-gh}\bs P_{[g,h]}\bs\varepsilon_{[h]}\bs\varepsilon_{[h]}'\bs P_{[h,g]}\bs\varepsilon_{[g]}\\
        &\quad-\frac{2}{k}\sum_{g=1}^G\sum_{h\neq g}\sum_{l\notin\{g,h\}}\bs\varepsilon_{[g]}'\bs P_{[g,h]}\bs\varepsilon_{[h]}\bs\varepsilon_{[l]}'\bs P_{[l,h],-gh}\bs P_{[h,g]}\bs\varepsilon_{[g]}\\
        &\quad+\frac{2}{k}\sum_{g=1}^G\sum_{h\neq g}\sum_{m,l\notin\{g,h\}}\bs\varepsilon_{[l]}'\bs P_{[l,g],-gh}\bs P_{[g,h]}\bs\varepsilon_{[h]}\bs\varepsilon_{[m]}'\bs P_{[m,h],-gh}\bs P_{[h,g]}\bs\varepsilon_{[g]}.
    \end{split}
\end{equation}
The first term is $\hat{V}^{AR}_{CLJ}(\bs\beta_0)$ and hence by \cref{thm:V cljar} consistent for $V^{AR}_{CLJ}$. The other three terms converge to zero in probability. I show this for the final term, the other two terms follow by the same arguments. A key step in the proof uses the decoupling result in Lemma G.6 of \citet{kolesar2026cluster}, which in turn is based on Theorem 3.1.1 of \citet{pena2012decoupling}.

First note that the final term has conditional expectation zero. The conditional variance is
\begin{equation}
    \begin{split}
        &\var\bigg(\frac{2}{k}\sum_{g=1}^G\sum_{h\neq g}\sum_{m,l\notin\{g,h\}}\bs\varepsilon_{[l]}'\bs P_{[l,g],-gh}\bs P_{[g,h]}\bs\varepsilon_{[h]}\bs\varepsilon_{[m]}'\bs P_{[k,m],-gh}\bs P_{[h,g]}\bs\varepsilon_{[g]}\bigg|\bs Z\bigg)\\
        &=\frac{4}{k^2}\E\bigg(\bigg[\sum_{g=1}^G\sum_{h\neq g}\sum_{k,l\notin\{g,h\}}\bs\varepsilon_{[l]}'\bs P_{[l,g],-gh}\bs P_{[g,h]}\bs\varepsilon_{[h]}\bs\varepsilon_{[m]}'\bs P_{[m,h],-gh}\bs P_{[h,g]}\bs\varepsilon_{[g]}\bigg]^2\bigg|\bs Z\bigg)\\
        &\leq\frac{C}{k^2}\E\bigg(\sum_{g=1}^G\sum_{h\neq g}\sum_{m,l\notin\{g,h\}}\bs\varepsilon_{[l]}'\bs P_{[l,g],-gh}\bs P_{[g,h]}\bs\varepsilon_{[h]}\bs\varepsilon_{[m]}'\bs P_{[m,h],-gh}\bs P_{[h,g]}\bs\varepsilon_{[g]}\\
        &\quad\bs\varepsilon_{[g]}'\bs P_{[g,h]}\bs P_{[h,m],-gh}\bs\varepsilon_{[m]}\bs\varepsilon_{[h]}'\bs P_{[h,g]}\bs P_{[h,l],-gh}\bs\varepsilon_{[l]}\bigg|\bs Z\bigg),
    \end{split}
\end{equation}
where the final line uses the decoupling result.

Consider two cases: $m\neq l$ and $m=l$. First
\begin{equation}\label{eq:first case}
    \begin{split}
        &\frac{C}{k^2}\E\bigg(\sum_{g=1}^G\sum_{h\neq g}\sum_{m\notin\{g,h\}}\sum_{l\notin\{g,h,m\}}\bs\varepsilon_{[g]}'\bs P_{[g,h]}\bs P_{[h,m],-gh}\bs\varepsilon_{[m]}\bs\varepsilon_{[h]}'\bs P_{[h,g]}\bs P_{[h,l],-gh}\bs\varepsilon_{[l]}\\
        &\quad\bs\varepsilon_{[l]}'\bs P_{[l,g],-gh}\bs P_{[g,h]}\bs\varepsilon_{[h]}\bs\varepsilon_{[m]}'\bs P_{[k,h],-gh}\bs P_{[h,g]}\bs\varepsilon_{[g]}\bigg|\bs Z\bigg)\\
        &\frac{C\nmax}{k^2}\E\bigg(\sum_{g=1}^G\sum_{h\neq g}\sum_{m\notin\{g,h\}}\sum_{l\notin\{g,h,m\}}\bs\varepsilon_{[g]}'\bs P_{[g,h]}\bs P_{[h,m],-gh}\bs\varepsilon_{[m]}\bs\varepsilon_{[h]}'\bs P_{[h,g]}\\
        &\quad\bs P_{[h,l],-gh}\bs P_{[l,g],-gh}\bs P_{[g,h]}\bs\varepsilon_{[h]}\bs\varepsilon_{[m]}'\bs P_{[k,h],-gh}\bs P_{[h,g]}\bs\varepsilon_{[g]}\bigg|\bs Z\bigg)
    \end{split}
\end{equation}
by \cref{assit:cljs eigval}.

Now note that for any vector $\bs v$ we have by Woodbury's matrix identity
\begin{equation}
    \begin{split}
        \sum_{l\notin\{g,h,k\}}\bs v_{[i]}'\bs P_{[i,l],-gh}\bs P_{[l,i],-gh}\bs v_{[i]}
        &\leq\sum_{l\notin\{g,h\}}\bs v_{[i]}'\bs P_{[i,l],-gh}\bs P_{[l,i],-gh}\bs v_{[i]}\\
        &=\bs v_{[i]}'\bs P_{[i,i],-gh}\bs v_{[i]}\\
        &=\bs v_{[i]}'(\bs P_{[i,i]}+\bs P_{[i,gh]}\bs M_{[gh,gh]}^{-1}\bs P_{[gh,i]})\bs v_{[i]},
    \end{split}
\end{equation}
where $[gh]$ is shorthand for $[g]\cup [h]$ such that for example $\bs P_{[i,gh]}=\bs Z_{[i]}(\bs Z'\bs Z)\bs Z_{[g]\cup [h]}$ and $\bs M_{[gh,gh]}=\bs I_{n_g+n_h}-\bs P_{[gh,gh]}$. Also,
\begin{equation}
    \begin{split}
        \lambdamax(\bs P_{[i,i]}+\bs P_{[i,gh]}\bs M_{[gh,gh]}^{-1}\bs P_{[gh,i]})&\leq\lambdamax(\bs P_{[i,i]})+\lambdamax(\bs M_{[gh,gh]}^{-1})\lambdamax(\bs P_{[i,gh]}\bs P_{[gh,i]})\\
        &\leq\lambdamax(\bs P_{[i,i]})+\lambdamax(\bs M_{[gh,gh]}^{-1})\lambdamax(\bs P_{[i,i]})\leq C,
    \end{split}
\end{equation}
by \cref{assit:cljar alt bound P}.

Then \cref{eq:first case} becomes by repeated applications
\begin{equation}
    \begin{split}
        &\frac{C\nmax}{k^2}\E\bigg(\sum_{g=1}^G\sum_{h\neq g}\sum_{m\notin\{g,h\}}\sum_{l\notin\{g,h,m\}}\bs\varepsilon_{[g]}'\bs P_{[g,h]}\bs P_{[h,m],-gh}\bs\varepsilon_{[m]}\bs\varepsilon_{[h]}'\bs P_{[hg]}\bs P_{[h,l],-gh}\\
        &\quad\bs P_{[lg],-gh}\bs P_{[g,h]}\bs\varepsilon_{[h]}\bs\varepsilon_{[m]}'\bs P_{[m,h],-gh}\bs P_{[h,g]}\bs\varepsilon_{[g]}\bigg|\bs Z\bigg)\\
        &\leq\frac{C\nmax}{k^2}\E\bigg(\sum_{g=1}^G\sum_{h\neq g}\sum_{m\notin\{g,h\}}\bs\varepsilon_{[g]}'\bs P_{[g,h]}\bs P_{[h,m],-gh}\bs\varepsilon_{[m]}\bs\varepsilon_{[h]}'\bs P_{[h,g]}\\
        &\quad\bs P_{[g,h]}\bs\varepsilon_{[h]}\bs\varepsilon_{[m]}'\bs P_{[m,h],-gh}\bs P_{[h,g]}\bs\varepsilon_{[g]}\bigg|\bs Z\bigg)\leq\frac{C\nmax^4k}{k^2}.
    \end{split}
\end{equation}

Now consider the case that $m=l$ and use that for any conformable matrices $\bs A, \bs B, \bs C,\bs D$ it holds that $\tr(\bs A\bs B\bs C\bs D)=\vec(\bs A')'(\bs D'\otimes\bs B)\vec(\bs C)$
\begin{equation}\label{eq:second case}
    \begin{split}
        &\frac{C}{k^2}\E\bigg(\sum_{g=1}^G\sum_{h\neq g}\sum_{l\notin\{g,h\}}\bs\varepsilon_{[g]}'\bs P_{[g,h]}\bs P_{[h,l],-gh}\bs\varepsilon_{[l]}\bs\varepsilon_{[l]}'\bs P_{[l,h],-gh}\bs P_{[h,g]}\bs\varepsilon_{[g]}\\
        &\quad\bs\varepsilon_{[h]}'\bs P_{[h,g]}\bs P_{[h,l],-gh}\bs\varepsilon_{[l]}\bs\varepsilon_{[l]}'\bs P_{[l,g],-gh}\bs P_{[g,h]}\bs\varepsilon_{[h]}\bigg|\bs Z\bigg)\\
        &=\frac{C}{k^2}\E(\sum_{g=1}^G\sum_{h\neq g}\sum_{l\notin\{g,h\}}\tr(\underbrace{\bs P_{[l,g],-gh}\bs P_{[g,h]}\bs\varepsilon_{[h]}\bs\varepsilon_{[g]}'\bs P_{[g,h]}\bs P_{[h,l],-gh}}_{\bs A}\underbrace{\bs\varepsilon_{[l]}\bs\varepsilon_{[l]}'}_{\bs B}\\
        &\quad\underbrace{\bs P_{[l,h],-gh}\bs P_{[h,g]}\bs\varepsilon_{[g]}\bs\varepsilon_{[h]}'\bs P_{[h,g]}\bs P_{[h,l],-gh}}_{\bs C}\underbrace{\bs\varepsilon_{[l]}\bs\varepsilon_{[l]}'}_{\bs D})|\bs Z)\\
        &\leq \frac{C\nmax^2}{k^2}\E\bigg(\sum_{g=1}^G\sum_{h\neq g}\sum_{l\notin\{g,h\}}\tr(\bs P_{[l,g],-gh}\bs P_{[g,h]}\bs\varepsilon_{[h]}\bs\varepsilon_{[g]}'\bs P_{[g,h]}\bs P_{[h,l],-gh}\\
        &\quad\bs P_{[l,h],-gh}\bs P_{[h,g]}\bs\varepsilon_{[g]}\bs\varepsilon_{[h]}'\bs P_{[h,g]}\bs P_{[h,l],-gh})\bigg|\bs Z\bigg)\\
        &\leq\frac{C\nmax^4k}{k^2},
    \end{split}
\end{equation}
since $\E\big(\lambdamax(\bs\varepsilon_{[l]}\bs\varepsilon_{[l]}'\otimes\bs\varepsilon_{[l]}\bs\varepsilon_{[l]}')\big|\bs Z\big)\leq C\nmax^2$. The conditional variance of the final term in \cref{eq:CF var AR decomp} hence converges in probability to zero because $\nmax^4k/k^2\to 0$.

\subsubsection{Variance of the score}
Next consider the variance of the score. Decompose it similarly as the variance of the AR
\begin{equation}\label{eq:CF var S decomp}
    \begin{split}
        \hat{\bs V}^{S}_{CF}(\bs\beta_0)&=\frac{1}{n}\sum_{g=1}^G\sum_{j=1}^G\sum_{h\notin\{g,j\}}\bs X_{[g]}'\bs P_{[g,h]}\bs\varepsilon_{[h]}\bs\varepsilon_{[h]}'\bs P_{[g,j]}\bs X_{[j]}\\
        &\quad+\sum_{g=1}^G\sum_{h\neq g}\bs X_{g}'\bs P_{[g,h]}\bs\varepsilon_{[h]}\bs\varepsilon_{[g]}'\bs P_{[g,h]}\bs X_{[h]}\\
        &\quad-\sum_{g=1}^G\sum_{j=1}^G\sum_{h\notin\{g,j\}}\sum_{l\neq h}\bs X_{[g]}'\bs P_{[g,h]}\bs P_{[h,l],-h}\bs\varepsilon_{[l]}\bs\varepsilon_{[h]}'\bs P_{[g,j]}\bs X_{[j]}\\
        &\quad-\sum_{g=1}^G\sum_{h\neq g}\sum_{l\notin\{g,h\}}\bs X_{[g]}'\bs P_{[g,h]}\bs\varepsilon_{[h]}\bs\varepsilon_{[l]}'\bs P_{[l,g],-gh}'\bs P_{[g,h]}\bs X_{[h]}.
    \end{split}
\end{equation}
The first two terms are consistent for $\bs V_{CLJ}^S$ by \cref{thm:V cljar}. The final two terms have expectation zero, so what remains is to show that the variance of each element of this covariance matrix converges to zero. Let $p_1,p_2\in\{1,\dots,p\}$ be given. We have
\begin{equation}\label{eq:VS 4}
    \begin{split}
        &\var\bigg(\frac{1}{n}\sum_{g=1}^G\sum_{j=1}^G\sum_{h\notin\{g,j\}}\sum_{l\neq h}\bs X_{(p_1),[g]}'\bs P_{[g,h]}\bs P_{[hl],-h}\bs\varepsilon_{[l]}\bs\varepsilon_{[h]}'\bs P_{[h,j]}\bs X_{(p_2),[j]}\bigg|\bs Z\bigg)\\
        &\leq\frac{C}{n^2}\bigg(\var\bigg(\sum_{g=1}^G\sum_{j=1}^G\sum_{h\notin\{g,j\}}\sum_{l\neq h}\bs\Pi_{(p_1)}'\bs Z_{[g]}'\bs P_{[g,h]}\bs P_{[h,l],-h}\bs\varepsilon_{[l]}\bs\varepsilon_{[h]}'\bs P_{[h,j]}\bs Z_{[j]}\bs\Pi_{(p_2)}\bigg|\bs Z\bigg)\\
        &\quad+\var\bigg(\sum_{g=1}^G\sum_{j=1}^G\sum_{h\notin\{g,j\}}\sum_{l\neq h}\bs\eta_{(p_1),[g]}'\bs P_{[g,h]}\bs P_{[h,l],-h}\bs\varepsilon_{l}\bs\varepsilon_{h}'\bs P_{[h,j]}\bs Z_{[j]}\bs\Pi_{(p_2)}\bigg|\bs Z\bigg)\\
        &\quad+\var\bigg(\sum_{g=1}^G\sum_{j=1}^G\sum_{h\notin\{g,j\}}\sum_{l\neq h}\bs\Pi_{(p_1)}'\bs Z_{[g]}'\bs P_{[g,h]}\bs P_{[h,l],-h}\bs\varepsilon_{[l]}\bs\varepsilon_{[h]}'\bs P_{[h,j]}\bs\eta_{(p_2),[j]}\bigg|\bs Z\bigg)\\
        &\quad+\var\bigg(\sum_{g=1}^G\sum_{j=1}^G\sum_{h\notin\{g,j\}}\sum_{l\neq h}\bs\eta_{(p_1),[g]}'\bs P_{[g,h]}\bs P_{[h,l],-h}\bs\varepsilon_{[l]}\bs\varepsilon_{[h]}'\bs P_{[h,j]}\bs\eta_{(p_2),[j]}\bigg|\bs Z\bigg)\bigg).
    \end{split}
\end{equation}

Consider the first term. By Lemma G.6 in \citet{kolesar2026cluster} we have
\begin{equation}
    \begin{split}
        &\frac{1}{n^2}\var\bigg(\sum_{g=1}^G\sum_{j=1}^G\sum_{h\notin\{g,j\}}\sum_{l\neq h}\bs\Pi_{(p_1)}'\bs Z_{[g]}'\bs P_{[g,h]}\bs P_{[h,l],-h}\bs\varepsilon_{[l]}\bs\varepsilon_{[h]}'\bs P_{[h,j]}\bs Z_{[j]}\bs\Pi_{(p_2)}|\bs Z\bigg)\\
        &\leq\frac{C}{n^2}\E\bigg(\sum_{g_1,g_2}\sum_{j_1,j_2}\sum_{h\notin\{g_1,g_2,j_1,j_2\}}\sum_{l\neq h}\bs\Pi_{(p_1)}'\bs Z_{[g_1]}'\bs P_{[g_1,h]}\bs P_{[h,l],-h}\bs\varepsilon_{[l]}\\
        &\quad\bs\varepsilon_{[l]}'\bs P_{[l,h],-h}\bs P_{[h,g_2]}\bs Z_{[g_2]}\bs\Pi_{(p_1)}\bs\Pi_{(p_2)}'\bs Z_{[j_1]}'\bs P_{[j_{1},h]}\bs\varepsilon_{[h]}\bs\varepsilon_{[h]}'\bs P_{[h,j_2]}\bs Z_{[j_2]}\bs\Pi_{(p_2)}\bigg|\bs Z\bigg)\\
        &\leq\frac{C\nmax^2}{n^2}\sum_{g_1,g_2}\sum_{j_1,j_2}\sum_{h\notin\{g_1,g_2,j_1,j_2\}}\sum_{l\neq h}\bs\Pi_{(p_1)}'\bs Z_{[g_1]}'\bs P_{[g_1,h]}\bs P_{[h,l],-h}\bs P_{[l,h],-h}\bs P_{[h,g_2]}\bs Z_{[g_2]}\bs\Pi_{(p_1)}\\
        &\quad\bs\Pi_{(p_2)}'\bs Z_{[j_1]}'\bs P_{[j_{1},h]}\bs P_{[h,j_2]}\bs Z_{[j_2]}\bs\Pi_{(p_2)}\\
        &\leq\frac{C\nmax^2}{n^2}\sum_{g_1,g_2}\sum_{j_1,j_2}\sum_{h\notin\{g_1,g_2,j_1,j_2\}}\bs\Pi_{(p_1)}'\bs Z_{[g_1]}'\bs P_{[g_1,h]}\bs P_{[h,g_2]}\bs Z_{[g_2]}\bs\Pi_{(p_1)}\\
        &\quad\bs\Pi_{(p_2)}'\bs Z_{[j_1]}'\bs P_{[j_{1},h]}\bs P_{[h,j_2]}\bs Z_{[j_2]}\bs\Pi_{(p_2)}\\
        &=\frac{C\nmax^2}{n^2}\sum_{h=1}^G\sum_{g_1\neq h}\sum_{g_2\neq h}\sum_{j_1\neq h}\sum_{j_2\neq h}\bs\Pi_{(p_1)}'\bs Z_{[g_1]}'\bs P_{[g_1,h]}\bs P_{[h,g_2]}\bs Z_{[g_2]}\bs\Pi_{(p_1)}\\
        &\quad\bs\Pi_{(p_2)}'\bs Z_{[j_1]}'\bs P_{[j_{1},h]}\bs P_{[h,j_2]}\bs Z_{[j_2]}\bs\Pi_{(p_2)}\\
        &=\frac{C\nmax^2}{n^2}\sum_{h=1}^G\bigg(\sum_{g\neq h}\bs\Pi_{(p_1)}'\bs Z_{[g]}'\bs P_{[g,h]}\bigg)\bigg(\sum_{g\neq h}\bs P_{[h,g]}\bs Z_{[g]}\bs\Pi_{(p_1)}\bigg)\\
        &\quad\bigg(\sum_{g\neq h}\bs\Pi_{(p_2)}'\bs Z_{[g]}'\bs P_{[g,h]}\bigg)\bigg(\sum_{g\neq h}\bs P_{[h,g]}\bs Z_{[g]}\bs\Pi_{(p_2)}\bigg),
    \end{split}
\end{equation}
and then follow same steps as in \cref{eq:PiZPHPZPi}.

Consider the second term of \cref{eq:VS 4}. Similarly, by Lemma G.6 in \citet{kolesar2026cluster}
\begin{equation}
    \begin{split}
        &\frac{1}{n^2}\var\bigg(\sum_{g=1}^G\sum_{j=1}^G\sum_{h\notin\{g,j\}}\sum_{l\neq h}\bs\eta_{(p_1),[g]}'\bs P_{[g,h]}\bs P_{[h,l],-h}\bs\varepsilon_{[l]}\bs\varepsilon_{[h]}'\bs P_{[h,j]}\bs Z_{[j]}\bs\Pi_{(p_2)}\bigg|\bs Z\bigg)\\
        &\leq\frac{C}{n^2}\sum_{g=1}^G\sum_{j_1,j_2}\sum_{h\notin\{g,j_1,j_2\}}\sum_{l\neq h}\E(\bs\eta_{(p_1),[g]}'\bs P_{[g,h]}\bs P_{[h,l],-h}\bs\varepsilon_{[l]}\bs\varepsilon_{[l]}'\bs P_{[l,h],-h}\bs P_{[h,g]}\bs\eta_{(p_1),[g]}\\
        &\quad\bs\Pi_{(p_2)}'\bs Z_{[j_1]}'\bs P_{[j_1,h]}\bs\varepsilon_{[h]}\bs\varepsilon_{[h]}'\bs P_{[h,j_2]}\bs Z_{[j_2]}\bs\Pi_{(p_2)}|\bs Z).
    \end{split}
\end{equation}
I discern two cases: $g=l$ and $g\neq l$. For the first case use again $\tr(\bs A\bs B\bs C\bs D)=\vec(\bs A')'(\bs D'\otimes\bs B)\vec(\bs C)$
\begin{equation}
    \begin{split}
        &\frac{1}{n^2}\sum_{g=1}^G\sum_{j_1,j_2}\sum_{h\notin\{g,j_1,j_2\}}\E(\bs\eta_{(p_1),[g]}'\bs P_{[g,h]}\bs P_{[h,g],-h}\bs\varepsilon_{[g]}\bs\varepsilon_{[g]}'\bs P_{[g,h],-h}\bs P_{[h,g]}\bs\eta_{(p_1),[g]}\\
        &\quad\bs\Pi_{(p_2)}'\bs Z_{[j]}'\bs P_{[j,h]}\bs\varepsilon_{[h]}\bs\varepsilon_{[h]}'\bs P_{[h,j]}\bs Z_{[j]}\bs\Pi_{(p_2)}|\bs Z)\\
        &=\frac{C}{n^2}\sum_{g=1}^G\sum_{h\neq g}\E\bigg(\tr(\underbrace{\bs P_{[g,h]}\bs P_{[h,g],-h}}_{\bs A}\underbrace{\bs\varepsilon_{[g]}\bs\varepsilon_{[g]}'}_{\bs B}\underbrace{\bs P_{[g,h],-h}\bs P_{[h,g]}}_{\bs C}\underbrace{\bs\eta_{(p_1),[g]}\bs\eta_{(p_1),[g]}'}_{\bs D})\\
        &\quad\bigg(\sum_{j\neq h}\bs\Pi_{(p_2)}'\bs Z_{[j]}'\bs P_{[j,h]}\bigg)\bs\varepsilon_{[h]}\bs\varepsilon_{[h]}'\bigg(\sum_{j\neq h}\bs P_{[h,j]}\bs Z_{[j]}\bs\Pi_{(p_2)}\bigg)\bigg|\bs Z\bigg)\\
        &\leq\frac{C\nmax^3}{n^2}\sum_{g=1}^G\sum_{h\neq g}\tr(\bs P_{[g,h]}\bs P_{[h,g],-h}\bs P_{[g,h],-h}\bs P_{[h,g]})\\
        &\quad\bigg(\sum_{j\neq h}\bs\Pi_{(p_2)}'\bs Z_{[j]}'\bs P_{[j,h]}\bigg)\bigg(\sum_{j\neq h}\bs P_{[h,j]}\bs Z_{[j]}\bs\Pi_{(p_2)}\bigg)\\
        &\leq\frac{C\nmax^3}{n^2}\sum_{g=1}^G\sum_{h=1}^G\tr(\bs P_{[g,h]}\bs P_{[h,g],-h}\bs P_{[g,h],-h}\bs P_{[h,g]})\\
        &\quad\bigg(\sum_{j=1}^G\bs\Pi_{(p_2)}'\bs Z_{[j]}'\bs P_{[j,h]}\bigg)\bigg(\sum_{j=1}^G\bs P_{[h,j]}\bs Z_{[j]}\bs\Pi_{(p_2)}\bigg)\\
        &\leq\frac{C\nmax^3}{n^2}\sum_{g=1}^G\sum_{h=1}^G\tr(\bs P_{[g,h]}\bs P_{[h,g]})\bs\Pi_{(p_2)}'\bs Z_{[h]}'\bs Z_{[h]}\bs\Pi_{(p_2)}\\
        &\leq\frac{C\nmax^3}{n^2}\sum_{h=1}^G\tr(\bs P_{[h,h]})\bs\Pi_{(p_2)}'\bs Z_{[h]}'\bs Z_{[h]}\bs\Pi_{(p_2)}\\
        &\leq\frac{C\nmax^4}{n^2}\sum_{h=1}^G\bs\Pi_{(p_2)}'\bs Z_{[h]}'\bs Z_{[h]}\bs\Pi_{(p_2)}\leq\frac{C\nmax^4n}{n^2}.
    \end{split}
\end{equation}

For the second case
\begin{equation}
    \begin{split}
        &\frac{C}{n^2}\sum_{g=1}^G\sum_{j_1,j_2}\sum_{h\notin\{g,j_1,j_2\}}\sum_{l\notin\{g, h\}}\E(\bs\eta_{(p_1),[g]}'\bs P_{[g,h]}\bs P_{[h,l],-h}\bs\varepsilon_{[l]}\bs\varepsilon_{[l]}'\bs P_{[l,h],-h}\bs P_{[h,g]}\bs\eta_{(p_1),[g]}\\
        &\quad\bs\Pi_{(p_2)}'\bs Z_{[j_1]}'\bs P_{[j_1,h]}\bs\varepsilon_{[h]}\bs\varepsilon_{[h]}'\bs P_{[h,j_2]}\bs Z_{[j_2]}\bs\Pi_{(p_2)}|\bs Z)\\
        &\leq\frac{C\nmax^3}{n^2}\sum_{g=1}^G\sum_{h\neq g}\sum_{l\notin\{g, h\}}\tr(\bs P_{[g,h]}\bs P_{[h,l],-h}\bs P_{[l,h],-h}\bs P_{[h,g]})\\
        &\quad\bigg(\sum_{j\neq h}\bs\Pi_{(p_2)}'\bs Z_{[j]}'\bs P_{[j,h]}\bigg)\bigg(\sum_{j\neq h}\bs P_{[h,j]}\bs Z_{[j]}\bs\Pi_{(p_2)}\bigg)\\
        &\leq\frac{C\nmax^3}{n^2}\sum_{g=1}^G\sum_{h=1}^G\sum_{l=1}^G\tr(\bs P_{[g,h]}\bs P_{[h,l],-h}\bs P_{[l,h],-h}\bs P_{[h,g]})\\
        &\quad\bigg(\sum_{j\neq h}\bs\Pi_{(p_2)}'\bs Z_{[j]}'\bs P_{[j,h]}\bigg)\bigg(\sum_{j\neq h}\bs P_{[h,j]}\bs Z_{[j]}\bs\Pi_{(p_2)}\bigg)\\
        &\leq\frac{C\nmax^4n}{n^2},
    \end{split}
\end{equation}
following the same steps as for the first case. The third term in \cref{eq:VS 4} follows analogously.

Using again Lemma G.6 in \citet{kolesar2026cluster}, the fourth term in \cref{eq:VS 4} becomes
\begin{equation}
    \begin{split}
        &\frac{1}{n^2}\var\bigg(\sum_{g=1}^G\sum_{j=1}^G\sum_{h\notin\{g,j\}}\sum_{l\neq h}\bs\eta_{(p_1),[g]}'\bs P_{[g,h]}\bs P_{[h,l],-h}\bs\varepsilon_{[l]}\bs\varepsilon_{[h]}'\bs P_{[h,j]}\bs\eta_{(p_2),[j]}\bigg|\bs Z\bigg)\\
        &\leq\frac{C}{n^2}\sum_{g=1}^G\sum_{j=1}^G\sum_{h\notin\{g,j\}}\sum_{l\neq h}\E(\bs\eta_{(p_1),[g]}'\bs P_{[g,h]}\bs P_{[h,l],-h}\bs\varepsilon_{[l]}\bs\varepsilon_{[l]}'\bs P_{[l,h],-h}\bs P_{[h,g]}\bs\eta_{(p_1),[g]}\\
        &\quad\bs\eta_{(p_2),[j]}'\bs P_{[j,h]}\bs\varepsilon_{[h]}\bs\varepsilon_{[h]}'\bs P_{[h,j]}\bs\eta_{(p_2),[j]}|\bs Z).
    \end{split}
\end{equation}
Distinguish five cases $g=j=l$, $g\neq j=l$, $g=j\neq l$, $g=l\neq j$, $g\neq j\neq l\neq g$.

First $g=j=l$
\begin{equation}
    \begin{split}
        &\frac{C}{n^2}\sum_{g=1}^G\sum_{h\neq g}\E(\bs\eta_{(p_1),[g]}'\bs P_{[g,h]}\bs P_{[h,g],-h}\bs\varepsilon_{[g]}\bs\varepsilon_{[g]}'\bs P_{[g,h],-h}\bs P_{[h,g]}\bs\eta_{(p_1),[g]}\\
        &\quad\bs\eta_{(p_2),[g]}'\bs P_{[g,h]}\bs\varepsilon_{[h]}\bs\varepsilon_{[h]}'\bs P_{[h,g]}\bs\eta_{(p_2),[g]}|\bs Z)\\
        &\leq\frac{C\nmax}{n^2}\sum_{g=1}^G\sum_{h\neq g}\E(\bs\eta_{(p_1),[g]}'\bs P_{[g,h]}\bs P_{[h,g],-h}\bs\varepsilon_{[g]}\bs\varepsilon_{[g]}'\bs P_{[g,h],-h}\bs P_{[h,g]}\bs\eta_{(p_1),[g]}\\
        &\quad\bs\eta_{(p_2),[g]}'\bs P_{[g,h]}\bs P_{[h,g]}\bs\eta_{(p_2),[g]}|\bs Z)\\
        &\leq\frac{C\nmax}{n^2}\sum_{g=1}^G\sum_{h\neq g}\E(\bs\eta_{(p_1),[g]}'\bs P_{[g,h]}\bs P_{[h,g],-h}\bs\varepsilon_{[g]}\bs\varepsilon_{[g]}'\bs P_{[g,h],-h}\bs P_{[h,g]}\bs\eta_{(p_1),[g]}\\
        &\quad\bs\eta_{(p_2),[g]}'\bs P_{[g,g]}\bs\eta_{(p_2),[g]}|\bs Z)\\
        &\leq\frac{C\nmax}{n^2}\sum_{g=1}^G\sum_{h\neq g}\E(\bs\eta_{(p_1),[g]}'\bs P_{[g,h]}\bs P_{[h,g],-h}\bs\varepsilon_{[g]}\bs\varepsilon_{[g]}'\bs P_{[g,h],-h}\bs P_{[h,g]}\bs\eta_{(p_1),[g]}\bs\eta_{(p_2),[g]}'\bs\eta_{(p_2),[g]}|\bs Z)\\
        &\leq\frac{C\nmax}{n^2}\sum_{g=1}^G\sum_{h\neq g}\E\big(\lambdamax(\bs\varepsilon_{[g]}\bs\varepsilon_{[g]}')\lambdamax(\bs\eta_{(p_1),[g]}\bs\eta_{(p_1),[g]}')\\
        &\quad\tr(\bs P_{[g,h]}\bs P_{[h,g],-h}\bs P_{[g,h],-h}\bs P_{[h,g]})\bs\eta_{(p_2),[g]}'\bs\eta_{(p_2),[g]}\big|\bs Z\big)\\
        &\leq\frac{C\nmax}{n^2}\sum_{g=1}^G\sum_{h\neq g}\E(\bs\varepsilon_{[g]}'\bs\varepsilon_{[g]}\bs\eta_{(p_1),[g]}'\bs\eta_{(p_1),[g]}\bs\eta_{(p_2),[g]}'\bs\eta_{(p_2),[g]}|\bs Z)\\
        &\quad\tr(\bs P_{[g,h]}\bs P_{[h,g],-h}\bs P_{[g,h],-h}\bs P_{[h,g]})\leq\frac{k\nmax^4}{n^2},
    \end{split}
\end{equation}
using that $abc\leq C\big(a^2+(bc)^2\big)\leq C(a^2+b^4+c^4)$, $\E\big((\bs\eta_{(p_1),[g]}'\bs\eta_{(p_1),[g]})^4\big|\bs Z\big)\leq\E\big((\sum_{i\in[g]}\eta_{(p_1),i}^2)^4\big|\bs Z\big)\leq n_g^3\E\big(\sum_{i=1}^{n_{g}}\eta_{(p_1),i}^8\big|\bs Z\big)\leq n_g^3 C$ by \cref{assit:cljar alt moment}, and $\sum_{g=1}^G\sum_{h\neq g}\tr(\bs P_{[g,h]}\bs P_{[h,g],-h}\bs P_{[g,h],-h}\bs P_{[h,g]})\leq C k$. The other four cases can be bounded similarly.

Element $(p_1,p_2)$ of the covariance matrix of the final term in \cref{eq:CF var S decomp} can be written as
\begin{equation}\label{eq:VS 42}
    \begin{split}
        &\frac{1}{n^2}\var\bigg(\sum_{g=1}^G\sum_{h\neq g}\sum_{l\notin\{g,h\}}\bs X_{(p_1),[g]}'\bs P_{[g,h]}\bs\varepsilon_{[h]}\bs\varepsilon_{[l]}'\bs P_{[l,g],-gh}\bs P_{[g,h]}\bs X_{(p_2),[h]}\bigg|\bs Z\bigg)\\
        &\leq\frac{C}{n^2}\bigg(\var\bigg(\sum_{g=1}^G\sum_{h\neq g}\sum_{l\notin\{g,h\}}\bs\Pi_{(p_1)}'\bs Z_{[g]}'\bs P_{[g,h]}\bs\varepsilon_{[h]}\bs\varepsilon_{[l]}'\bs P_{[l,g],-gh}\bs P_{[g,h]}\bs Z_{[h]}\bs\Pi_{(p_2)}\bigg|\bs Z\bigg)\\
        &\quad+\var\bigg(\sum_{g=1}^G\sum_{h\neq g}\sum_{l\notin\{g,h\}}\bs\eta_{(p_1),[g]}'\bs P_{[g,h]}\bs\varepsilon_{[h]}\bs\varepsilon_{[l]}'\bs P_{[l,g],-gh}\bs P_{[g,h]}\bs Z_{[h]}\bs\Pi_{(p_2)}\bigg|\bs Z\bigg)\\
        &\quad+\var\bigg(\sum_{g=1}^G\sum_{h\neq g}\sum_{l\notin\{g,h\}}\bs\Pi_{(p_1)}'\bs Z_{[g]}'\bs P_{[g,h]}\bs\varepsilon_{[h]}\bs\varepsilon_{[l]}'\bs P_{[l,g],-gh}\bs P_{[g,h]}\bs\eta_{(p_2),[h]}\bigg|\bs Z\bigg)\\
        &\quad+\var\bigg(\sum_{g=1}^G\sum_{h\neq g}\sum_{l\notin\{g,h\}}\bs\eta_{(p_1),[g]}'\bs P_{[g,h]}\bs\varepsilon_{[h]}\bs\varepsilon_{[l]}'\bs P_{[l,g],-gh}\bs P_{[g,h]}\bs\eta_{(p_2),[h]}\bigg|\bs Z\bigg)\bigg).
    \end{split}
\end{equation}

For the first term use Lemma G.6 of \citet{kolesar2026cluster} to write
\begin{equation}
    \begin{split}
        &\frac{1}{n^2}\var\bigg(\sum_{g=1}^G\sum_{h\neq g}\sum_{l\notin\{g,h\}}\bs\Pi_{(p_1)}'\bs Z_{[g]}'\bs P_{[g,h]}\bs\varepsilon_{[h]}\bs\varepsilon_{[l]}'\bs P_{[l,g],-gh}\bs P_{[g,h]}\bs Z_{[h]}\bs\Pi_{(p_2)}\bigg|\bs Z\bigg)\\
        &\leq\frac{C}{n^2}\E\bigg(\sum_{g_1,g_2}\sum_{h\notin\{g_1,g_2\}}\sum_{l\notin\{g_1,g_2,h\}}\bs\Pi_{(p_1)}'\bs Z_{[g_1]}'\bs P_{[g_1,h]}\bs\varepsilon_{[h]}\bs\varepsilon_{[h]}'\bs P_{[h,g]_2}\bs Z_{[g_2]}\bs\Pi_{(p_1)}\\
        &\quad\bs\Pi_{(p_2)}'\bs 
        Z_{[h]}'\bs P_{[h,g_1]}\bs P_{[g_1,l],-g_1h}\bs\varepsilon_{[l]}\bs\varepsilon_{[l]}'\bs P_{[l,g_2],-g_2h}\bs P_{[g_2,h]}\bs Z_{[h]}\bs\Pi_{(p_2)}\bigg|\bs Z\bigg)\\
        &\leq\frac{C}{n^2}\sum_{g_1,g_2}\sum_{h\notin\{g_1,g_2\}}\sum_{l\notin\{g_1,g_2,h\}}\bs\Pi_{(p_1)}'\bs Z_{[g_1]}'\bs P_{[g_1,h]}\bs\Sigma_{h}\bs P_{[h,g_2]}\bs Z_{[g_2]}\bs\Pi_{(p_1)}\\
        &\quad\bs\Pi_{(p_2)}'\bs Z_{[h]}'\bs P_{[h,g_1]}\bs P_{[g_1,l],-g_1h}\bs\Sigma_{l}\bs P_{[l,g_2],-g_2h}\bs P_{[g_2,h]}\bs Z_{[h]}\bs\Pi_{(p_2)}\\
        &\leq\frac{C}{n^2}\sum_{g_1,g_2}\sum_{h\notin\{g_1,g_2\}}\bs\Pi_{(p_1)}'\bs Z_{[g_1]}'\bs P_{[g_1,h]}\bs\Sigma_{h}\bs P_{[h,g_2]}\bs Z_{[g_2]}\bs\Pi_{(p_1)}\\
        &\quad\bs\Pi_{(p_2)}'\bs Z_{[h]}'\bs P_{[h,g_1]}[\bs P\bs\Sigma_{-g_1g_2h}\bs P]_{[g_1,g_2]}\bs P_{[g_2,h]}\bs Z_{[h]}\bs\Pi_{(p_2)}\\
        &\leq\frac{C}{n^2}\sum_{g_1,g_2}\sum_{h\notin\{g_1,g_2\}}(\bs\Pi_{(p_1)}'\bs Z_{[g_1]}'\bs P_{[g_1,h]}\bs\Sigma_{h}\bs P_{[h,g_2]}\bs Z_{[g_2]}\bs\Pi_{(p_1)})^2\\
        &\quad+\big(\bs\Pi_{(p_2)}'\bs Z_{[h]}'\bs P_{[h,g_1]}[\bs P\bs\Sigma_{-g_1g_2h}\bs P]_{[g_1,g_2]}\bs P_{[g_2,h]}\bs Z_{[h]}\bs\Pi_{(p_2)}\big)^2\leq\frac{C\nmax^4k}{n^2},
    \end{split}
\end{equation}
where $\bs\Sigma_{-g_1g_2h}$ is the $n\times n$ matrix $\bs\Sigma$ with the blocks corresponding to clusters $g_1$, $g_2$ and $h$ set to zero.

Similarly, the second term in \cref{eq:VS 42} can be bounded by
\begin{equation}
    \begin{split}
        &\frac{1}{n^2}\var\bigg(\sum_{g=1}^G\sum_{h\neq g}\sum_{l\notin\{g,h\}}\bs\eta_{(p_1),[g]}'\bs P_{[g,h]}\bs\varepsilon_{[h]}\bs\varepsilon_{[l]}'\bs P_{[l,g],-gh}\bs P_{[g,h]}\bs Z_{[h]}\bs\Pi_{(p_2)}\bigg|\bs Z\bigg)\\
        &\leq\frac{C}{n^2}\sum_{g=1}^G\sum_{h\neq g}\sum_{l\notin\{g,h\}}\E(\bs\eta_{(p_1),[g]}'\bs P_{[g,h]}\bs\varepsilon_{[h]}\bs\varepsilon_{[h]}'\bs P_{[h,g]}\bs\eta_{(p_1),[g]}\\
        &\quad\bs\Pi_{(p_2)}'\bs 
        Z_{[h]}'\bs P_{[h,g]}\bs P_{[g,l],-gh}\bs\varepsilon_{[l]}\bs\varepsilon_{[l]}'\bs P_{[l,g],-gh}\bs P_{[g,h]}\bs Z_{[h]}\bs\Pi_{(p_2)}|\bs Z)\\
        &\leq\frac{C\nmax^4k}{n^2}.
    \end{split}
\end{equation}

The third term in \cref{eq:VS 42}
\begin{equation}\label{eq:VS 423}
    \begin{split}
        &\frac{1}{n^2}\var\bigg(\sum_{g=1}^G\sum_{h\neq g}\sum_{l\notin\{g,h\}}\bs\Pi_{(p_1)}'\bs Z_{[g]}'\bs P_{[g,h]}\bs\varepsilon_{[h]}\bs\varepsilon_{[l]}'\bs P_{[l,g],-gh}\bs P_{[g,h]}\bs\eta_{(p_2),[h]}\bigg|\bs Z\bigg)\\
        &\leq\frac{C}{n^2}\sum_{g_1,g_2}\sum_{h\notin\{ g_1,g_2\}}\sum_{l\notin\{g_1,g_2,h\}}\E(\bs\Pi_{(p_1)}'\bs Z_{[g_1]}'\bs P_{[g_1,h]}\bs\varepsilon_{[h]}\bs\varepsilon_{[h]}'\bs P_{[h,g_2]}\bs Z_{[g_2]}\bs\Pi_{(p_1)}\\
        &\quad\bs\eta_{(p_2),[h]}'\bs P_{[h,g_1]}\bs P_{[g_1,l],-g_1h}\bs\varepsilon_{[l]}\bs\varepsilon_{[l]}'\bs P_{[l,g_2],-g_2h}\bs P_{[g_2,h]}\bs\eta_{(p_2),[h]}|\bs Z)\\
        &\leq\frac{C}{n^2}\sum_{g_1,g_2}\sum_{h\notin\{ g_1,g_2\}}\sum_{l\notin\{g_1,g_2,h\}}\E(\bs\Pi_{(p_1)}'\bs Z_{[g_1]}'\bs P_{[g_1,h]}\bs\varepsilon_{[h]}\bs\eta_{(p_2),[h]}'\bs P_{[h,g_2]}\bs P_{[g_2,l],-g_2h}\bs\varepsilon_{[l]}\\
        &\quad\bs\varepsilon_{[l]}'\bs P_{[l,g_2],-g_2h}\bs P_{[g_2,h]}\bs\eta_{(p_2),[h]}\bs\varepsilon_{[h]}'\bs P_{[h,g_1]}\bs Z_{[g_1]}\bs\Pi_{(p_1)}|\bs Z)\\
        &\quad+\E(\bs\Pi_{(p_1)}'\bs Z_{[g_2]}'\bs P_{[g_2,h]}\bs\varepsilon_{[h]}\bs\eta_{(p_2),[h]}'\bs P_{[h,g_1]}\bs P_{[g_1,l],-g_1h}\bs\varepsilon_{[l]}\\
        &\quad\bs\varepsilon_{[l]}'\bs P_{[l,g_1],-g_1h}\bs P_{[h,g_1]}\bs\eta_{(p_2),[h]}\bs\varepsilon_{[h]}'\bs P_{[h,g_2]}\bs Z_{[g_2]}\bs\Pi_{(p_1)}|\bs Z)\\
        &=\frac{C}{n^2}\sum_{g_1,g_2}\sum_{h\notin\{ g_1,g_2\}}\sum_{l\notin\{g_1,g_2,h\}}\E(\bs\Pi_{(p_1)}'\bs Z_{[g_1]}'\bs P_{[g_1,h]}\bs\varepsilon_{[h]}\bs\eta_{(p_2),[h]}'\bs P_{[h,g_2]}\bs P_{[g_2,l],-g_2h}\bs\varepsilon_{[l]}\\
        &\quad\bs\varepsilon_{[l]}'\bs P_{[l,g_2],-g_2h}\bs P_{[g_2,h]}\bs\eta_{(p_2),[h]}\bs\varepsilon_{[h]}'\bs P_{[h,g_1]}\bs Z_{[g_1]}\bs\Pi_{(p_1)}|\bs Z)\\
        &\leq\frac{C\nmax}{n^2}\sum_{g_1,g_2}\sum_{h\notin\{ g_1,g_2\}}\sum_{l\notin\{g_2,h\}}\E(\bs\Pi_{(p_1)}'\bs Z_{[g_1]}'\bs P_{[g_1,h]}\bs\varepsilon_{[h]}\bs\eta_{(p_2),[h]}'\bs P_{[h,g_2]}\bs P_{[g_2,l],-g_2h}\\
        &\quad\bs P_{[l,g_2],-g_2h}\bs P_{[g_2,h]}\bs\eta_{(p_2),[h]}\bs\varepsilon_{[h]}'\bs P_{[h,g_1]}\bs Z_{[g_1]}\bs\Pi_{(p_1)}|\bs Z)\\
        &\leq\frac{C\nmax}{n^2}\sum_{g_1,g_2}\sum_{h\notin\{ g_1,g_2\}}\E(\bs\Pi_{(p_1)}'\bs Z_{[g_1]}'\bs P_{[g_1,h]}\bs\varepsilon_{[h]}\bs\eta_{(p_2),[h]}'\bs P_{[h,g_2]}\\
        &\quad\bs P_{[g_2,g_2],-g_2h}\bs P_{[g_2,h]}\bs\eta_{(p_2),[h]}\bs\varepsilon_{[h]}'\bs P_{[h,g_1]}\bs Z_{[g_1]}\bs\Pi_{(p_1)}|\bs Z)\\
        &\leq\frac{C\nmax}{n^2}\sum_{g_1,g_2}\sum_{h\notin\{ g_1,g_2\}}\E(\bs\Pi_{(p_1)}'\bs Z_{[g_1]}'\bs P_{[g_1,h]}\bs\varepsilon_{[h]}\bs\eta_{(p_2),[h]}'\bs P_{[h,g_2]}\bs P_{[g_2,h]}\bs\eta_{(p_2),[h]}\\
        &\quad\bs\varepsilon_{[h]}'\bs P_{[h,g_1]}\bs Z_{[g_1]}\bs\Pi_{(p_1)}|\bs Z)\\
        &\leq\frac{C\nmax}{n^2}\sum_{g_1,g_2,h}\E(\bs\Pi_{(p_1)}'\bs Z_{[g_1]}'\bs P_{[g_1,h]}\bs\varepsilon_{[h]}\bs\eta_{(p_2),[h]}'\bs P_{[h,g_2]}\bs P_{[g_2,h]}\bs\eta_{(p_2),[h]}\bs\varepsilon_{[h]}'\bs P_{[h,g_1]}\bs Z_{[g_1]}\bs\Pi_{(p_1)}|\bs Z)\\
        &\leq\frac{C\nmax}{n^2}\sum_{g_1,h}\E(\bs\Pi_{(p_1)}'\bs Z_{[g_1]}'\bs P_{[g_1,h]}\bs\varepsilon_{[h]}\bs\eta_{(p_2),[h]}'\bs P_{[h,h]}\bs\eta_{(p_2),[h]}\bs\varepsilon_{[h]}'\bs P_{[h,g_1]}\bs Z_{[g_1]}\bs\Pi_{(p_1)}|\bs Z)\\
        &\leq\frac{C\nmax^4k}{n^2}.
    \end{split}
\end{equation}

For the fourth term in \cref{eq:VS 42}
\begin{equation}\label{eq:VS 424}
    \begin{split}
        &\frac{1}{n^2}\var\bigg(\sum_{g=1}^G\sum_{h\neq g}\sum_{l\notin\{g,h\}}\bs\eta_{(p_1),[g]}'\bs P_{[g,h]}\bs\varepsilon_{[h]}\bs\varepsilon_{[l]}'\bs P_{[l,g],-gh}\bs P_{[g,h]}\bs\eta_{(p_2),[h]}\bigg|\bs Z\bigg)\\
        &\leq\frac{C}{n^2}\sum_{g=1}^G\sum_{h\neq g}\sum_{l\notin\{g,h\}}\E(\bs\eta_{(p_1),[g]}'\bs P_{[g,h]}\bs\varepsilon_{[h]}\bs\varepsilon_{[h]}'\bs P_{[h,g]}\bs\eta_{(p_1),[g]}\\
        &\quad\bs\eta_{(p_2),[h]}'\bs P_{[h,g]}\bs P_{[g,l],-gh}\bs\varepsilon_{[l]}\bs\varepsilon_{[l]}'\bs P_{[l,g],-gh}\bs P_{[g,h]}\bs\eta_{(p_2),[h]}|\bs Z)\\
        &\leq\frac{C}{n^2}\sum_{g=1}^G\sum_{h\neq g}\sum_{l\notin\{g,h\}}\E(\bs\varepsilon_{[h]}'\bs P_{[h,g]}\bs\Omega_{g}^{(p_1,p_1)}\bs P_{[g,h]}\bs\varepsilon_{[h]}\\
        &\quad\bs\eta_{(p_2),[h]}'\bs P_{[h,g]}\bs P_{[g,l],-gh}\bs\Sigma_{l}\bs P_{[l,g],-gh}\bs P_{[g,h]}\bs\eta_{(p_2),[h]})\leq\frac{C\nmax^4k}{n^2}.
    \end{split}
\end{equation}
Hence, given that $\nmax^4k/n^2\to0$, \cref{eq:VS 42} converges to zero and the variance estimator of the score is consistent.

\subsubsection{Covariance of AR and the score}
Finally, consider the covariance between the AR and the score statistic.
\begin{equation}\label{eq:C}
    \begin{split}
        \hat{\bs C}_{CF}(\bs\beta_0)&=\frac{2}{\sqrt{nk}}\bigg(\sum_{g=1}^G\sum_{h\neq g}\bs X_{[g]}'\bs P_{[g,h]}\bs\varepsilon_{[h]}\bs\varepsilon_{[g]}'\bs P_{[g,h]}\bs\varepsilon_{[h]}\\
        &\quad-\sum_{g=1}^G\sum_{h\neq g}\sum_{l\notin\{g,h\}}\bs X_{[g]}'\bs P_{[g,h]}\bs P_{[h,l],-gh}\bs\varepsilon_{[l]}\bs\varepsilon_{[g]}'\bs P_{[g,h]}\bs\varepsilon_{[h]}\\
        &\quad-\sum_{g=1}^G\sum_{h\neq g}\sum_{l\notin\{g,h\}}\bs X_{[g]}'\bs P_{[g,h]}\bs\varepsilon_{[h]}\bs\varepsilon_{[l]}'\bs P_{[l,g],-gh}\bs P_{[g,h]}\bs\varepsilon_{[h]}\\
        &\quad+\sum_{g=1}^G\sum_{h\neq g}\sum_{l\notin\{g,h\}}\sum_{m\notin\{g,h\}}\bs X_{[g]}'\bs P_{[g,h]}\bs P_{[h,l],-gh}\bs\varepsilon_{[l]}\bs\varepsilon_{[m]}'\bs P_{[m,g],-gh}\bs P_{[g,h]}\bs\varepsilon_{[h]}\bigg).
    \end{split}
\end{equation}
The first term is consistent for the covariance by \cref{thm:V cljar}. Note that the final three terms have expectation zero, hence it remains to be shown that their variance converges to zero.

Start with element $p_1$ of the first remainder term
\begin{equation}\label{eq:C 1}
    \begin{split}
        &\frac{C}{nk}\var\bigg(\sum_{g=1}^G\sum_{h\neq g}\sum_{l\notin\{g,h\}}\bs X_{(p_1),[g]}'\bs P_{[g,h]}\bs P_{[h,l],-gh}\bs\varepsilon_{[l]}\bs\varepsilon_{[g]}'\bs P_{[g,h]}\bs\varepsilon_{[h]}\bigg|\bs Z\bigg)\\
        &\leq\frac{C}{nk}\bigg(\var\bigg(\sum_{g=1}^G\sum_{h\neq g}\sum_{l\notin\{g,h\}}\bs\Pi_{(p_1)}'\bs Z_{[g]}'\bs P_{[g,h]}\bs P_{[h,l],-gh}\bs\varepsilon_{[l]}\bs\varepsilon_{[g]}'\bs P_{[g,h]}\bs\varepsilon_{[h]}\bigg|\bs Z\bigg)\\
        &\quad+\var\bigg(\sum_{g=1}^G\sum_{h\neq g}\sum_{l\notin\{g,h\}}\bs\eta_{(p_1),[g]}'\bs P_{[g,h]}\bs P_{[h,l],-gh}\bs\varepsilon_{[l]}\bs\varepsilon_{[g]}'\bs P_{[g,h]}\bs\varepsilon_{[h]}\bigg|\bs Z\bigg)\bigg).
    \end{split}
\end{equation}

By Lemma G.6 in \citet{kolesar2026cluster} the first term can be bounded as
\begin{equation}
    \begin{split}
        &\frac{1}{nk}\var\bigg(\sum_{g=1}^G\sum_{h\neq g}\sum_{l\notin\{g,h\}}\bs\Pi_{(p_1)}'\bs Z_{[g]}'\bs P_{[g,h]}\bs P_{[h,l],-gh}\bs\varepsilon_{[l]}\bs\varepsilon_{[g]}'\bs P_{[g,h]}\bs\varepsilon_{[h]}\bigg|\bs Z\bigg)\\
        &\leq\frac{C}{nk}\sum_{g=1}^G\sum_{h\neq g}\sum_{l\notin\{g,h\}}\E(\bs\Pi_{(p_1)}'\bs Z_{[g]}'\bs P_{[g,h]}\bs P_{[h,l],-gh}\bs\varepsilon_{[l]}\bs\varepsilon_{[l]}'\bs P_{[l,h],-gh}\bs P_{[h,g]}\bs Z_{[g]}\bs\Pi_{(p_1)}\\
        &\quad\bs\varepsilon_{[g]}'\bs P_{[g,h]}\bs\varepsilon_{[h]}\bs\varepsilon_{[h]}'\bs P_{[h,g]}\bs\varepsilon_{[g]}|\bs Z)\\
        &\leq\frac{C}{nk}\sum_{g=1}^G\sum_{h\neq g}\sum_{l\notin\{g,h\}}\bs\Pi_{(p_1)}'\bs Z_{[g]}'\bs P_{[g,h]}\bs P_{[h,l],-gh}\bs\Sigma_{l}\bs P_{[l,h],-gh}\bs P_{[h,g]}\bs Z_{[g]}\bs\Pi_{(p_1)}\\
        &\quad\tr(\bs\Sigma_{g}\bs P_{[g,h]}\bs\Sigma_{h}\bs P_{[h,g]})\leq\frac{C\nmax^4 k}{nk}.
    \end{split}
\end{equation}

Similarly, the second term becomes
\begin{equation}
    \begin{split}
        &\frac{1}{nk}\var\bigg(\sum_{g=1}^G\sum_{h\neq g}\sum_{l\notin\{g,h\}}\bs\eta_{(p_1),[g]}'\bs P_{[g,h]}\bs P_{[h,l],-gh}\bs\varepsilon_{[l]}\bs\varepsilon_{g}'\bs P_{[g,h]}\bs\varepsilon_{[h]}\bigg|\bs Z\bigg)\\
        &\leq\frac{C}{nk}\sum_{g=1}^G\sum_{h\neq g}\sum_{l\notin\{g,h\}}\E(\bs\eta_{(p_1),[g]}'\bs P_{[g,h]}\bs P_{[h,l],-gh}\bs\varepsilon_{[l]}\bs\varepsilon_{[l]}'\bs P_{[l,h],-gh}\bs P_{[h,g]}\bs\eta_{(p_1),[g]}\\
        &\quad\bs\varepsilon_{[g]}'\bs P_{[g,h]}\bs\varepsilon_{[h]}\bs\varepsilon_{[h]}'\bs P_{[h,g]}\bs\varepsilon_{[g]}|\bs Z)\\
        &\leq\frac{C}{nk}\nmax^2\sum_{g=1}^G\sum_{h\neq g}\sum_{l\notin\{g,h\}}\E(\bs\eta_{(p_1),[g]}'\bs P_{[g,h]}\bs P_{[h,l],-gh}\bs P_{[l,h],-gh}\bs P_{[h,g]}\bs\eta_{(p_1),[g]}\\
        &\quad\bs\varepsilon_{[g]}'\bs P_{[g,h]}\bs P_{[h,g]}\bs\varepsilon_{[g]}|\bs Z)\leq\frac{C\nmax^4 k}{nk}.
    \end{split}
\end{equation}

The variance of the other remainder terms can be bounded similarly. Hence, since $\nmax^4k/nk\to0$ also the covariance between the AR and score statistics can be estimated consistently.

\subsection{Proof of \texorpdfstring{\cref{lem:cluster LA2}}{Lemma A.1}}\label{app:cluster LA2}
\begin{proof}
The proof of \cref{lem:cluster LA2} requires the following generalisations of parts of Lemmas B1 to B3 in \citet{chao2012asymptotic}.

\begin{lemma}\label{lem:sums}
    Under \cref{it:P} and for any subset $\mathcal{I}_2$ of $\{(g,h)^{G}_{g,h=1}\}$ and any subset $\mathcal{I}_3$ of $\{(g,h,i)_{g,h,i=1}^{G}\}$ it holds that $\sum_{g,h\in \mathcal{I}_2}\tr(\bs P_{[g,h]}\bs P_{[h,g]}\bs P_{[g,h]}\bs P_{[h,g]})\leq k$ and $\sum_{g,h,i\in \mathcal{I}_{3}}\tr(\bs P_{[g,h]}\bs P_{[h,i]}\allowbreak\bs P_{[i,h]}\bs P_{[h,g]})\leq k$.
\end{lemma}

\begin{proof}
    For the first part write 
        \begin{equation}
            \begin{split}
                \sum_{g,h\in I_2}\tr(\bs P_{[g,h]}\bs P_{[h,g]}\bs P_{[g,h]}\bs P_{[h,g]})&\leq\sum_{g,h=1}^G\tr(\bs P_{[g,h]}\bs P_{[h,g]}\bs P_{[g,h]}\bs P_{[h,g]})\\
                &\leq\sum_{g_1,g_2,h=1}^G\tr(\bs P_{[g_1,h]}\bs P_{[h,g_2]}\bs P_{[g_2,h]}\bs P_{[h,g_1]})\\
                &\leq\sum_{g,h=1}^G\tr(\bs P_{[g,h]}\bs P_{[h,h]}\bs P_{[h,g]})\\
                &\leq\sum_{g,h=1}^G\tr(\bs P_{[g,h]}\bs P_{[h,g]})\\
                &\leq\sum_{g=1}^G\tr(\bs P_{[g,g]})=k.
            \end{split}
        \end{equation}
        Similarly, for the second part
        \begin{equation}
            \begin{split}
                \sum_{g,h,i\in I_{3}}\tr(\bs P_{[g,h]}\bs P_{[h,i]}\bs P_{[i,h]}\bs P_{[h,g]})&\leq\sum_{g,h,i=1}^G\tr(\bs P_{[g,h]}\bs P_{[h,i]}\bs P_{[i,h]}\bs P_{[h,g]})\\
                &\leq\sum_{g,h=1}^G\tr(\bs P_{[g,h]}\bs P_{[h,h]}\bs P_{[h,g]})\leq k.
            \end{split}
        \end{equation}
\end{proof}

For the next lemma, let $\bs L$ be the lower block triangular matrix of $\bs P$ defined by $L_{ij}=P_{Z,ij}\1\{\langle i\rangle <\langle j\rangle\}$, where $\langle i\rangle$ is the number of the cluster observation $i$ is in.

\begin{lemma}\label{lem:LL}
    Under \cref{it:P} it holds that $\lambdamax(\bs L\bs L')\leq\|\bs L\bs L'\|_F\leq C\sqrt{k}$.
\end{lemma}
\begin{proof}
First note that
\begin{equation}\label{eq:tr P-B}
    \begin{split}
        \tr\big((\bs P-\bs B_{P})^4\big)&\leq C\big[\tr(\bs P^4)+\tr(\bs B_{P}^4)\big]\leq C\big[\tr(\bs P)+\tr(\bs B_{P}\bs B_{P}^2\bs B_{P})\big]\\
        &\leq C\big[k+\tr(\bs B_{P}^2)\big]\leq C\big[k+\tr(\bs B_{P})\big]\leq Ck,
    \end{split}
\end{equation}
by \cref{it:P}.

Also since $\bs P-\bs B_{P}=\bs L+\bs L'$, 
\begin{equation}
    \begin{split}
        \tr\big((\bs P-\bs B_{P})^4\big)&=\tr\big((\bs L+\bs L')^4\big)\\
        &=2\tr(\bs L^4)+8\tr(\bs L^3\bs L')+4\tr(\bs L^2\bs L^{\prime 2})+2\tr(\bs L'\bs L\bs L'\bs L),
    \end{split}
\end{equation}
with
\begin{equation}
    \begin{split}
        \tr(\bs L^4)&=\sum_{g,h,i,j=1}^G\tr\big(\bs P_{[g,h]}\bs P_{[h,i]}\bs P_{[i,j]}\bs P_{[j,g]}\1\{g>j\}\1\{h>i\}\1\{i>j\}\1\{j>g\}\big)=0,
    \end{split}
\end{equation}
\begin{equation}
    \begin{split}
        \tr(\bs L^3\bs L')&=\sum_{g,h,i,j=1}^G\tr\big(\bs P_{[g,h]}\bs P_{[h,i]}\bs P_{[i,j]}\bs P_{[j,g]}\1\{g>h\}\1\{h>i\}\1\{i>j\}\1\{g>j\}\big)\\
        &=\sum_{g>h>i>j}\tr(\bs P_{[g,h]}\bs P_{[h,i]}\bs P_{[i,j]}\bs P_{[j,g]})\\
        &=\sum_{j<i<h<g}\tr(\bs P_{[g,h]}\bs P_{[h,i]}\bs P_{[i,j]}\bs P_{[j,g]}),
    \end{split}
\end{equation}
\begin{equation}
    \begin{split}
        \tr(\bs L^2\bs L^{\prime 2})&=\sum_{g,h,i,j=1}^G\tr\big(\bs P_{[g,h]}\bs P_{[h,i]}\bs P_{[i,j]}\bs P_{[j,g]}\1\{g>h\}\1\{h>i\}\1\{j>i\}\1\{g>j\}\big)\\
        &=\sum_{g>\{h,j\}>i}\tr(\bs P_{[g,h]}\bs P_{[h,i]}\bs P_{[i,j]}\bs P_{[j,g]})\\
        &=\sum_{g>h>i}\tr(\bs P_{[g,h]}\bs P_{[h,i]}\bs P_{[i,h]}\bs P_{[h,g]})+\sum_{g>h>j>i}\tr(\bs P_{[g,h]}\bs P_{[h,i]}\bs P_{[i,j]}\bs P_{[j,g]})\\
        &\quad+\sum_{g>j>h>i}\tr(\bs P_{[g,h]}\bs P_{[h,i]}\bs P_{[i,j]}\bs P_{[j,g]})\\
        &=\sum_{i<h<g}\tr(\bs P_{[g,h]}\bs P_{[h,i]}\bs P_{[i,h]}\bs P_{[h,g]})+\sum_{i<j<h<g}\tr(\bs P_{[g,h]}\bs P_{[h,i]}\bs P_{[i,j]}\bs P_{[j,g]})\\
        &\quad+\sum_{i<h<j<g}\tr(\bs P_{[g,h]}\bs P_{[h,i]}\bs P_{[i,j]}\bs P_{[j,g]})\\
        &=\sum_{i<h<g}\tr(\bs P_{[g,h]}\bs P_{[h,i]}\bs P_{[i,h]}\bs P_{[h,g]})+2\sum_{i<j<h<g}\tr(\bs P_{[g,h]}\bs P_{[h,i]}\bs P_{[i,j]}\bs P_{[j,g]}),
    \end{split}
\end{equation}
and
\begin{equation}
    \begin{split}
        \tr(\bs L\bs L'\bs L\bs L')&=\sum_{g,h,i,j=1}^G\tr\big(\bs P_{[g,h]}\bs P_{[h,i]}\bs P_{[i,j]}\bs P_{[j,g]}\1\{g>h\}\1\{i>h\}\1\{i>j\}\1\{g>j\}\big)\\
        &=\sum_{h<g}\tr(\bs P_{[g,h]}\bs P_{[h,g]}\bs P_{[g,h]}\bs P_{[h,g]})+\sum_{h<i<g}\tr(\bs P_{[g,h]}\bs P_{[h,i]}\bs P_{[i,h]}\bs P_{[h,g]})\\
        &\quad+\sum_{h<g<i}\tr(\bs P_{[g,h]}\bs P_{[h,i]}\bs P_{[i,h]}\bs P_{[h,g]})+\sum_{h<j<g}\tr(\bs P_{[g,h]}\bs P_{[h,g]}\bs P_{[g,j]}\bs P_{[j,g]})\\
        &\quad+\sum_{j<h<g}\tr(\bs P_{[g,h]}\bs P_{[h,g]}\bs P_{[g,j]}\bs P_{[j,g]})\\
        &\quad+\bigg[\sum_{j<h<i<g}+\sum_{h<j<i<g}+\sum_{j<h<g<i}+\sum_{h<j<g<i}\bigg]\times\tr(\bs P_{[g,h]}\bs P_{[h,i]}\bs P_{[i,j]}\bs P_{[j,g]})\\
        &=\sum_{h<g}\tr(\bs P_{[g,h]}\bs P_{[h,g]}\bs P_{[g,h]}\bs P_{[h,g]})+\sum_{h<i<g}\tr(\bs P_{[g,h]}\bs P_{[h,i]}\bs P_{[i,h]}\bs P_{[h,g]})\\
        &\quad+\sum_{h<g<i}\tr(\bs P_{[g,h]}\bs P_{[h,i]}\bs P_{[i,h]}\bs P_{[h,g]})+\sum_{h<j<g}\tr(\bs P_{[g,h]}\bs P_{[h,g]}\bs P_{[g,j]}\bs P_{[j,g]})\\
        &\quad+\sum_{j<h<g}\tr(\bs P_{[g,h]}\bs P_{[h,g]}\bs P_{[g,j]}\bs P_{[j,g]})+4\sum_{j<h<i<g}\tr(\bs P_{[g,h]}\bs P_{[h,i]}\bs P_{[i,j]}\bs P_{[j,g]}).
    \end{split}
\end{equation}
Combining we find
\begin{equation}\label{eq:LLLL}
    \begin{split}
        &\tr\big((\bs P-\bs B_{P})^4\big)\\
        &=2\sum_{h<g}\tr(\bs P_{[g,h]}\bs P_{[h,g]}\bs P_{[g,h]}\bs P_{[h,g]})\\
        &\quad+4\sum_{i<h<g}\tr(\bs P_{[g,h]}\bs P_{[h,i]}\bs P_{[i,h]}\bs P_{[h,g]})+4\sum_{h<i<g}\tr(\bs P_{[g,h]}\bs P_{[h,i]}\bs P_{[i,h]}\bs P_{[h,g]})
        \\
        &\quad+4\sum_{h<j<g}\tr(\bs P_{[g,h]}\bs P_{[h,g]}\bs P_{[g,j]}\bs P_{[j,g]})\\
        &\quad+8\sum_{j<i<h<g}\tr(\bs P_{[g,h]}\bs P_{[h,i]}\bs P_{[i,j]}\bs P_{[j,g]})+8\sum_{i<j<h<g}\tr(\bs P_{[g,h]}\bs P_{[h,i]}\bs P_{[i,j]}\bs P_{[j,g]})\\
        &\quad+8\sum_{j<h<i<g}\tr(\bs P_{[g,h]}\bs P_{[h,i]}\bs P_{[i,j]}\bs P_{[j,g]}).
    \end{split}
\end{equation}
Let $S=\sum_{j<i<h<g}\tr(\bs P_{[g,h]}\bs P_{[h,i]}\bs P_{[i,j]}\bs P_{[j,g]})+\sum_{i<j<h<g}\tr(\bs P_{[g,h]}\bs P_{[h,i]}\bs P_{[i,j]}\bs P_{[j,g]})+\sum_{j<h<i<g}\allowbreak\tr(\bs P_{[g,h]}\bs P_{[h,i]}\bs P_{[i,j]}\bs P_{[j,g]})$, then rearranging above expression, taking absolute values and applying the triangle inequality yields
\begin{equation}\label{eq:S}
    \begin{split}
        8|S|=&\big|\tr\big((\bs P-\bs B_{P})^4\big)\big|+2\bigg|\sum_{h<g}\tr(\bs P_{[g,h]}\bs P_{[h,g]}\bs P_{[g,h]}\bs P_{[h,g]})\bigg|\\
        &\quad+4\bigg|\sum_{i<h<g}\tr(\bs P_{[g,h]}\bs P_{[h,i]}\bs P_{[i,h]}\bs P_{[h,g]})+4\sum_{h<i<g}\tr(\bs P_{[g,h]}\bs P_{[h,i]}\bs P_{[i,h]}\bs P_{[h,g]})\bigg|
        \\
        &\quad+4\bigg|\sum_{h<j<g}\tr(\bs P_{[g,h]}\bs P_{[h,g]}\bs P_{[g,j]}\bs P_{[j,g]})\bigg|\geq Ck,
    \end{split}
\end{equation}
by \cref{lem:sums,eq:tr P-B}.

Next, let $\bs r$ be a $n$ vector of independent mean zero, unit variance random variables and define the random quantities
\begin{equation}
    \begin{split}
        \Delta_{1}&=\sum_{g<h<i}\bs r_{[h]}'\bs P_{[h,g]}\bs P_{[g,i]}\bs r_{[i]}+\bs r_{[g]}'\bs P_{[g,h]}\bs P_{[h,i]}\bs r_{[i]}+\bs r_{[g]}'\bs P_{[g,i]}\bs P_{[i,h]}\bs r_{[h]}\\
        \Delta_{2}&=\sum_{g<h<i}\bs r_{[h]}'\bs P_{[h,g]}\bs P_{[g,i]}\bs r_{[i]}+\bs r_{[g]}'\bs P_{[g,h]}\bs P_{[h,i]}\bs r_{[i]}\\
        \Delta_{3}&=\sum_{g<h<i}\bs r_{[g]}'\bs P_{[g,i]}\bs P_{[i,h]}\bs r_{[h]}.
    \end{split}
\end{equation}
Then
\begin{equation}
    \begin{split}
        \E(\Delta_{3}^2|\mathcal{Z})&=\E\bigg(\bigg[\sum_{g<h<i}\bs r_{[g]}'\bs P_{[g,i]}\bs P_{[i,h]}\bs r_{[h]}\bigg]\bigg[\sum_{l<m<q}\bs r_{[l]}'\bs P_{[l,q]}\bs P_{[q,m]}\bs r_{[m]}\bigg]\bigg|\mathcal{Z}\bigg),
    \end{split}
\end{equation}
which has nonzero expectation only when $g=l<h=m<\{i,q\}$. Hence,
\begin{equation}\label{eq:delta3}
    \begin{split}
        &\E\bigg(\sum_{g<h<\{i,q\}}\sum_{g<h<i}\bs r_{[g]}'\bs P_{[g,i]}\bs P_{[i,h]}\bs r_{[h]}\bs r_{[h]}'\bs P_{[h,q]}\bs P_{[q,g]}\bs r_{[g]}\bigg|\mathcal{Z}\bigg)\\
        &=\sum_{g<h<\{i,q\}}\tr(\bs P_{[g,i]}\bs P_{[i,h]}\bs P_{[h,q]}\bs P_{[q,g]})\\
        &=\sum_{g<h<i}\tr(\bs P_{[g,i]}\bs P_{[i,h]}\bs P_{[h,i]}\bs P_{[i,g]})+2\sum_{g<h<i<q}\tr(\bs P_{[g,i]}\bs P_{[i,h]}\bs P_{[h,q]}\bs P_{[q,g]})\\
        &\leq Ck+2\sum_{g<h<i<q}\tr(\bs P_{[g,i]}\bs P_{[i,h]}\bs P_{[h,q]}\bs P_{[q,g]}),
    \end{split}
\end{equation}
by \cref{lem:sums} and because $\E(\bs r\bs r'|\mathcal{Z})=\bs I$ and $\E(\bs r'\bs A\bs r|\mathcal{Z})=\tr(\bs A)$ for any square matrix $\bs A$.

Also,
\begin{equation}
    \begin{split}
        \E(\Delta_{2}\Delta_{3}|\mathcal{Z})&=\E\bigg(\bigg[\sum_{g<h<i}\bs r_{[h]}'\bs P_{[h,g]}\bs P_{[g,i]}\bs r_{[i]}+\bs r_{[g]}'\bs P_{[g,h]}\bs P_{[h,i]}\bs r_{[i]}\bigg]\\
        &\quad\bigg[\sum_{l<m<q}\bs r_{[l]}'\bs P_{[l,q]}\bs P_{[q,m]}\bs r_{[m]}\bigg]\bigg|\mathcal{Z}\bigg).
    \end{split}
\end{equation}
Expand and note that the first term only has nonzero expectation when $g<h=l<i=m<q$. The second term only has nonzero expectation when $g=l<h<i=m<q$. Hence,
\begin{equation}
    \begin{split}
        &\E\bigg(\bigg[\sum_{g<h<i}\bs r_{[h]}'\bs P_{[h,g]}\bs P_{[g,i]}\bs r_{[i]}+\bs r_{[g]}'\bs P_{[g,h]}\bs P_{[h,i]}\bs r_{[i]}\bigg]\bigg[\sum_{l<m<q}\bs r_{[l]}'\bs P_{[l,q]}\bs P_{[q,m]}\bs r_{[m]}\bigg]\bigg|\mathcal{Z}\bigg)\\
        &=\E\bigg(\sum_{g<h<i<q}\bs r_{[h]}'\bs P_{[h,g]}\bs P_{[g,i]}\bs r_{[i]}\bs r_{[i]}'\bs P_{[i,q]}\bs P_{[q,h]}\bs r_{[h]}\\
        &\quad+\bs r_{[g]}'\bs P_{[g,h]}\bs P_{[h,i]}\bs r_{[i]}\bs r_{[i]}'\bs P_{[i,q]}\bs P_{[q,g]}\bs r_{[g]}\bigg|\mathcal{Z}\bigg)\\
        &=\sum_{g<h<i<q}\tr(\bs P_{[h,g]}\bs P_{[g,i]}\bs P_{[i,q]}\bs P_{[q,h]})+\tr(\bs P_{[g,h]}\bs P_{[h,i]}\bs P_{[i,q]}\bs P_{[q,g]}).
    \end{split}
\end{equation}
Similarly,
\begin{equation}
    \begin{split}
        \E(\Delta_{2}^2|\mathcal{Z})&=\E\bigg(\bigg[\sum_{g<h<i}\bs r_{[h]}'\bs P_{[h,g]}\bs P_{[g,i]}\bs r_{[i]}+\bs r_{[g]}'\bs P_{[g,h]}\bs P_{[h,i]}\bs r_{[i]}\bigg]\\
        &\quad\bigg[\sum_{l<m<q}\bs r_{[m]}'\bs P_{[m,l]}\bs P_{[l,q]}\bs r_{[q]}+\bs r_{[l]}'\bs P_{[l,m]}\bs P_{[m,q]}\bs r_{[q]}\bigg]\bigg|\mathcal{Z}\bigg).
    \end{split}
\end{equation}
Expand and note that the first term has nonzero expectation only when $\{g,l\}<h=m<i=q$, the second when $g<h=l<m<q=i$, the third when $l<g=m<h<i=q$ and the fourth when $g=l<\{h,m\}<i=q$. Hence,
\begin{equation}
    \begin{split}
        &\E\bigg(\bigg[\sum_{g<h<i}\bs r_{[h]}'\bs P_{[h,g]}\bs P_{[g,i]}\bs r_{[i]}+\bs r_{[g]}'\bs P_{[g,h]}\bs P_{[h,i]}\bs r_{[i]}\bigg]\\
        &\quad\bigg[\sum_{l<m<q}\bs r_{[m]}'\bs P_{[m,l]}\bs P_{[l,q]}\bs r_{[q]}+\bs r_{[l]}'\bs P_{[l,m]}\bs P_{[m,q]}\bs r_{[q]}]\bigg|\mathcal{Z}\bigg)\\
        &=\E\bigg(\sum_{\{g,l\}<h<i}\bs r_{[h]}'\bs P_{[h,g]}\bs P_{[g,i]}\bs r_{[i]}\bs r_{[i]}'\bs P_{[i,l]}\bs P_{[l,h]}\bs r_{[h]}\\
        &\quad+\sum_{g<h<m<i}\bs r_{[h]}'\bs P_{[h,g]}\bs P_{[g,i]}\bs r_{[i]}\bs r_{[i]}'\bs P_{[i,m]}\bs P_{[m,g]}\bs r_{[g]}\\
        &\quad+\sum_{l<g<h<i}\bs r_{[g]}'\bs P_{[g,h]}\bs P_{[h,i]}\bs r_{[i]}\bs r_{[i]}'\bs P_{[i,l]}\bs P_{[l,h]}\bs r_{[h]}\\
        &\quad+\sum_{g<\{h,m\}<i}\bs r_{[g]}'\bs P_{[g,h]}\bs P_{[h,i]}\bs r_{[i]}\bs r_{[i]}'\bs P_{[i,m]}\bs P_{[m,h]}\bs r_{[h]}\bigg|\mathcal{Z}\bigg)\\
        &=\sum_{\{g,l\}<h<i}\tr(\bs P_{[h,g]}\bs P_{[g,i]}\bs P_{[i,l]}\bs P_{[l,h]})+\sum_{g<h<m<i}\tr(\bs P_{[h,g]}\bs P_{[g,i]}\bs P_{[i,m]}\bs P_{[m,g]})\\
        &\quad+\sum_{l<g<h<i}\tr(\bs P_{[g,h]}\bs P_{[h,i]}\bs P_{[i,l]}\bs P_{[l,h]})+\sum_{g<\{h,m\}<i}\tr(\bs P_{[g,h]}\bs P_{[h,i]}\bs P_{[i,m]}\bs P_{[m,h]})\\
        &=\sum_{g<h<i}\tr(\bs P_{[h,g]}\bs P_{[g,i]}\bs P_{[i,g]}\bs P_{[g,h]})+\sum_{g<l<h<i}\tr(\bs P_{[h,g]}\bs P_{[g,i]}\bs P_{[i,l]}\bs P_{[l,h]})\\
        &\quad+\sum_{l<g<h<i}\tr(\bs P_{[h,g]}\bs P_{[g,i]}\bs P_{[i,l]}\bs P_{[l,h]})
        +\sum_{g<h<m<i}\tr(\bs P_{[h,g]}\bs P_{[g,i]}\bs P_{[i,m]}\bs P_{[m,g]})\\
        &\quad+\sum_{l<g<h<i}\tr(\bs P_{[g,h]}\bs P_{[h,i]}\bs P_{[i,l]}\bs P_{[l,h]})+\sum_{g<h<i}\tr(\bs P_{[g,h]}\bs P_{[h,i]}\bs P_{[i,h]}\bs P_{[h,h]})\\
        &\quad+\sum_{g<h<m<i}\tr(\bs P_{[g,h]}\bs P_{[h,i]}\bs P_{[i,m]}\bs P_{[m,h]})+\sum_{g<m<h<i}\tr(\bs P_{[g,h]}\bs P_{[h,i]}\bs P_{[i,m]}\bs P_{[m,h]})\\
        &=\sum_{g<h<i}\tr(\bs P_{[h,g]}\bs P_{[g,i]}\bs P_{[i,g]}\bs P_{[g,h]})+\sum_{g<h<i}\tr(\bs P_{[g,h]}\bs P_{[h,i]}\bs P_{[i,h]}\bs P_{[h,g]})+2S\\
        &\leq Ck+2|S|\leq Ck,
    \end{split}
\end{equation}
by \cref{lem:sums} and \cref{eq:S}.

By $\Delta_{1}=\Delta_{2}+\Delta_{3}$ it holds that $\E(\Delta_{1}^2|\mathcal{Z})=\E(\Delta_{2}^2|\mathcal{Z})+\E(\Delta_{3}^2|\mathcal{Z})+2\E(\Delta_{2}\Delta_{3}|\mathcal{Z})\leq Ck+2S\leq Ck$. Then, rearranging \cref{eq:delta3}, applying the triangle inequality and using the expressions above yields
\begin{equation}
    \begin{split}
        &\bigg|\sum_{g<h<i<q}\tr(\bs P_{[g,i]}\bs P_{[i,h]}\bs P_{[h,q]}\bs P_{[q,g]})\bigg|\leq\E(\Delta_{3}^2|\mathcal{Z})+Ck\\
        &=\E\big([\Delta_{1}-\Delta_{2}]^2|\mathcal{Z}\big)+Ck\leq 2\E(\Delta_{1}^2|\mathcal{Z})-2\E(\Delta_{2}^2|\mathcal{Z})+Ck\leq Ck.
    \end{split}
\end{equation}
Hence, by \cref{lem:sums} and the triangle inequality it follows from \cref{eq:LLLL} that $\big|\tr(\bs L\bs L'\bs L\bs L')\big|\leq Ck$ and $\lambdamax(\bs L\bs L')\leq\|\bs L\bs L'\|_F=\sqrt{\tr(\bs L\bs L'\bs L\bs L')}\leq C\sqrt{k}$.
\end{proof}

    Proceed with the proof of \cref{lem:cluster LA2}. For clarity of exposition, I follow \citet{chao2012asymptotic} as closely as possible in this proof.

    Let $\bs b_{1n}=\bs c_{1n}\Xi_{n}^{-1/2}$ and $\bs b_{2n}=\bs c_{2n}\Xi_{n}^{-1/2}$ and note that these are bounded in $n$ because $\Xi_{n}$ is bounded away from zero by hypothesis. Let $w_{gn}=\bs b_{1n}'\bs W_{[g],n}'\bs\iota$ and $u_{i}=\bs b_{2n}'\bs U_{i}$, where I suppress the $n$ subscript on $u_{i}$ for notational convenience. Then, $Y_{n}=w_{1n}+\sum_{g=2}^Gy_{gn}$, $y_{gn}=w_{gn}+\bar{y}_{gn}$, $\bar{y}_{gn}=\sum_{h<g}(\bs u_{[h]}'\bs P_{[h,g]}\bar{\bs\varepsilon}_{[g]}+\bs u_{[g]}'\bs P_{[g,h]}\bar{\bs\varepsilon}_{[h]})/\sqrt{k}$.

    Also, $\E(\|w_{1n}\|^4|\mathcal{Z})\leq\sum_{g=1}^G\E(\|w_{gn}\|^4|\mathcal{Z})\leq C\sum_{g=1}^G\E(\|\bs W_{[g],n}'\bs\iota\|^4|\mathcal{Z})\toas 0$, so by a conditional version of Markov's inequality, I deduce that for any $v>0$, $\Pr(|w_{1n}|\geq v|\mathcal{Z})\to 0$ as $n\to\infty$; i.e. $w_{1n}\toprob 0$ unconditionally. Hence $Y_{n}=\sum_{g=2}^Gy_{gn}+o_p(1)$.

    Now, I will show that $Y_{n}\todist N(0,1)$ by first showing that, conditional on $\mathcal{Z}$, $\sum_{g=2}^Gy_{gn}\todist N(0,1)$, a.s. To proceed, let $\mathcal{X}_{g}=\{\bs W_{[g],n},\bs U_{[g]},\bar{\bs\varepsilon}_{[g]}\}$ for $g=1,\dots, G$. Define the $\sigma$-fields $F_{g,n}=\sigma(\mathcal{X}_{1},\dots,\mathcal{X}_{g})$ for $g=1,\dots,G$. Note that, by construction $F_{g-1,n}\subseteq F_{g,n}$. Moreover, it is straightforward to verify that, conditional on $\mathcal{Z}$, $\{y_{gn},F_{g,n}, 1\leq g\leq G, G\geq 2\}$ is a martingale difference array, and I can apply the martingale central limit theorem. Let $\bs\Sigma_{g}=\E(\bar{\bs\varepsilon}_{[g]}\bar{\bs\varepsilon}_{[g]}'|\mathcal{Z})$, $\bs\Omega_{g}=\bs\Omega_{gn}(\mathcal{Z})=\E(\bs u_{[g]}\bs u_{[g]}'|\mathcal{Z})$, and $\bs\Gamma_{g}=\E(\bs u_{[g]}\bar{\bs\varepsilon}_{[g]}'|\mathcal{Z})$, where to simplify notation I suppress the dependence of $\bs\Sigma_{g}$ on $\mathcal{Z}$ and of $\bs\Omega_{g}$ and $\bs\Gamma_{g}$ on $\mathcal{Z}$ and $n$. Now note that $\E(w_{gn}\bar{y}_{hn}|\mathcal{Z})=0$ for all $g$ and $h$ and that
    \begin{equation}
        \begin{split}
            &\E\big[(\bar{y}_{gn})^2\big|\mathcal{Z}\big]\\
            &=\sum_{h<g}\sum_{i<g}\E\big[(\bs u_{[h]}'\bs P_{[h,g]}\bar{\bs\varepsilon}_{[g]}+\bs u_{[g]}'\bs P_{[g,h]}\bar{\bs\varepsilon}_{[h]})(\bs u_{[i]}'\bs P_{[i,g]}\bar{\bs\varepsilon}_{[g]}+\bs u_{[g]}'\bs P_{[g,i]}\bar{\bs\varepsilon}_{[i]})\big|\mathcal{Z}\big]/k\\
            &=\sum_{h<g}\big[\tr(\bs\Omega_{h}\bs P_{[h,g]}\bs\Sigma_{g}\bs P_{[g,h]})+2\tr(\bs\Gamma_{h}'\bs P_{[h,g]}\bs\Gamma_{g}'\bs P_{[g,h]})+\tr(\bs\Omega_{g}\bs P_{[g,h]}\bs\Sigma_{h}\bs P_{[h,g]})\big]/k.
        \end{split}
    \end{equation}
    Thus,
    \begin{equation}
        \begin{split}
            s^2_{n}(\mathcal{Z})&=\E\bigg[\bigg(\sum_{g=2}^Gy_{gn}\bigg)^2\bigg|\mathcal{Z}\bigg]=\sum_{g=2}^G(\E(w_{gn}^2|\mathcal{Z})+\E(y_{gn}^2|\mathcal{Z}))\\
            &=\bs b_{1n}'\bs D_{n}\bs b_{1n}-\E(w_{1n}^2|\mathcal{Z})+\sum_{g=2}^G\sum_{h<g}\big[\tr(\bs\Omega_{h}\bs P_{[h,g]}\bs\Sigma_{g}\bs P_{[g,h]})\\
            &\quad+2\tr(\bs\Gamma_{h}'\bs P_{[h,g]}\bs\Gamma_{g}'\bs P_{[g,h]})+\tr(\bs\Omega_{g}\bs P_{[g,h]}\bs\Sigma_{h}\bs P_{[h,g]})\big]/k\\
            &=\bs b_{1n}'\bs D_{n}\bs b_{1n}+\bs b_{2n}'\bar{\bs\Sigma}_{n}\bs b_{2n}+o_{a.s.}(1)\\
            &=\Xi^{-1/2}(\bs c_{1n}'\bs D_{n}\bs c_{1n}+\bs c_{2n}'\bar{\bs\Sigma}_{n}\bs c_{2n})\Xi^{-1/2}+o_{a.s.}(1)\\
            &=\Xi^{-1/2}\Xi\Xi^{-1/2}+o_{a.s.}(1)=1+o_{a.s.}(1)\toas 1,
        \end{split}
    \end{equation}
    where $\bs D_{n}=\bs D_{n}(\mathcal{Z})=\sum_{g=1}^G\sum_{i,j\in[g]}\E(\bs W_{i,n}\bs W_{j,n}'|\mathcal{Z})$ and
    \begin{equation}
        \bar{\bs\Sigma}_{n}=\bar{\bs\Sigma}_{n}(\mathcal{Z})=\sum_{g\neq h}\E(\bs U_{[g]}'\bs P_{[g,h]}\bar{\bs\varepsilon}_{[h]}\bar{\bs\varepsilon}_{[g]}'\bs P_{[g,h]}\bs U_{[h]}+\bs U_{[g]}'\bs P_{[g,h]}\bar{\bs\varepsilon}_{[h]}\bar{\bs\varepsilon}_{[h]}'\bs P_{[h,g]}\bs U_{[g]}|\mathcal{Z})/k.
    \end{equation}
    Thus, $s_{n}^2(\mathcal{Z})$ is bounded and bounded away from zero a.s. Also, $\sum_{g=2}^G\E(y^4_{gn}|\mathcal{Z})\leq C\sum_{g=1}^G\allowbreak\E(\|\bs W_{[g],n}'\bs\iota\|^4|\mathcal{Z})+C\sum_{g=2}^G\E(\bar{y}_{gn}^4|\mathcal{Z})$. By \cref{it:w}, $\sum_{g=1}^G\E(\|\bs W_{[g],n}'\bs\iota\|^4|\mathcal{Z})\to 0$. Let $\bar{y}^{\varepsilon}_{gn}=\sum_{h<g}\bs u_{[h]}'\bs P_{[h,g]}\bar{\bs\varepsilon}_{[g]}/\sqrt{k}$ and $\bar{y}^{u}_{gn}=\sum_{h<g}\bs u_{[g]}'\bs P_{[g,h]}\bar{\bs\varepsilon}_{[h]}/\sqrt{k}$. It holds that
    \begin{equation}\label{eq:lyapunov}
        \begin{split}
            &\sum_{g=2}^G\E\big[(\bar{y}^{\varepsilon}_{gn})^4\big|\mathcal{Z}\big]\\
            \quad&\leq\frac{C}{k^2}\sum_{g=2}^G\sum_{h_1,h_2,h_3,h_4<g}\E(\bs u_{[h_1]}\bs P_{[h_1,g]}\bar{\bs\varepsilon}_{[g]}\bar{\bs\varepsilon}_{[g]}'\bs P_{[g,h_2]}\bs u_{[h_2]}\bs u'_{[h_3]}\bs P_{[h_3,g]}\bar{\bs\varepsilon}_{[g]}\bar{\bs\varepsilon}_{[g]}'\bs P_{[g,h_4]}\bs u_{[h_4]}|\mathcal{Z})
        \end{split}
    \end{equation}
    This term is only non-zero when $h_1=\dots=h_4$, or when $h_{i}=h_{j}\neq h_{k}=h_{l}$ for $i,j,k,l$ all permutations of $\{1,2,3,4\}$. The terms corresponding to the first case can be written as
    \begin{equation}\label{eq:lyapunov1}
        \begin{split}
            &\frac{C}{k^2}\sum_{g=2}^G\sum_{h<g}\E(\bs u'_{[h]}\bs P_{[h,g]}\bar{\bs\varepsilon}_{[g]}\bar{\bs\varepsilon}_{[g]}'\bs P_{[g,h]}\bs u_{[h]}\bs u'_{[h]}\bs P_{[h,g]}\bar{\bs\varepsilon}_{[g]}\bar{\bs\varepsilon}_{[g]}'\bs P_{[g,h]}\bs u_{[h]}|\mathcal{Z})\\
            &=\frac{C}{k^2}\sum_{g=2}^G\sum_{h<g}\E\big(\tr(\bs P_{[h,g]}\bar{\bs\varepsilon}_{[g]}\bar{\bs\varepsilon}_{[g]}'\bs P_{[g,h]}\bs u_{[h]}\bs u'_{[h]})\tr(\bs P_{[h,g]}\bar{\bs\varepsilon}_{[g]}\bar{\bs\varepsilon}_{[g]}'\bs P_{[g,h]}\bs u_{[h]}\bs u'_{[h]})\big|\mathcal{Z}\big)\\
            &=\frac{C}{k^2}\sum_{g=2}^G\sum_{h<g}\E\big(\vec(\bs u_{[h]}\bs u'_{[h]})'(\bs P_{[h,g]}\otimes\bs P_{[h,g]})\vec(\bar{\bs\varepsilon}_{[g]}\bar{\bs\varepsilon}_{[g]}')\\
            &\quad\vec(\bar{\bs\varepsilon}_{[g]}\bar{\bs\varepsilon}_{[g]}')'(\bs P_{[g,h]}\otimes\bs P_{[g,h]})\vec(\bs u_{[h]}\bs u'_{[h]})\big|\mathcal{Z}\big).
        \end{split}
    \end{equation}
    Similar to \cref{eq:eigbound kappa}, $\lambdamax\big[\E\big(\vec(\bs u_{[g]}\bs u_{[g]}')\vec(\bs u_{[g]}\bs u_{[g]}')'\big|\mathcal{Z}\big)\big]\leq\nmax^2C$. Hence, \cref{eq:lyapunov1} can be written as
    \begin{equation}\label{eq:PPPP}
        \begin{split}
            &\frac{1}{k^2}\sum_{g=2}^G\sum_{h<g}\E\big(\vec(\bs u_{[h]}\bs u'_{[h]})'(\bs P_{[h,g]}\otimes\bs P_{[h,g]})\vec(\bar{\bs\varepsilon}_{[g]}\bar{\bs\varepsilon}_{[g]}')\\
            &\quad\vec(\bar{\bs\varepsilon}_{[g]}\bar{\bs\varepsilon}_{[g]}')'(\bs P_{[g,h]}\otimes\bs P_{[g,h]})\vec(\bs u_{[h]}\bs u'_{[h]})\big|\mathcal{Z}\big)\\
            &\leq\frac{C\nmax^2}{k^2}\sum_{g=2}^G\sum_{h<g}\E\big(\vec(\bs u_{[h]}\bs u'_{[h]})'(\bs P_{[h,g]}\otimes\bs P_{[h,g]})(\bs P_{[g,h]}\otimes\bs P_{[g,h]})\vec(\bs u_{[h]}\bs u'_{[h]})\big|\mathcal{Z}\big)\\
            &=\frac{C\nmax^2}{k^2}\sum_{g=2}^G\sum_{h<g}\E\big(\vec(\bs u_{[h]}\bs u'_{[h]})'(\bs P_{[h,g]}\bs P_{[g,h]}\otimes\bs P_{[h,g]}\bs P_{[g,h]})\vec(\bs u_{[h]}\bs u'_{[h]})\big|\mathcal{Z}\big)\\
            &=\frac{C\nmax^2}{k^2}\sum_{g=2}^G\sum_{h<g}\E\big(\vec(\bs P_{[h,g]}\bs P_{[g,h]})'(\bs u_{[h]}\bs u'_{[h]}\otimes\bs u_{[h]}\bs u'_{[h]})\vec(\bs P_{[h,g]}\bs P_{[g,h]})\big|\mathcal{Z}\big)\\
            &\leq\frac{C\nmax^4}{k^2}\sum_{g=2}^G\sum_{h<g}\tr(\bs P_{[g,h]}\bs P_{[h,g]}\bs P_{[g,h]}\bs P_{[h,g]})\\
            &\leq\frac{C\nmax^4}{k^2}\sum_{g_1,g_2,h=1}^G\tr(\bs P_{[g_1,h]}\bs P_{[h,g_2]}\bs P_{[g_2,h]}\bs P_{[h,g_1]})\\
            &\leq\frac{C\nmax^4}{k^2}\sum_{g_1,h=1}^G\tr(\bs P_{[g_1,h]}\bs P_{[h,h]}\bs P_{[h,g_1]})\\
            &\leq\frac{C\nmax^4}{k^2}\sum_{g_1,h=1}^G\tr(\bs P_{[g_1,h]}\bs P_{[h,g_1]})\\
            &\leq\frac{C\nmax^4}{k^2}\sum_{g_1}^G\tr(\bs P_{[g_1,g_1]})\\
            &\leq\frac{C\nmax^4k}{k^2}\toas0,
        \end{split}
    \end{equation}
    by \cref{it:P,it:rate}.
    
    Next, consider the terms corresponding to the second case following \cref{eq:lyapunov}.
    \begin{equation}
        \begin{split}
            &\frac{1}{k^2}\sum_{g=2}^G\E\bigg(\sum_{\substack{h_1, h_2<g\\h_1\neq h_2}}\bar{\bs\varepsilon}'_{[g]}\bs P_{[g,h_1]}\bs u_{[h_1]}\bs u_{[h_1]}'\bs P_{[h_1,g]}\bar{\bs\varepsilon}_{[g]}\bar{\bs\varepsilon}'_{[g]}\bs P_{[g,h_2]}\bs u_{[h_2]}\bs u_{[h_2]}'\bs P_{[h_2,g]}\bar{\bs\varepsilon}_{[g]}\bigg|\mathcal{Z}\bigg)\\
            &=\frac{1}{k^2}\sum_{g=2}^G\E\bigg(\sum_{\substack{h_1, h_2<g\\h_1\neq h_2}}\bar{\bs\varepsilon}'_{[g]}\bs P_{[g,h_1]}\bs\Omega_{h_1}\bs P_{[h_1,g]}\bar{\bs\varepsilon}_{[g]}\bar{\bs\varepsilon}'_{[g]}\bs P_{[g,h_2]}\bs\Omega_{h_2}\bs P_{[h_2,g]}\bar{\bs\varepsilon}_{[g]}\bigg|\mathcal{Z}\bigg)\\
            &=\frac{1}{k^2}\sum_{g=2}^G\E\bigg(\sum_{\substack{h_1, h_2<g\\h_1\neq h_2}}\vec(\bar{\bs\varepsilon}_{[g]}\bar{\bs\varepsilon}'_{[g]})'(\bs P_{[g,h_1]}\otimes\bs P_{[g,h_1]})\vec(\bs\Omega_{h_1})\\
            &\quad\vec(\bs\Omega_{h_2})'(\bs P_{[h_2,g]}\otimes\bs P_{[h_2,g]})\vec(\bar{\bs\varepsilon}_{[g]}\bar{\bs\varepsilon}'_{[g]})\bigg|\mathcal{Z}\bigg)\\
            &=\frac{1}{k^2}\sum_{g=2}^G\sum_{\substack{h_1, h_2<g\\h_1\neq h_2}}\tr\big[(\bs P_{[g,h_1]}\otimes\bs P_{[g,h_1]})\vec(\bs\Omega_{h_1})\\
            &\quad\vec(\bs\Omega_{h_2})'(\bs P_{[h_2,g]}\otimes\bs P_{[h_2,g]})\E\big(\vec(\bar{\bs\varepsilon}_{[g]}\bar{\bs\varepsilon}'_{[g]})\vec(\bar{\bs\varepsilon}_{[g]}\bar{\bs\varepsilon}'_{[g]})'\big|\mathcal{Z}\big)\big]\\
            &\leq\frac{1}{k^2}\sum_{g=2}^G\sum_{\substack{h_1, h_2<g\\h_1\neq h_2}}\big[\tr\big[(\bs P_{[g,h_1]}\otimes\bs P_{[g,h_1]})\vec(\bs\Omega_{h_1})\vec(\bs\Omega_{h_1})'(\bs P_{[h_1,g]}\otimes\bs P_{[h_1,g]})\big]\big]^{1/2}\\
            &\quad\big[\tr\big[\E\big(\vec(\bar{\bs\varepsilon}_{[g]}\bar{\bs\varepsilon}'_{[g]})\vec(\bar{\bs\varepsilon}_{[g]}\bar{\bs\varepsilon}'_{[g]})\big|\mathcal{Z}\big)(\bs P_{[g,h_2]}\otimes\bs P_{[g,h_2]})\vec(\bs\Omega_{h_2})\\&\quad\vec(\bs\Omega_{h_2})'(\bs P_{[h_2,g]}\otimes\bs P_{[h_2,g]})\E\big(\vec(\bar{\bs\varepsilon}_{[g]}\bar{\bs\varepsilon}'_{[g]})\vec(\bar{\bs\varepsilon}_{[g]}'\bar{\bs\varepsilon}'_{[g]})'\big|\mathcal{Z}\big)\big]\big]^{1/2},
        \end{split}
    \end{equation}
    by the Cauchy--Schwarz inequality. Results similar to \cref{eq:eigbound kappa} for $\E\big(\vec(\bar{\bs\varepsilon}_{[g]}\bar{\bs\varepsilon}'_{[g]})\allowbreak\vec(\bar{\bs\varepsilon}_{[g]}\bar{\bs\varepsilon}'_{[g]})'\big|\mathcal{Z}\big)$ and $\vec(\bs\Omega_{h})\vec(\bs\Omega_{h})'$ yield
    \begin{equation}
        \begin{split}
            &\frac{1}{k^2}\sum_{g=2}^G\sum_{\substack{h_1, h_2<g\\h_1\neq h_2}}\big[\tr\big[(\bs P_{[g,h_1]}\otimes\bs P_{[g,h_1]})\vec(\bs\Omega_{h_1})\vec(\bs\Omega_{h_1})'(\bs P_{[h_1,g]}\otimes\bs P_{[h_1,g]})\big]\big]^{1/2}\\
            &\quad\big[\tr\big[\E(\vec(\bar{\bs\varepsilon}_{[g]}\bar{\bs\varepsilon}'_{[g]})\vec(\bar{\bs\varepsilon}_{[g]}\bar{\bs\varepsilon}'_{[g]})|\mathcal{Z})(\bs P_{[g,h_2]}\otimes\bs P_{[g,h_2]})\vec(\bs\Omega_{h_2})\\&\quad\vec(\bs\Omega_{h_2})'(\bs P_{[h_2,g]}\otimes\bs P_{[h_2,g]})\E\big(\vec(\bar{\bs\varepsilon}_{[g]}\bar{\bs\varepsilon}'_{[g]})\vec(\bar{\bs\varepsilon}_{[g]}\bar{\bs\varepsilon}'_{[g]})'\big|\mathcal{Z}\big)\big]\big]^{1/2}\\
            &\leq\frac{C\nmax^4}{k^2}\sum_{g=2}^G\sum_{\substack{h_1, h_2<g\\h_1\neq h_2}}\big[\tr\big[(\bs P_{[g,h_1]}\otimes\bs P_{[g,h_1]})(\bs P_{[h_1,g]}\otimes\bs P_{[h_1,g]})\big]\big]^{1/2}\\
            &\quad\big[\tr\big[(\bs P_{[g,h_2]}\otimes\bs P_{[g,h_2]})(\bs P_{[h_2,g]}\otimes\bs P_{[h_2,g]})\big]\big]^{1/2}\\
            &\leq\frac{C\nmax^4}{k^2}\sum_{g,h_1,h_2=1}^G\tr(\bs P_{[g,h_1]}\bs P_{[h_1,g]})\tr(\bs P_{[g,h_2]}\bs P_{[h_2,g]})\\
            &=\frac{C\nmax^4}{k^2}\sum_{g=1}^G\tr^2(\bs P_{[g,g]})\toas 0,
        \end{split}
    \end{equation}
    by similar steps as in \cref{eq:PPPP}. Then by the triangle inequality $\sum_{g=2}^G\E(y_{gn}^4|\mathcal{Z})\toas 0$.
    
    Conditional on $\mathcal{Z}$, to apply the martingale central limit theorem, it suffices to show that for any $\epsilon>0$
    \begin{equation}\label{eq:pcondvar}
        \begin{split}
            \Pr\bigg(\bigg|\sum_{g=2}^G\E(y_{gn}^2|\mathcal{X}_1,\dots,\mathcal{X}_{g-1},\mathcal{Z})-s_{n}^2(\mathcal{Z})\bigg|\geq\epsilon\bigg|\mathcal{Z}\bigg)\to 0.
        \end{split}
    \end{equation}
    Now note that $\E(w_{gn}\bar{y}_{gn}|\mathcal{Z})=0$ a.s.\ and thus 
    \begin{equation}\label{eq:condvar}
        \begin{split}
            &\sum_{g=2}^G\E(y_{gn}^2|\mathcal{X}_1,\dots,\mathcal{X}_{g-1},\mathcal{Z})-s_{n}^2(\mathcal{Z})=\sum_{g=2}^G\big[\E(w_{gn}^2|\mathcal{X}_1,\dots,\mathcal{X}_{g-1},\mathcal{Z})-\E(w_{gn}^2|\mathcal{Z})\big]\\
            &\quad+2\sum_{g=2}^G\E(w_{gn}\bar{y}_{gn}|\mathcal{X}_1,\dots,\mathcal{X}_{g-1},\mathcal{Z})+\sum_{g=2}^G\big[\E(\bar{y}_{gn}^2|\mathcal{X}_1,\dots,\mathcal{X}_{g-1},\mathcal{Z})-\E(y_{gn}^2|\mathcal{Z})\big].
        \end{split}
    \end{equation}
    I will show that each term on the right-hand side of \eqref{eq:condvar} converges to zero a.s. To proceed, note first that by independence of $\bs W_{[1],n},\dots,\bs W_{[G],g}$ conditional on $\mathcal{Z}$, $\E(w_{gn}^2|\mathcal{X}_1,\dots,\mathcal{X}_{g-1},\mathcal{Z})=\E(w_{gn}^2|\mathcal{Z})$ a.s. Next, note that $\E(w_{gn}\bar{y}_{gn}|\mathcal{X}_1,\dots,\mathcal{X}_{g-1},\mathcal{Z})\allowbreak=\E(w_{gn}\bs u_{[g]}'|\mathcal{Z})\sum_{h<g}\bs P_{[g,h]}\bar{\bs\varepsilon}_{[h]}/\sqrt{k}+\E(w_{gn}\bar{\bs\varepsilon}_{[g]}'|\mathcal{Z})\sum_{h<g}\bs P_{[g,h]}\bs u_{[h]}/\sqrt{k}$. Let $\bs\delta_{[g]}=\bs\delta_{[g]}(\mathcal{Z})=\E(w_{gn}\bs u_{[g]}|\mathcal{Z})$ and consider the first term, $\bs\delta_{[g]}'\sum_{h<g}\bs P_{[g,h]}\bar{\bs\varepsilon}_{[h]}/\sqrt{k}$. Let $\bar{\bs P}$ be the upper block triangular matrix with $\bar{\bs P}_{[g,h]}=\bs P_{[g,h]}$ for $h>g$ and $\bar{\bs P}_{[g,h]}=\bs 0$ for $g\leq h$, and let $\bs\delta=(\bs\delta_{[1]}',\dots,\bs\delta_{[G]}')'$. Then, $\sum_{g=2}^G\sum_{h<g}\bs\delta_{[g]}\bs P_{[g,h]}\bar{\bs\varepsilon}_{[h]}/\sqrt{k}=\bs\delta'\bar{\bs P}\bar{\bs\varepsilon}/\sqrt{k}$. By the Cauchy--Schwarz inequality, $\bs\delta'\bs\delta=\sum_{g=1}^G\sum_{i\in[g]}\big[\E(w_{gn}u_{i}|\mathcal{Z})\big]^2\leq\sum_{g=1}^G\sum_{i\in[g]}\allowbreak\E(w_{gn}^2|\mathcal{Z})\E(u_{i}^2|\mathcal{Z})\leq C\nmax\bs b_{1n}'\bs D_{n}\bs b_{1n}\leq C\nmax$ a.s. By \cref{lem:LL}, $\|\bar{\bs P}'\bar{\bs P}\|_{F}\leq C\sqrt{k}$ a.s., which in turn implies that $\lambdamax(\bar{\bs P}'\bar{\bs P})\leq C\sqrt{k}$ a.s. It then follows given $\E(\bar{\bs\varepsilon}_{[g]}\bar{\bs\varepsilon}_{[g]}'|\mathcal{Z})\leq C\nmax$ a.s.\ that $\E\big[(\bs\delta'\bar{\bs P}'\bar{\bs\varepsilon}/\sqrt{k})^2\big|\mathcal{Z}\big]\leq C\nmax\bs\delta'\bar{\bs P}'\bar{\bs P}\bs\delta/k\allowbreak\leq C\nmax\|\bs\delta\|^2/\sqrt{k}\leq C\nmax^2/\sqrt{k}\toas 0$, so that by the conditional Markov's inequality we have for any $\epsilon>0$, $\Pr\big(\big|\bs\delta(\mathcal{Z})'\bar{\bs P}'\bar{\bs\varepsilon}/\sqrt{k}\big|\geq\epsilon\big|\mathcal{Z}\big)\toas 0$. Similarly, we have $\sum_{g=2}^G\E(w_{gn}\bar{\bs\varepsilon}_{[g]}'|\mathcal{Z})\sum_{h<g}\bs P_{[g,h]}\bs u_{[h]}/\sqrt{k}\toas 0$. Therefore, it follows by the triangle inequality that, for any $\epsilon>0$, $\Pr\big(\big|\sum_{g=2}^G\E(w_{gn}\bar{y}_{gn}|\mathcal{X}_{1},\dots,\mathcal{X}_{g-1},\mathcal{Z})\big|\geq\epsilon\big|\mathcal{Z}\big)\toas0$.
    
    To finish showing that equation \cref{eq:pcondvar} is satisfied, it only remains to show that, for any $\epsilon>0$,
    \begin{equation}\label{eq:pcondvar last}
        \begin{split}
            \Pr\bigg(\bigg|\sum_{g=2}^G\big[\E(\bar{y}_{gn}^2|\mathcal{X}_{1},\dots,\mathcal{X}_{g-1},\mathcal{Z})-\E(\bar{y}_{gn}^2|\mathcal{Z})\big]\bigg|\geq\epsilon\bigg|\mathcal{Z}\bigg)\toas 0.
        \end{split}
    \end{equation}
    Now, write
    \begin{equation}\label{eq:condvar y}
        \begin{split}
            &\sum_{g=2}^G\big[\E(\bar{y}_{gn}^2|\mathcal{X}_{1},\dots,\mathcal{X}_{g-1},\mathcal{Z})-\E(\bar{y}_{gn}^2|\mathcal{Z})\big]\\
            &=\sum_{g=2}^G\bigg[\sum_{h<g}\tr(\bs\Omega_{g}\bs P_{[g,h]}(\bar{\bs\varepsilon}_{[h]}\bar{\bs\varepsilon}_{[h]}'-\bs\Sigma_{h})\bs P_{[g,h]})/k+2\sum_{i<h<g}\tr(\bs\Omega_{g}\bs P_{[g,h]}\bar{\bs\varepsilon}_{[h]}\bar{\bs\varepsilon}_{[i]}'\bs P_{[i,g]})/k\\
            &\quad+\sum_{h<g}\tr(\bs\Sigma_{g}\bs P_{[g,h]}(\bs u_{[h]}\bs u_{[h]}'-\bs\Omega_{h})\bs P_{[g,h]})/k+2\sum_{i<h<g}\tr(\bs\Sigma_{g}\bs P_{[g,h]}\bs u_{[h]}\bs u_{[i]}'\bs P_{[i,g]})/k\\
            &\quad+2\sum_{h<g}\tr(\bs\Gamma_{g}\bs P_{[g,h]}(\bs u_{[h]}\bar{\bs\varepsilon}_{[h]}'-\bs\Gamma_{h})\bs P_{[g,h]})/k+2\sum_{i<h<g}\tr(\bs\Gamma_{g}\bs P_{[g,h]}\bs u_{[h]}\bar{\bs\varepsilon}_{[i]}'\bs P_{[i,g]})/k\\
            &\qquad+2\sum_{i<h<g}\tr(\bs\Gamma_{g}\bs P_{[g,i]}\bs u_{[i]}\bar{\bs\varepsilon}_{[h]}'\bs P_{[h,g]})/k\bigg].
        \end{split}
    \end{equation}

    With slight abuse of notation, define
    \begin{equation}
        \begin{split}
            \bs v_{[g]}'=\begin{bmatrix}
                \bar{\bs\varepsilon}_{[g]}' & \bs u_{[g]}'
            \end{bmatrix},\ \bs M_g=\begin{bmatrix}
                \bs\Sigma_{g} & \bs\Gamma_{g}' \\
                \bs\Gamma_{g} & \bs\Omega_{g}
            \end{bmatrix}\text{ and }\bs P_{[g,h]}^*=\begin{bmatrix}
                \bs 0 & \bs P_{[g,h]} \\
                \bs P_{[g,h]} & \bs 0
            \end{bmatrix}.
        \end{split}
    \end{equation}
    For later purposes also define
    \begin{equation}
        \begin{split}
            \bs P_{[g]}^*=\begin{bmatrix}
                \bs 0 & \bs P_{[g]} \\
                \bs P_{[g]} & \bs 0
                \end{bmatrix},\ \bs P^*=\begin{bmatrix}
                    \bs 0 & \bs P \\
                    \bs P & \bs 0
                \end{bmatrix},\ \bs M=\begin{bmatrix}
                    \bs\Sigma & \bs\Gamma' \\
                    \bs\Gamma & \bs\Omega
                \end{bmatrix},\text{ and } \bs T_{g}^*=\begin{bmatrix}
                \bs T_{g} & \bs 0\\
                \bs 0 & \bs T_{g}
            \end{bmatrix},
        \end{split}
    \end{equation}
    where $\bs T_g$ is a $n\times n$ diagonal matrix with ones on the diagonal elements corresponding to indices in clusters $h\geq g$ and zeros elsewhere.
    
    Then \cref{eq:condvar y} can be written as
    \begin{equation}\label{eq:vM}
        \begin{split}
            \sum_{g=2}^G\bigg[\sum_{h<g}\tr(\bs M_{g}\bs P^*_{[g,h]}(\bs v_{[h]}\bs v_{[h]}'-\bs M_{h})\bs P_{[h,g]}^*)/k+2\sum_{i<h<g}\tr(\bs M_{g}\bs P^*_{[g,h]}\bs v_{[h]}\bs v_{[i]}'\bs P_{[i,g]}^*)/k\bigg].
        \end{split}
    \end{equation}
    Consider first the first term and write
    \begin{equation}
        \begin{split}
            &\E\bigg(\bigg[\sum_{g=2}^G\sum_{h<g}\tr\big(\bs M_{g}\bs P^*_{[g,h]}(\bs v_{[h]}\bs v_{[h]}'-\bs M_{h})\bs P_{[h,g]}^*\big)/k\bigg]^2\bigg|\mathcal{Z}\bigg)\\
            &=\frac{1}{k^2}\E\bigg(\sum_{g_1,g_2=2}^G\sum_{h_1<g_1,h_2<g_2}\tr\big(\bs M_{g_1}\bs P^*_{[g_1,h_1]}(\bs v_{[h_1]}\bs v_{[h_1]}'-\bs M_{h_1})\bs P_{[h_1,g_1]}^*\big)\\
            &\quad\tr\big(\bs M_{g_2}\bs P^*_{[g_2,h_2]}(\bs v_{[h_2]}\bs v_{[h_2]}'-\bs M_{h_2})\bs P_{[h_2,g_2]}^*\big)\bigg|\mathcal{Z}\bigg)\\
            &=\frac{1}{k^2}\E\bigg(\sum_{g_1,g_2=2}^G\sum_{h<\min\{g_1,g_2\}}\tr\big(\bs M_{g_1}\bs P^*_{[g_1,h]}(\bs v_{[h]}\bs v_{[h]}'-\bs M_{h})\bs P_{[h,g_1]}^*\big)\\
            &\quad\tr\big(\bs M_{g_2}\bs P^*_{[g_2,h]}(\bs v_{[h]}\bs v_{[h]}'-\bs M_{h})\bs P_{[h,g_2]}^*\big)\bigg|\mathcal{Z}\bigg),
        \end{split}
    \end{equation}
    by the zero expectation of $\bs v_{[h_1]}\bs v_{[h_1]}'-\bs M_{h_1}$. Next, write 
    \begin{equation}
        \begin{split}
            &\frac{1}{k^2}\E\bigg(\sum_{g_1,g_2=2}^G\sum_{h<\min\{g_1,g_2\}}\tr\big(\bs M_{g_1}\bs P^*_{[g_1,h]}(\bs v_{[h]}\bs v_{[h]}'-\bs M_{h})\bs P_{[h,g_1]}^*\big)\\
            &\quad\tr\big(\bs M_{g_2}\bs P^*_{[g_2,h]}(\bs v_{[h]}\bs v_{[h]}'-\bs M_{h})\bs P_{[h,g_2]}^*\big)\bigg|\mathcal{Z}\bigg)\\
            &=\frac{1}{k^2}\E\bigg(\sum_{h=1}^{G-1}\sum_{g_1,g_2>h}\tr\big(\bs M_{g_1}\bs P^*_{[g_1,h]}(\bs v_{[h]}\bs v_{[h]}'-\bs M_{h})\bs P_{[h,g_1]}^*\big)\\
            &\quad\tr\big(\bs M_{g_2}\bs P^*_{[g_2,h]}(\bs v_{[h]}\bs v_{[h]}'-\bs M_{h})\bs P_{[h,g_2]}^*\big)\bigg|\mathcal{Z}\bigg)\\
            &=\frac{1}{k^2}\E\bigg(\sum_{h=1}^{G-1}\tr\big(\bs M\bs T_{h+1}^*\bs P^{*\prime}_{[h]}(\bs v_{[h]}\bs v_{[h]}'-\bs M_{h})\bs P_{[h]}^*\bs T_{h+1}^*\big)\\
            &\quad\tr\big(\bs M\bs T_{h+1}^*\bs P^{*\prime}_{[h]}(\bs v_{[h]}\bs v_{[h]}'-\bs M_{h})\bs P_{[h]}^*\bs T_{h+1}^*\big)\bigg|\mathcal{Z}\bigg)\\
            &=\frac{1}{k^2}\E\bigg(\sum_{h=1}^{G-1}\vec(\bs P_{[h]}^*\bs T_{h+1}^*\bs M\bs T_{h+1}^*\bs P^{*\prime}_{[h]})'\vec(\bs v_{[h]}\bs v_{[h]}'-\bs M_{h})\\
            &\quad\vec(\bs v_{[h]}\bs v_{[h]}'-\bs M_{h})'\vec(\bs P_{[h]}^*\bs T_{h+1}^*\bs M\bs T_{h+1}^*\bs P^{*\prime}_{[h]})\bigg|\mathcal{Z}\bigg)\\
            &\leq\frac{C\nmax^2}{k^2}\sum_{h=1}^{G-1}\vec(\bs P_{[h]}^*\bs T_{h+1}^*\bs M\bs T_{h+1}^*\bs P^{*\prime}_{[h]})'\vec(\bs P_{[h]}^*\bs T_{h+1}^*\bs M\bs T_{h+1}^*\bs P^{*\prime}_{[h]}),
        \end{split}
    \end{equation}
    by a result similar to \cref{eq:eigbound kappa}.
    \begin{equation}\label{eq:vM 1 as}
        \begin{split}
            &\frac{C\nmax^2}{k^2}\sum_{h=1}^{G-1}\vec(\bs P_{[h]}^*\bs T_{h+1}^*\bs M\bs T_{h+1}^*\bs P^{*\prime}_{[h]})'\vec(\bs P_{[h]}^*\bs T_{h+1}^*\bs M\bs T_{h+1}^*\bs P^{*\prime}_{[h]})\\
            &=\frac{C\nmax^2}{k^2}\sum_{h=1}^{G-1}\tr(\bs P_{[h]}^*\bs T_{h+1}^*\bs M\bs T_{h+1}^*\bs P^{*\prime}_{[h]}\bs P_{[h]}^*\bs T_{h+1}^*\bs M\bs T_{h+1}^*\bs P^{*\prime}_{[h]})\toas 0,
        \end{split}
    \end{equation}
    by eigenvalue bounds on $\bs P^{*\prime}_{[h]}\bs P^*_{[h]}$, $\bs M\bs M$, $\sum_{h=1}^{G-1}\tr(\bs P^{*\prime}_{[h]}\bs P^*_{[h]})\leq2k$ and \cref{it:rate}.

    For the second term in \cref{eq:vM}, write
    \begin{equation}\label{eq:vM 2 as}
        \begin{split}
            &\E\bigg(\bigg[2\sum_{i<h<g}\tr(\bs M_{g}\bs P^*_{[g,h]}\bs v_{[h]}\bs v_{[i]}'\bs P_{[i,g]}^*)/k\bigg]^2\bigg|\mathcal{Z}\bigg)\\
            &=\frac{C}{k^2}\E\bigg(\bigg[\sum_{g=2}^G\sum_{\substack{h,l<g\\h\neq l}}\bs v_{[h]}^{\prime}\bs P^*_{[h,g]}\bs M_{g}\bs P^*_{[g,l]}\bs v_{[l]}\bigg]^2\bigg|\mathcal{Z}\bigg)\\
            &=\frac{1}{k^2}\E\bigg(\sum_{g_1,g_2=2}^G\sum_{\substack{h_1,l_1<g_1\\h_1\neq l_1}}\sum_{\substack{h_2,l_2<g_2\\h_2\neq l_2}}\bs v_{[h_1]}^{\prime}\bs P^*_{[h_1,g_1]}\bs M_{g_1}\bs P^*_{[g_1,l_1]}\bs v_{[l_1]}\\
            &\quad \bs v_{[h_2]}^{\prime}\bs P^*_{[h_2,g_2]}\bs M_{g_2}\bs P^*_{[g_2,l_2]}\bs v_{[l_2]}\bigg|\mathcal{Z}\bigg)\\
            &=\frac{C}{k^2}\E\bigg(\sum_{g_1,g_2=2}^G\sum_{\substack{h,l<\min\{g_1,g_2\}\\h\neq l}}\bs v_{[h]}^{\prime}\bs P^*_{[h,g_1]}\bs M_{g_1}\bs P^*_{[g_1,l]}\bs v_{[l]}\bs v_{[h]}^{\prime}\bs P^*_{[h,g_2]}\bs M_{g_2}\bs P^*_{Z,[g_2,l]}\bs v_{[l]}\bigg|\mathcal{Z}\bigg)\\
            &=\frac{C}{k^2}\sum_{g_1,g_2=2}^G\sum_{\substack{h,l<\min\{g_1,g_2\}\\h\neq l}}\tr(\bs M_{h}\bs P^*_{Z,[h,g_1]}\bs M_{g_1}\bs P^*_{[g_1,l]}\bs M_{l}\bs P^*_{[l,g_2]}\bs M_{g_2}\bs P^*_{[g_2,h]})\\
            &=\frac{C}{k^2}\sum_{\substack{h,l=1\\h\neq l}}^{G-1}\sum_{g_1,g_2>\max\{h,l\}}\tr(\bs M_{h}\bs P^*_{[h,g_1]}\bs M_{g_1}\bs P^*_{[g_1,l]}\bs M_{l}\bs P^*_{[l,g_2]}\bs M_{g_2}\bs P^*_{[g_2,h]})\\
            &=\frac{C}{k^2}\sum_{\substack{h,l=1\\h\neq l}}^{G-1}\tr\bigg(\bs M_{h}\bigg[\sum_{g>\max\{h,l\}}\bs P^*_{[h,g]}\bs M_{g}\bs P^*_{[g,l]}\bigg]\bs M_{l}\bigg[\sum_{g>\max\{h,l\}}\bs P^*_{[l,g]}\bs M_{g}\bs P^*_{[g,h]}\bigg]\bigg)\\
            &\leq\frac{C}{k^2}\sum_{\substack{h,l=1\\h\neq l}}^{G-1}\tr(\bs P^*_{[h]}\bs T^*_{\max\{h,l\}+1}\bs M\bs T^*_{\max\{h,l\}+1}\bs P^{*\prime}_{[l]}\bs P^*_{[l]}\bs T^*_{\max\{h,l\}+1}\bs M\bs T^*_{\max\{h,l\}+1}\bs P^{*\prime}_{[h]})\\
            &=\frac{C}{k^2}\sum_{h=1}^{G-1}\sum_{l<h}\tr(\bs P^*_{[h]}\bs T^*_{h+1}\bs M\bs T^*_{h+1}\bs P^{*\prime}_{[l]}\bs P^*_{[l]}\bs T^*_{h+1}\bs M\bs T^*_{h+1}\bs P^{*\prime}_{[h]})\toas 0,
        \end{split}
    \end{equation}
    since $\lambdamax(\sum_{l<h}\bs P^{*\prime}_{[l]}\bs P^*_{[l]})=\lambdamax(\bs P^*\bs S_{h-1}^*\bs P^*)=\lambdamax(\bs P^*)\leq C$, where $\bs S^*_{h}$ is similar to $\bs T^*_{h}$ but then with $\bs S_{h}$ instead of $\bs T_{h}$ and $\bs S_g$ is a $n\times n$ diagonal matrix with ones on the diagonal elements corresponding to indices in clusters $h\leq g$, and $\lambdamax(\bs M\bs M)\leq C$.

    Then combining the triangle and Markov's inequalities with \cref{eq:vM 1 as,eq:vM 2 as} I conclude that \cref{eq:pcondvar last} and hence \cref{eq:pcondvar} hold.

    The preceding argument shows that as $n\to\infty$, $\Pr(Y_n\leq y|\mathcal{Z})\to\Phi(y)$ a.s.\ $\Pr_{\mathcal{Z}}$, for every real number $y$, where $\Phi(y)$ denotes the cumulative distribution function of a standard normal distribution. Moreover, it is clear that, for some $\epsilon>0$, $\sup_{n}\E\big[\Pr(Y_n\leq y|\mathcal{Z})\big]\to\E\big(\Phi(y)\big)=\Phi(y)$, which gives the desired conclusion.  
\end{proof}


\subsection{Proof of \texorpdfstring{\cref{thm:V cljar}}{Lemma A.2}}\label{app:plug-in variance}
The proof uses the same approach as \citet{crudu2021inference}.
\subsubsection{Unbiasedness}
Consider the conditional expectation of the three components of the variance estimator in \cref{eq:Vhat clj}
\begin{equation}
    \begin{split}
        \E\big(\hat{V}_{CLJ}^{AR}(\bs\beta_0)\big|\bs Z\big)
        &=\E\bigg(\frac{2}{k}\sum_{g\neq h}\bs\varepsilon_{[g]}'\bs P_{Z,[g,h]}\bs\varepsilon_{[h]}\bs\varepsilon_{[h]}'\bs P_{Z,[h,g]}\bs\varepsilon_{[g]}\bigg|\bs Z\bigg)\\
        &=\frac{2}{k}\sum_{g\neq h}\tr(\bs\Sigma_{g}\bs P_{Z,[g,h]}\bs\Sigma_{h}\bs P_{Z,[h,g]});\\
        \E\big(\hat{\bs V}_{CLJ}^{S}(\bs\beta_0)\big|\bs Z\big)&=\frac{1}{n}\E\bigg(\bs X'\ddot{\bs P}_{Z}\bs B_{\varepsilon\varepsilon'}\ddot{\bs P}_{Z}\bs X+\sum_{g\neq h}\bs X_{[g]}'\bs P_{Z,[g,h]}\bs\varepsilon_{[h]}\bs\varepsilon_{[g]}'\bs P_{Z,[g,h]}\bs X_{[h]}\bigg|\bs Z\bigg)\\
        &=\frac{1}{n}\E\bigg(\bs\Pi'\bs Z'\ddot{\bs P}_{Z}\bs B_{\varepsilon\varepsilon'}\ddot{\bs P}_{Z}\bs Z\bs\Pi+\sum_{g\neq h\neq j}\bs\eta_{[g]}'\bs P_{Z,[g,h]}\bs\varepsilon_{[h]}\bs\varepsilon_{[h]}'\bs P_{Z,[h,j]}\bs\eta_{[j]}\\
        &\quad+\sum_{g\neq h}(\bs Z\bs\Pi+\bs\eta)_{[g]}'\bs P_{Z,[g,h]}\bs\varepsilon_{[h]}\bs\varepsilon_{[g]}'\bs P_{Z,[g,h]}(\bs Z\bs\Pi+\bs\eta)_{[h]}\bigg|\bs Z\bigg)\\
        &=\frac{1}{n}\bigg[\bs\Pi'\bs Z'\ddot{\bs P}_{Z}\bs\Sigma\ddot{\bs P}_{Z}\bs Z\bs\Pi+\E\bigg(\sum_{g\neq h}\bs\eta_{[g]}'\bs P_{Z,[g,h]}\bs\varepsilon_{[h]}\bs\varepsilon_{[h]}'\bs P_{Z,[h,g]}\bs\eta_{[g]}\\
        &\quad+\sum_{g\neq h}\bs\eta_{[g]}'\bs P_{Z,[g,h]}\bs\varepsilon_{[h]}\bs\varepsilon_{[g]}'\bs P_{Z,[g,h]}\bs\eta_{[h]}\bigg|\bs Z\bigg)\bigg];\\
        \E\big(\hat{\bs C}_{CLJ}(\bs\beta_0)\big|\bs Z\big)&=\E\bigg(\frac{2}{\sqrt{nk}}\sum_{g\neq h}\bs X_{[g]}'\bs P_{Z,[g,h]}\bs\varepsilon(\bs\beta)_{[h]}\bs\varepsilon(\bs\beta)_{[g]}'\bs P_{Z,[g,h]}\bs\varepsilon(\bs\beta)_{[h]}\bigg|\bs Z\bigg)\\
        &=\E\bigg(\frac{2}{\sqrt{nk}}\sum_{g\neq h}\bs\eta_{[g]}'\bs P_{Z,[g,h]}\bs\varepsilon(\bs\beta)_{[h]}\bs\varepsilon(\bs\beta)_{[g]}'\bs P_{Z,[g,h]}\bs\varepsilon(\bs\beta)_{[h]}\bigg|\bs Z\bigg).
    \end{split}
\end{equation}
Comparing these expressions with those in \cref{eq:V clj AR,eq:V clj S,eq:C clj} shows that $\hat{\bs V}_{CLJ}(\bs\beta_0)$ is conditionally unbiased.

\subsubsection{Consistency}
First consider the variance estimator of the cluster jackknife AR statistic. Write it as
\begin{equation}
    \begin{split}
        \hat{V}_{CLJ}^{AR}(\bs\beta_0)=\frac{2}{k}\tr(\bs B_{\varepsilon\varepsilon'}\ddot{\bs P}_{Z}\bs B_{\varepsilon\varepsilon'}\ddot{\bs P}_{Z}).
    \end{split}
\end{equation}

Define $\bs H=\bs\Sigma-\bs B_{\varepsilon\varepsilon'}$ and $\bs H_{g}=\bs\Sigma_{g}-\bs\varepsilon_{[g]}\bs\varepsilon_{[g]}'$, such that \begin{equation}\label{eq:V-Vhat}
    \begin{split}
        V_{CLJ}^{AR}-\hat{V}_{CLJ}^{AR}(\bs\beta_0)=\frac{2}{k}\tr(\bs H\ddot{\bs P}_{Z}\bs H\ddot{\bs P}_{Z}+\bs H\ddot{\bs P}_{Z}\bs\Sigma\ddot{\bs P}_{Z}+\bs\Sigma\ddot{\bs P}_{Z}\bs H\ddot{\bs P}_{Z}).
    \end{split}
\end{equation}

Now, since $\E(\bs H_{g}|\bs Z)=\bs 0$,
\begin{equation}\label{eq:HPHP}
    \begin{split}
        &\E\bigg(\bigg[\frac{2}{k}\sum_{g\neq h}\tr(\bs H_{g}\bs P_{Z,[g,h]}\bs H_{h}\bs P_{Z,[h,g]})\bigg]^2\bigg|\bs Z\bigg)\\
        &=\frac{8}{k^2}\E\bigg(\sum_{g\neq h}\big[\tr(\bs H_{g}\bs P_{Z,[g,h]}\bs H_{h}\bs P_{Z,[h,g]})\big]^2\bigg|\bs Z\bigg)\\
        &\leq\frac{8}{k^2}\E\bigg(\sum_{g\neq h}\tr(\bs H_{g}\bs P_{Z,[g,h]}\bs P_{Z,[h,g]}\bs H_{g})\tr(\bs H_{h}\bs P_{Z,[h,g]}\bs P_{Z,[g,h]}\bs H_{h})\bigg|\bs Z\bigg)\\
        &\leq\frac{C\nmax^4}{k^2}\sum_{g\neq h}\tr^2(\bs P_{Z,[h,g]}\bs P_{Z,[g,h]})\toas 0,
    \end{split}
\end{equation}
by the Cauchy--Schwarz inequality on the Frobenius inner product, the result below and steps similar to those in \cref{eq:PPPP} below.
\begin{equation}\label{eq:HH}
    \begin{split}
        &\E\big(\lambdamax(\bs H_{g}\bs H_{g})\big|\bs Z\big)\\
        &\leq\lambdamax(\bs\Sigma_{g}\bs\Sigma_{g})+2\E\big(\lambdamax(\bs\Sigma_{g}\bs\varepsilon_{[g]}\bs\varepsilon_{[g]}')\big|\bs Z\big)+\E\big(\lambdamax(\bs\varepsilon_{[g]}\bs\varepsilon_{[g]}'\bs\varepsilon_{[g]}\bs\varepsilon_{[g]}')\big|\bs Z\big)\\
        &\leq 3\lambdamax(\bs\Sigma_{g}\bs\Sigma_{g})+\E(\bs\varepsilon_{[g]}'\bs\varepsilon_{[g]}\bs\varepsilon_{[g]}'\bs\varepsilon_{[g]})\leq n_g^2 C,
    \end{split}
\end{equation}
by \cref{assit:cljar eigval,assit:cljar moment}.

The expectation of the other terms in \cref{eq:V-Vhat} squared converge by the same arguments. Finally, the triangle and Markov's inequality lead to the conclusion that $\hat{V}_{CLJ}^{AR}(\bs\beta_0)\toprob V_{CLJ}^{AR}(\bs\beta_0)$.

For the consistency of the covariance estimator $\hat{\bs C}_{CLJ}(\bs\beta)$ consider its $i^{\text{th}}$ element, $i=1,\dots,p$ and write $\bs J^{(i)}=\bs\Xi^{(i)}-\bs B_{\eta_{(i)}\varepsilon'}$ for $\bs\Xi^{(i)}=\E(\bs\eta_{(i)}\bs\varepsilon'|\bs Z)$. Let $\bs\Xi_{g}^{(i)}$ be the $g^{\text{th}}$ diagonal block of $\bs\Xi^{(i)}$. Then
\begin{equation}\label{eq:C-Chat}
    \begin{split}
        \bs C_{CLJ,i}-\hat{\bs C}_{CLJ}(\bs\beta_0)_{i}=\frac{2}{\sqrt{nk}}\tr(\bs J^{(i)}\ddot{\bs P}_{Z}\bs H\ddot{\bs P}_{Z}+\bs J^{(i)}\ddot{\bs P}_{Z}\bs\Sigma\ddot{\bs P}_{Z}+\bs\Xi^{(i)}\ddot{\bs P}_{Z}\bs H\ddot{\bs P}_{Z}),
    \end{split}
\end{equation}
and consistency follows by similar arguments as for \cref{eq:V-Vhat}.

Finally, consider the $(i,j)^{\text{th}}$ element of the variance estimator of the score. Write it as
\begin{equation}\label{eq:Vhat S consistency}
    \begin{split}
        &\hat{V}^S_{CLJ}(\bs\beta_0)_{ij}\\
        &=\frac{1}{n}\bigg(\bs\Pi_{(i)}'\bs Z'\ddot{\bs P}_{Z}\bs B_{\varepsilon\varepsilon'}\ddot{\bs P}_{Z}\bs Z\bs\Pi_{(j)}+\sum_{g\neq h}\bs\eta_{(i),[g]}'\bs P_{Z,[g,h]}\bs\varepsilon_{[h]}\bs\varepsilon_{[h]}'\bs P_{Z,[h,g]}\bs\eta_{(j),[g]}\\
        &\quad+\sum_{g\neq h}\bs\eta_{(i),[g]}'\bs P_{Z,[g,h]}\bs\varepsilon_{[h]}\bs\varepsilon_{[g]}'\bs P_{Z,[g,h]}\bs\eta_{(j),[h]}+\bs\Pi'_{(i)}\bs Z'\ddot{\bs P}_{Z}\bs B_{\varepsilon\varepsilon'}\ddot{\bs P}_{Z}\bs\eta_{(j)}\\
        &\quad+\bs\eta_{(i)}'\ddot{\bs P}_{Z}\bs B_{\varepsilon\varepsilon'}\ddot{\bs P}_{Z}\bs Z\bs\Pi_{(j)}+\sum_{g\neq h\neq l\neq g}\bs\eta_{(i),[g]}'\bs P_{Z,[g,h]}\bs\varepsilon_{[h]}\bs\varepsilon_{[h]}'\bs P_{Z,[h,l]}\bs\eta_{(j),[l]}\\
        &\quad+\sum_{g\neq h}\bs\Pi_{(i)}'\bs Z_{[g]}'\bs P_{Z,[g,h]}\bs\varepsilon_{[h]}\bs\varepsilon_{[g]}'\bs P_{Z,[g,h]}\bs Z_{[h]}\bs\Pi_{(j)}\\
        &\quad+\sum_{g\neq h}\bs\eta_{(i),[g]}'\bs P_{Z,[g,h]}\bs\varepsilon_{[h]}\bs\varepsilon_{[g]}'\bs P_{Z,[g,h]}\bs Z_{[g]}\bs\Pi_{(j)}\\
        &\quad+\sum_{g\neq h}\bs\Pi_{(i)}'\bs Z_{[g]}'\bs P_{Z,[g,h]}\bs\varepsilon_{[h]}\bs\varepsilon_{[g]}'\bs P_{Z,[g,h]}\bs\eta_{(j),[g]}\bigg),
    \end{split}
\end{equation}
and write the $(i,j)^\text{th}$ element of the variance as
\begin{equation}\label{eq:V S consistency}
    \begin{split}
        V^{S}_{CLJ,ij}&=\frac{1}{n}\bigg[\bs\Pi_{(i)}'\bs Z'\ddot{\bs P}_{Z}\bs B_{\varepsilon\varepsilon'}\ddot{\bs P}_{Z}\bs Z\bs\Pi_{(j)}+\sum_{g\neq h}\tr(\bs\Omega^{(i,j)}_{g}\bs P_{Z,[g,h]}\bs\Sigma_{h}\bs P_{Z,[h,g]})+\\
        &\quad\sum_{g\neq h}\tr(\bs\Xi^{(i)}_{g}\bs P_{Z,[g,h]}\bs\Xi^{(j)}_{h}\bs P_{Z,[h,g]})\bigg],
    \end{split}
\end{equation}
where $\bs\Omega^{(i,j)}_{g}=\E(\bs\eta_{(i)}\bs\eta_{(j)}'|\bs Z)$.

Note the similarity of the first three terms in \cref{eq:Vhat S consistency} and \cref{eq:V S consistency}. Take $\bs H$ as before, then the expectation of the squared difference between the first terms is
\begin{equation}\label{eq:PiZPHPZPi}
    \begin{split}
        &\frac{1}{n^2}\E\big([\bs\Pi_{(i)}'\bs Z'\ddot{\bs P}_{Z}\bs H\ddot{\bs P}_{Z}\bs Z\bs\Pi_{(j)}]^2\big|\bs Z\big)\\
        &=\frac{1}{n^2}\E\bigg(\sum_{h=1}^G\bigg[\bigg(\sum_{g\neq h}\bs\Pi_{(i)}'\bs Z_{[g]}'\bs P_{Z,[g,h]}\bigg)\bs H_h\bigg(\sum_{g\neq h}\bs P_{Z,[h,g]}\bs Z_{[g]}\bs\Pi_{(j)}\bigg)\bigg]^2\bigg|\bs Z\bigg)\\
        &\leq\frac{1}{n^2}\E\bigg(\sum_{h=1}^G\bigg(\sum_{g\neq h}\bs\Pi_{(i)}'\bs Z_{[g]}'\bs P_{Z,[g,h]}\bigg)\bigg(\sum_{g\neq h}\bs P_{Z,[h,g]}\bs Z_{[g]}\bs\Pi_{(i)}\bigg)\\
        &\quad\bigg(\sum_{g\neq h}\bs\Pi_{(j)}'\bs Z'_{[g]}\bs P_{Z,[g,h]}\bigg)\bs H_{h}\bs H_h\bigg(\sum_{g\neq h}\bs P_{Z,[h,g]}\bs Z_{[g]}\bs\Pi_{(j)}\bigg)\bigg|\bs Z\bigg)\\
        &\leq\frac{C\nmax^2}{n^2}\sum_{h=1}^G\bigg(\sum_{g=1}^G\bs\Pi_{(i)}'\bs Z_{[g]}'\bs P_{Z,[g,h]}-\bs\Pi_{(i)}'\bs Z_{[h]}'\bs P_{Z,[h,h]}\bigg)\\
        &\quad\bigg(\sum_{g=1}^G\bs P_{Z,[h,g]}\bs Z_{[g]}\bs\Pi_{(i)}-\bs P_{Z,[h,h]}\bs Z_{[h]}\bs\Pi_{(i)}\bigg)\bigg(\sum_{g=1}^G\bs\Pi_{(j)}'\bs Z'_{[g]}\bs P_{Z,[g,h]}-\bs\Pi_{(j)}'\bs Z'_{[h]}\bs P_{Z,[h,h]}\bigg)\\
        &\quad\bigg(\sum_{g=1}^G\bs P_{Z,[h,g]}\bs Z_{[g]}\bs\Pi_{(j)}-\bs P_{Z,[h,h]}\bs Z_{[h]}\bs\Pi_{(j)}\bigg)\toas 0,
    \end{split}
\end{equation}
by the Cauchy--Schwarz inequality, \cref{eq:HH}, \cref{assit:cljar bound P,assit:cljs Z,assit:cljar nmax}.

Convergence of the expectation of the squared difference between the second and third terms in \cref{eq:Vhat S consistency,eq:V S consistency} follows by similar arguments as for the variance of the AR statistic and the covariance.

Next, consider the expectation of the fourth term in \cref{eq:Vhat S consistency} squared.
\begin{equation}\label{eq:PiZPBeePeta}
    \begin{split}
        &\frac{1}{n^2}\E\big([\bs\Pi'_{(i)}\bs Z'\ddot{\bs P}_{Z}\bs B_{\varepsilon\varepsilon'}\ddot{\bs P}_{Z}\bs\eta_{(j)}]^2\big|\bs Z\big)\\
        &=\frac{1}{n^2}\E\bigg(\sum_{\substack{g_1\neq h_1\neq l\\g_2\neq h_2\neq l}}\bs\Pi'_{(i)}\bs Z_{[g_1]}'\bs P_{Z,[g_1,h_1]}\bs\varepsilon_{[h_1]}\bs\varepsilon_{[h_1]}'\bs P_{Z,[h_1,l]}\bs\eta_{(j),[l]}\\
        &\quad\bs\eta_{(j),[l]}'\bs P_{Z,[l,h_2]}\bs\varepsilon_{[h_2]}\bs\varepsilon_{[h_2]}'\bs P_{Z,[h_2,g_2]}\bs Z_{[g_2]}\bs\Pi\bigg|\bs Z\bigg)\\
        &=\frac{1}{n^2}\E\bigg(\sum_{\substack{g_1\neq h\neq l\\g_2\neq h}}\bs\Pi'_{(i)}\bs Z_{[g_1]}'\bs P_{Z,[g_1,h]}\bs\varepsilon_{[h]}\bs\varepsilon_{[h]}'\bs P_{Z,[h,l]}\bs\Omega^{(j,j)}_{l}\bs P_{Z,[l,h]}\bs\varepsilon_{[h]}\bs\varepsilon_{[h]}'\bs P_{Z,[h,g_2]}\bs Z_{[g_2]}\bs\Pi_{(i)}\\
        &\quad+\sum_{\substack{g_1\neq h_1\neq l\\g_2\neq h_2\neq l\\
        h_1\neq h_2}}\bs\Pi'_{(i)}\bs Z_{[g_1]}'\bs P_{Z,[g_1,h_1]}\bs\Sigma_{h_1}\bs P_{Z,[h_1,l]}\bs\Omega^{(j,j)}_{l}\bs P_{Z,[l,h_2]}\bs\Sigma_{h_2}\bs P_{Z,[h_2,g_2]}\bs Z_{[g_2]}\bs\Pi_{(i)}\bigg|\bs Z\bigg).
    \end{split}
\end{equation}
The first term can be bounded as
\begin{equation}
    \begin{split}
        &\frac{1}{n^2}\E\bigg(\sum_{\substack{g_1\neq h\neq l\\g_2\neq h}}\bs\Pi'_{(i)}\bs Z_{[g_1]}'\bs P_{Z,[g_1,h]}\bs\varepsilon_{[h]}\bs\varepsilon_{[h]}'\bs P_{Z,[h,l]}\bs\Omega^{(j,j)}_{l}\bs P_{Z,[l,h]}\bs\varepsilon_{[h]}\bs\varepsilon_{[h]}'\bs P_{Z,[h,g_2]}\bs Z_{[g_2]}\bs\Pi_{(i)}\bigg|\bs Z\bigg)\\
        &=\frac{1}{n^2}\E\bigg(\sum_{h\neq l}\bigg(\sum_{g\neq h}\bs\Pi'_{(i)}\bs Z_{[g]}'\bs P_{Z,[g,h]}\bigg)\bs\varepsilon_{[h]}\bs\varepsilon_{[h]}'\bs P_{Z,[h,l]}\bs\Omega^{(j,j)}_{l}\\
        &\quad\bs P_{Z,[l,h]}\bs\varepsilon_{[h]}\bs\varepsilon_{[h]}'\bigg(\sum_{g\neq h}\bs P_{Z,[h,g]}\bs Z_{[g]}\bs\Pi_{(i)}\bigg)\bigg|\bs Z\bigg)\\
        &\leq\frac{C}{n^2}\E\bigg(\sum_{h,l=1}^G\bigg(\sum_{g\neq h}\bs\Pi'_{(i)}\bs Z_{[g]}'\bs P_{Z,[g,h]}\bigg)\bs\varepsilon_{[h]}\bs\varepsilon_{[h]}'\bs P_{Z,[h,l]}\\
        &\quad\bs P_{Z,[l,h]}\bs\varepsilon_{[h]}\bs\varepsilon_{[h]}'\bigg(\sum_{g\neq h}\bs P_{Z,[h,g]}\bs Z_{[g]}\bs\Pi_{(i)}\bigg)\bigg|\bs Z\bigg)\\
        &\leq\frac{C}{n^2}\E\bigg(\sum_{h=1}^G\bigg(\sum_{g\neq h}\bs\Pi'_{(i)}\bs Z_{[g]}'\bs P_{Z,[g,h]}\bigg)\bs\varepsilon_{[h]}\bs\varepsilon_{[h]}'\bs\varepsilon_{[h]}\bs\varepsilon_{[h]}'\bigg(\sum_{g\neq h}\bs P_{Z,[h,g]}\bs Z_{[g]}\bs\Pi_{(i)}\bigg)\bigg|\bs Z\bigg)\toas 0,
    \end{split}
\end{equation}
by \cref{assit:cljs eigval,assit:cljar nmax}, \cref{eq:HH} and the same arguments as for \cref{eq:PiZPHPZPi}. The second term in \cref{eq:PiZPBeePeta} follows by similar arguments. The fifth term in \cref{eq:Vhat S consistency} follows by symmetry and the sixth uses largely the same arguments.

The expectation of the seventh term is
\begin{equation}
    \begin{split}
        &\frac{1}{n}\E\bigg(\bigg[\sum_{g\neq h}\bs\Pi_{(i)}'\bs Z_{[g]}'\bs P_{Z,[g,h]}\bs\varepsilon_{[h]}\bs\varepsilon_{[g]}'\bs P_{Z,[g,h]}\bs Z_{[h]}\bs\Pi_{(j)}\bigg]^2\bigg|\bs Z\bigg)\\
        &=\frac{1}{n^2}\sum_{g\neq h}\E(\bs\Pi_{(i)}'\bs Z_{[g]}'\bs P_{Z,[g,h]}\bs\varepsilon_{[h]}\bs\varepsilon_{[g]}'\bs P_{Z,[g,h]}\bs Z_{[h]}\bs\Pi_{(j)}\bs\Pi_{(i)}'\bs Z_{[g]}'\bs P_{Z,[g,h]}\bs\varepsilon_{[h]}\bs\varepsilon_{[g]}'\bs P_{Z,[g,h]}\bs Z_{[h]}\bs\Pi_{(j)}\\
        &\quad+\bs\Pi_{(i)}'\bs Z_{[g]}'\bs P_{Z,[g,h]}\bs\varepsilon_{[h]}\bs\varepsilon_{[g]}'\bs P_{Z,[g,h]}\bs Z_{[h]}\bs\Pi_{(j)}\bs\Pi_{(j)}'\bs Z_{[g]}'\bs P_{Z,[g,h]}\bs\varepsilon_{[h]}\bs\varepsilon_{[g]}'\bs P_{Z,[g,h]}\bs Z_{[h]}\bs\Pi_{(i)}|\bs Z)\\
        &=\frac{1}{n^2}\sum_{g\neq h}\bs\Pi_{(i)}'\bs Z_{[g]}'\bs P_{Z,[g,h]}\bs\Sigma_{h}\bs P_{Z,[h,g]}\bs Z_{[g]}\bs\Pi_{(i)}\bs\Pi_{(j)}'\bs Z_{[h]}'\bs P_{Z,[h,g]}\bs\Sigma_{g}\bs P_{Z,[g,h]}\bs Z_{[h]}\bs\Pi_{(j)}\\
        &\quad+\bs\Pi_{(i)}'\bs Z_{[g]}'\bs P_{Z,[g,h]}\bs\Sigma_{h}\bs P_{Z,[h,g]}\bs Z_{[g]}\bs\Pi_{(j)}\bs\Pi_{(j)}'\bs Z_{[h]}'\bs P_{Z,[h,g]}\bs\Sigma_{g}\bs P_{Z,[g,h]}\bs Z_{[h]}\bs\Pi_{(i)}
    \end{split}
\end{equation}
The first term consists of two quadratic parts and therefore can be bounded using the eigenvalue bounds from \cref{assit:cljs eigval} and the results from above. Write the second term as
\begin{equation}
    \begin{split}
        &\frac{1}{n^2}\sum_{g\neq h}\bs\Pi_{(i)}'\bs Z_{[g]}'\bs P_{Z,[g,h]}\bs\Sigma_{h}\bs P_{Z,[h,g]}\bs Z_{[g]}\bs\Pi_{(j)}\bs\Pi_{(j)}'\bs Z_{[h]}'\bs P_{Z,[h,g]}\bs\Sigma_{g}\bs P_{Z,[g,h]}\bs Z_{[h]}\bs\Pi_{(i)}\\
        &\leq\frac{1}{n^2}\sum_{g\neq h}\big[\bs\Pi_{(i)}'\bs Z_{[g]}'\bs P_{Z,[g,h]}\bs\Sigma_{h}\bs P_{Z,[h,g]}\bs Z_{[g]}\bs\Pi_{(i)}\bs\Pi_{(j)}'\bs Z_{[h]}'\bs P_{Z,[h,g]}\bs\Sigma_{g}\bs P_{Z,[g,h]}\bs Z_{[h]}\bs\Pi_{(j)}\\
        &\quad\bs\Pi_{(j)}'\bs Z_{[g]}'\bs P_{Z,[g,h]}\bs\Sigma_{h}\bs P_{Z,[h,g]}\bs Z_{[g]}\bs\Pi_{(j)}\bs\Pi_{(i)}'\bs Z_{[h]}'\bs P_{Z,[h,g]}\bs\Sigma_{g}\bs P_{Z,[g,h]}\bs Z_{[h]}\bs\Pi_{(i)}\big]^{1/2}\\
        &\leq\frac{C}{n^2}\sum_{g\neq h}\big[\tr(\bs P_{Z,[h,g]}\bs Z_{[g]}\bs\Pi_{(i)}\bs\Pi_{(i)}'\bs Z_{[g]}'\bs P_{Z,[g,h]})\tr(\bs P_{Z,[g,h]}\bs Z_{[h]}\bs\Pi_{(j)}\bs\Pi_{(j)}'\bs Z_{[h]}'\bs P_{Z,[h,g]})\\
        &\quad\tr(\bs P_{Z,[h,g]}\bs Z_{[g]}\bs\Pi_{(j)}\bs\Pi_{(j)}'\bs Z_{[g]}'\bs P_{Z,[g,h]})\tr(\bs P_{Z,[g,h]}\bs Z_{[h]}\bs\Pi_{(i)}\bs\Pi_{(i)}'\bs Z_{[h]}'\bs P_{Z,[h,g]})\big]^{1/2}\\
        &\leq\frac{C\nmax^4}{n^2}\sum_{g\neq h}\big[\tr(\bs P_{Z,[h,g]}\bs P_{Z,[g,h]})\tr(\bs P_{Z,[g,h]}\bs P_{Z,[h,g]})\\
        &\quad\tr(\bs P_{Z,[h,g]}\bs P_{Z,[g,h]})\tr(\bs P_{Z,[g,h]}\bs P_{Z,[h,g]})\big]^{1/2}\toas 0,
    \end{split}
\end{equation}
by \cref{assit:cljs eigval,assit:cljs Z,assit:cljar nmax} and steps similar to those in \cref{eq:PPPP} below.

The eighth term in \cref{eq:Vhat S consistency} is
\begin{equation}
    \begin{split}
        &\frac{1}{n^2}\E\bigg(\bigg[\sum_{g\neq h}\bs\eta_{(i),[g]}'\bs P_{Z,[g,h]}\bs\varepsilon_{[h]}\bs\varepsilon_{[g]}'\bs P_{Z,[g,h]}\bs Z_{[g]}\bs\Pi_{(j)}\bigg]^2\bigg|\bs Z\bigg)\\
        &=\frac{1}{n^2}\sum_{g\neq h}\E(\bs\eta_{(i),[g]}'\bs P_{Z,[g,h]}\bs\varepsilon_{[h]}\bs\varepsilon_{[g]}'\bs P_{Z,[g,h]}\bs Z_{[g]}\bs\Pi_{(j)}\bs\eta_{(i),[g]}'\bs P_{Z,[g,h]}\bs\varepsilon_{[h]}\bs\varepsilon_{[g]}'\bs P_{Z,[g,h]}\bs Z_{[g]}\bs\Pi_{(j)}\\
        &\quad+\bs\eta_{(i),[g]}'\bs P_{Z,[g,h]}\bs\varepsilon_{[h]}\bs\varepsilon_{[g]}'\bs P_{Z,[g,h]}\bs Z_{[g]}\bs\Pi_{(j)}\bs\eta_{(i),[h]}'\bs P_{Z,[h,g]}\bs\varepsilon_{[g]}\bs\varepsilon_{[h]}'\bs P_{Z,[h,g]}\bs Z_{[g]}\bs\Pi_{(j)}|\bs Z)\\
        &=\frac{1}{n^2}\sum_{g\neq h}\E(\bs\eta_{(i),[g]}'\bs P_{Z,[g,h]}\bs\Sigma_{h}\bs P_{Z,[h,g]}\bs\eta_{(i),[g]}\bs\varepsilon_{[g]}'\bs P_{Z,[g,h]}\bs Z_{[h]}\bs\Pi_{(j)}\bs\Pi_{(j)}'\bs Z_{[h]}'\bs P_{Z,[h,g]}\bs\varepsilon_{[g]}\\
        &\quad+\bs\eta_{(i),[g]}'\bs P_{Z,[g,h]}\bs\varepsilon_{[h]}\bs\varepsilon_{[g]}'\bs P_{Z,[g,h]}\bs Z_{[h]}\bs\Pi_{(j)}\bs\eta_{(i),[h]}'\bs P_{Z,[h,g]}\bs\varepsilon_{[g]}\bs\varepsilon_{[h]}'\bs P_{Z,[h,g]}\bs Z_{[g]}\bs\Pi_{(j)}|\bs Z),
    \end{split}
\end{equation}
whose first term has a quadratic form and therefore can be bounded by the usual eigenvalue bounds and a result similar to \cref{eq:eigbound kappa}. Use the Cauchy--Schwarz inequality to write the second term as
\begin{equation}
    \begin{split}
        &\frac{1}{n^2}\sum_{g\neq h}\E(\bs\eta_{(i),[g]}'\bs P_{Z,[g,h]}\bs\varepsilon_{[h]}\bs\varepsilon_{[g]}'\bs P_{Z,[g,h]}\bs Z_{[h]}\bs\Pi_{(j)}\bs\eta_{(i),[h]}'\bs P_{Z,[h,g]}\bs\varepsilon_{[g]}\bs\varepsilon_{[h]}'\bs P_{Z,[h,g]}\bs Z_{[g]}\bs\Pi_{(j)}|\bs Z)\\
        &=
        \frac{1}{n^2}\sum_{g\neq h}\big[\E(\bs\eta_{(i),[g]}'\bs P_{Z,[g,h]}\bs\varepsilon_{[h]}\bs\varepsilon_{[g]}'\bs P_{Z,[g,h]}\bs Z_{[h]}\bs\Pi_{(j)}\bs\Pi_{(j)}'\bs Z_{[h]}'\bs P_{Z,[h,g]}\bs\varepsilon_{[g]}\bs\varepsilon_{[h]}'\bs P_{Z,[h,g]}\bs\eta_{(i),[g]}|\bs Z)\\
        &\quad\E(\bs\eta_{(i),[h]}'\bs P_{Z,[h,g]}\bs\varepsilon_{[g]}\bs\varepsilon_{[h]}'\bs P_{Z,[h,g]}\bs Z_{[g]}\bs\Pi_{(j)}\bs\Pi_{(j)}'\bs Z_{[g]}'\bs P_{Z,[g,h]}\bs\varepsilon_{[h]}\bs\varepsilon_{[g]}'\bs P_{Z,[h,g]}\bs\eta_{(j),[g]}|\bs Z)\big]^{1/2}\toas 0,
    \end{split}
\end{equation}
by \cref{assit:cljar nmax,assit:cljs eigval,eq:PPPP} below. The  ninth term follows by symmetry. 

The triangle and Markov's inequality lead to the conclusion that $\hat{\bs V}_{CLJ}^{S}(\bs\beta_0)\toprob\bs V_{CLJ}^{S}(\bs\beta_0)$.

\section{Additional simulation results}\label{app:extra MC}
\begin{figure}[tp]
    \centering
    \caption{Size of tests with and without cross-fit variance.}
    \label{fig:size CF}
    \includegraphics{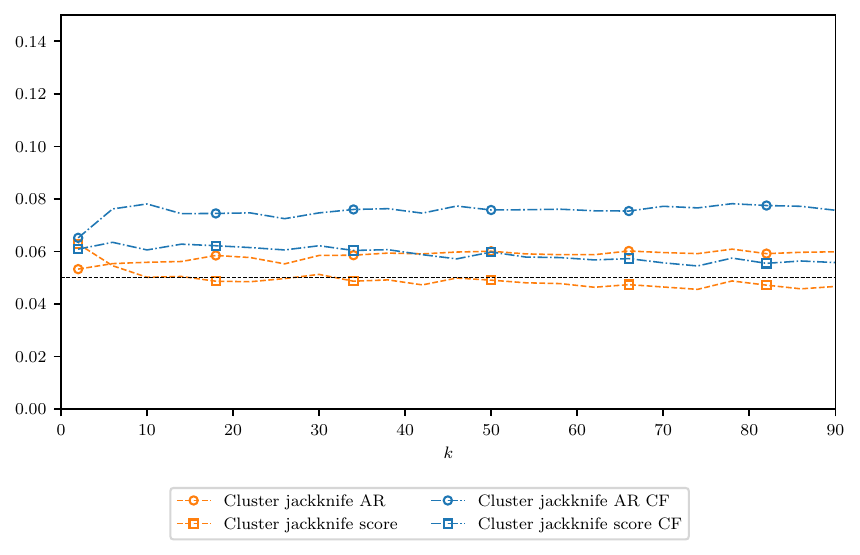}
    \begin{minipage}{0.95\textwidth}
    \footnotesize
    \textit{Note:} This figure shows the rejection rates of the cluster jackknife AR and cluster jackknife score tests with and without cross-fit variance (CF) when testing $\beta=0$ at $5\%$ significance level. The data are clustered with first stage coefficient $\pi=0.1$, the number of clusters $G=100$ and the cluster balancedness governed by $\gamma=1$. $k$ is the number of instruments. The DGP is described in \cref{sec:MC}.
    \end{minipage}
\end{figure}

\Cref{fig:size} in \cref{sec:MC} showed that the cluster jackknife AR test with cross-fit variance is slightly oversized in the simulations. This appendix compares the tests with cross-fit variance to those without to see whether this is a variance estimation problem. \Cref{fig:size CF} gives the results. The DGP is the same as for \cref{fig:size}. The figure shows that the tests without cross-fit variance are size correct, hence confirming the hypothesis that there is a variance estimation problem.

\Cref{sec:model} argued that to determine whether many instruments can be an issue, one should compare the number of instruments with the number of clusters. To test this hypothesis, I varied the number of clusters in the DGP from \cref{sec:MC} while keeping the total number of observations constant, and calculated the size of the cluster AR test, cluster score test, cluster jackknife AR test with and without cross-fit variance, cluster jackknife score test with and without cross-fit variance and a $t$-test based on CJIVE and clustered standard errors. In particular, I calculated the rejection rates when testing $\beta=0$ at $5\%$ significance level in 10\,000 data sets from DGPs with $G$ between $100$ and $750$, $k=30$, $\pi=0.1$ and $\gamma=1$ and the other parameters as for \cref{fig:size}.

\begin{figure}[tp]
    \centering
    \caption{Effect of the number of clusters and cluster balance on size.}
    \label{fig:size clusters}
    \includegraphics{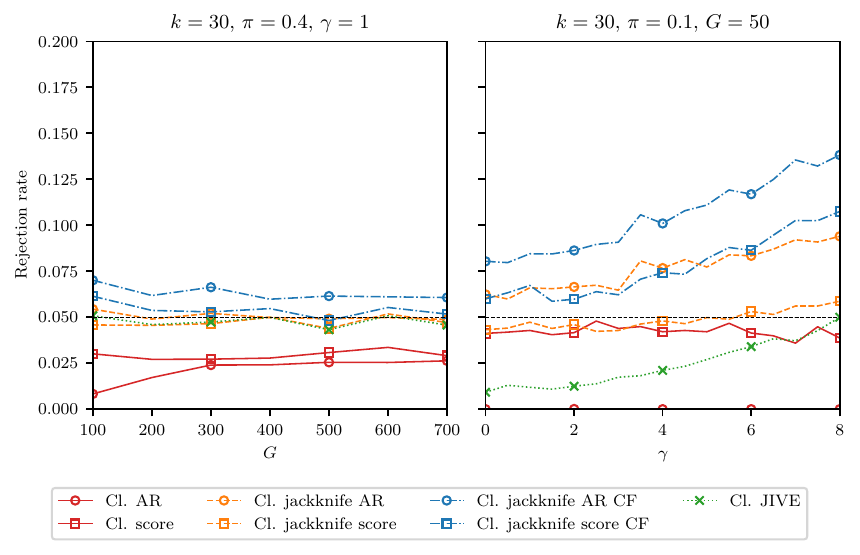}
    \begin{minipage}{0.95\textwidth}
    \footnotesize
    \textit{Note:} Rejection rates of the cluster AR test, cluster score test, cluster jackknife AR test with and without cross-fit variance, cluster jackknife score test with and without cross-fit variance and a $t$-test based on CJIVE and clustered standard errors when testing $\beta=0$ at $5\%$ significance level. The left panel shows the effect of the number of clusters $G$. The data are clustered. The right panel shows the effect increasing the cluster unbalancedness as governed by $\gamma$. $\pi=0.1$ is the first stage coefficient and $k$ the number of instruments. The DGP is described in \cref{sec:MC}.
    \end{minipage}
\end{figure}

The left panel of \cref{fig:size clusters} shows the results for $10\,000$ draws. The figure confirms the hypothesis in the sense that when the ratio of instruments to clusters becomes smaller, while the ratio of instruments to observations stays constant, the cluster AR test becomes less conservative.

Next, I study the effect of a dominant cluster. This could be a violation of \cref{assit:cljar nmax} which is required for the limiting distribution of the cluster jackknife tests. I generate data from the same DGP as before, but increase the cluster unbalancedness parameter $\gamma$ from $0$ to $8$. I also decrease the number of clusters to $G=50$, such that the cluster sizes can be more dispersed. When $\gamma=0$ each cluster has 40 observations. When $\gamma=8$ the smallest cluster contains 1 observation and the largest cluster contains 120 observations.

The right panel of \cref{fig:size clusters} shows the size of the cluster AR test, cluster score test, cluster jackknife AR test with and without cross-fit variance, cluster jackknife score test with and without cross-fit variance and a $t$-test based on CJIVE and clustered standard errors over $10\,000$ data sets. First note that the cluster AR test has zero rejection rates. The rejection rates of the other tests are as before for small $\gamma$ and, with the exception of the cluster score and the cluster jackknife score without cross-fit variance, increase with $\gamma$.

\begin{figure}[tp]
    \centering
    \caption{Power of tests with and without cross-fit variance.}
    \label{fig:power clustered CF}
    \includegraphics{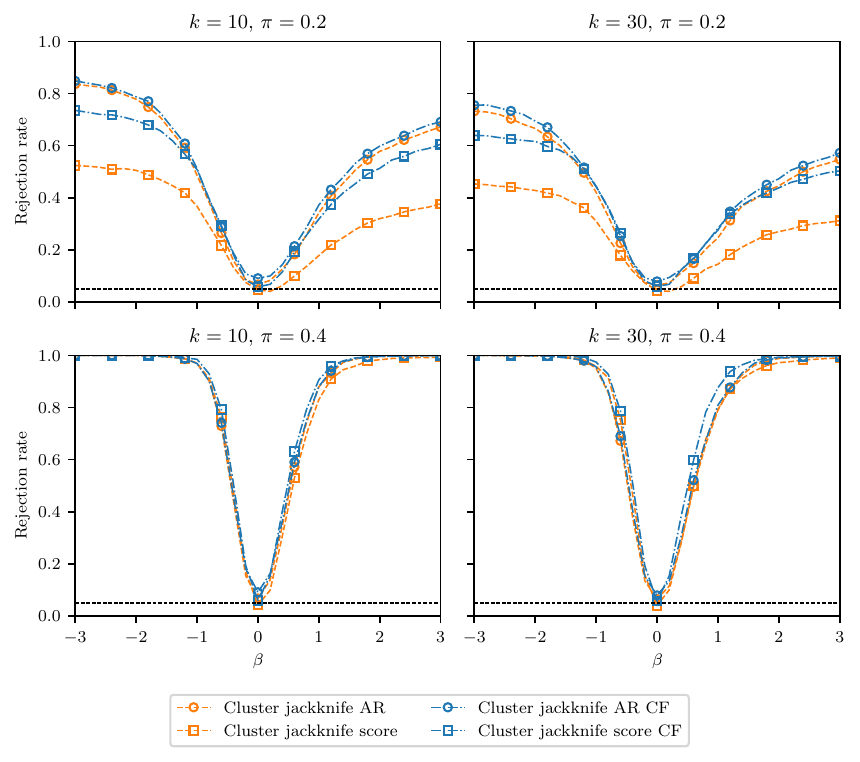}
    \begin{minipage}{0.95\textwidth}
    \footnotesize
    \textit{Note:} Rejection rates of the cluster jackknife AR test with and without cross-fit (CF) variance and cluster jackknife score test with and without cross-fit variance when testing $\beta=0$ at $5\%$ significance level when the true value of $\beta$ in the DGP varies. $k$ is the number of instruments and $\pi$ is the first stage coefficient. The DGP is described in \cref{sec:MC}.
    \end{minipage}
\end{figure}

Finally, \cref{fig:power clustered CF} compares the power of the cluster jackknife tests with and without cross-fit variance. The power gain is substantial when the instruments are weak and, as expected, when the tested $\beta$ is far from the true value. For stronger instruments the difference in power is small. Furthermore, note that especially the cluster jackknife score test benefits from a cross-fit variance estimator.

\section{Details of the application to \texorpdfstring{\citet{dube2020queens}}{Dube and Harish (2020)}}\label{app:application}
\Cref{tab:DH20} shows the exact bounds of the confidence sets for the effect of queenly reign on the probability that a polity is at war relative to polities led by a king. 

\begin{landscape}
\begin{table}[ht!]
\centering
\caption{95\% confidence set for the effect of queenly reign on war.}
\label{tab:DH20}
\begin{tabular}{lrrrrrrrr}
\toprule
 & \multicolumn{2}{l}{Cl. AR} & \multicolumn{2}{l}{Cl. score} & \multicolumn{2}{l}{Cl. jack. AR CF} & \multicolumn{2}{l}{Cl. jack. score CF} \\
 & Lower & Upper & Lower & Upper & Lower & Upper & Lower & Upper \\
\midrule
FBM, Sis & 0.107 & 0.829 & 0.086 & $\infty$ & 0.142 & 0.815 & 0.107 & 0.842 \\
FBM, Sis, Sis $\times$ No children &  \multicolumn{2}{c}{Empty}  & -0.070 & $\infty$ & 0.251 & 0.511 & 0.100 & 0.658 \\
FBM, Sis, Sis $\times$ FMB & 0.119 & 0.443 & -$\infty$ & $\infty$ & 0.140 & 0.576 & 0.026 & 0.656 \\
FBM, Sis, FBM $\times$ Two children & 0.049 & 0.931 & 0.056 & $\infty$ & 0.092 & 0.736 & 0.100 & 0.773 \\
Full & 0.017 & 0.454 & 0.004 & $\infty$ & 0.043 & 0.501 & 0.041 & 0.514 \\
\bottomrule
\end{tabular}\vspace{0.2cm}
\begin{minipage}{0.95\textwidth}
 \footnotesize \textit{Note}: 95\% confidence interval for the effect of queenly reign on the probability of being at war relative to polities led by a king. The confidence intervals are based on inverting the cluster AR test, the cluster score test, the cluster jackknife AR, the cluster jackknife score, both with cross-fit (CF) variance. The data come from \citet{dube2020queens} and are grouped into 176 clusters. The instruments are whether the previous sovereign had a firstborn male child (FBM), whether the previous sovereign had a sister (Sis) and interactions of these with each other and indicators of whether the previous sovereign had no or two children. The control variables are the same as in columns 1, 2, 4 and 5 of Table A5 in \citet{dube2020queens}. The full model pools the instruments and control variables of the other models.
 \end{minipage}
\end{table}
\end{landscape}

\end{appendices}

\bibliographystyle{chicago}
\bibliography{Bib}

\end{document}